\PassOptionsToPackage{dvipsnames,table}{xcolor}
\documentclass[sigplan,nonacm]{acmart}

\usepackage{xcolor}
\usepackage{multirow}
\usepackage{pifont}
\usepackage[most]{tcolorbox}
\usepackage{enumitem}
\usepackage{subcaption}
\begin{document}

\newcommand{\iu}{{j}}

\newcommand{\myparatight}[1]{\vspace{0.5ex}\noindent\textbf{#1.~~}}

\newcommand\name{\mbox{\sc AIR-LLM}}
\newcommand\namebf{\mbox{\sc\textbf{AIR-LLM}}}

\definecolor{colorEtoile}{HTML}{1F77B4}
\definecolor{colorMunich}{HTML}{9467BD}

\newcommand{\sceneCell}[2]{\textbf{\textcolor{#1}{#2}}}
\newcommand{\cellEtoile}[1]{\sceneCell{colorEtoile}{#1}}
\newcommand{\cellMunich}[1]{\sceneCell{colorMunich}{#1}}

\newcommand{\sceneMethod}[2]{\textbf{{\namebf} @ \textcolor{#1}{#2}}}
\newcommand{\methodEtoile}{\sceneMethod{colorEtoile}{Paris}}
\newcommand{\methodMunich}{\sceneMethod{colorMunich}{Munich}}

\newcommand{\OptOrig}{{\textsf{OPT-Orig}}}
\newcommand{\OptChan}{{\textsf{OPT-Chan}}}
\newcommand{\OptApprox}{{\textsf{OPT-Approx}}}

\newcommand{\myCeil}[1]{\left\lceil{#1}\right\rceil}
\newcommand{\myFloor}[1]{\left\lfloor{#1}\right\rfloor}
\newcommand{\myTrans}[1]{{#1}^{\top}}
\newcommand{\myGauss}{\mathcal{N}}

\newcommand{\specIdx}{k}
\newcommand{\freqSub}{\Delta f}
\newcommand{\band}{B}
\newcommand{\antTxNum}{A^{t}}
\newcommand{\antTxIdx}{a^{t}}
\newcommand{\antRxNum}{A^{r}}
\newcommand{\antRxIdx}{a^{r}}
\newcommand{\signalSpecTx}{S^{t}}
\newcommand{\signalSpecRx}{S^{r}}
\newcommand{\fftSize}{N_{\textrm{fft}}}
\newcommand{\channel}{H}

\newcommand{\weightVec}{\mathbf{w}}
\newcommand{\inputVec}{\mathbf{x}}
\newcommand{\outputVec}{\mathbf{y}}
\newcommand{\blockSizeWeight}{L_w}
\newcommand{\blockSizeInput}{L_x}
\newcommand{\blockSizeOutput}{L}
\newcommand{\weightSpec}{W}
\newcommand{\inputSpec}{X}
\newcommand{\outputSpec}{Y}
\newcommand{\blockNum}{D}
\newcommand{\blockIdx}{d}

\newcommand{\weightFC}{\mathbf{W}}
\newcommand{\inputFC}{\mathbf{X}}
\newcommand{\outputFC}{\mathbf{Y}}
\newcommand{\inputSize}{N}
\newcommand{\outputSize}{M}
\newcommand{\outputIdx}{m}
\newcommand{\weightCONV}{\mathbf{W}^{\textrm{(CONV)}}}
\newcommand{\inputCONV}{\mathbf{X}^{\textrm{(CONV)}}}
\newcommand{\outputCONV}{\mathbf{Y}^{\textrm{(CONV)}}}
\newcommand{\kernelSize}{K}
\newcommand{\imageSize}{I}
\newcommand{\convRatio}{\eta}

\newcommand{\weightMat}{\mathbf{W}}
\newcommand{\outputMat}{\mathbf{Y}}

\newcommand{\antNum}{A}
\newcommand{\antIdx}{a}
\newcommand{\precoder}{P_{\textrm{pre}}}
\newcommand{\postcoder}{P_{\textrm{post}}}
\newcommand{\weightMatRx}{\mathbf{W}^{\prime}}
\newcommand{\weightVecRx}{\mathbf{w}^{\prime}}
\newcommand{\inputVecPre}{\mathbf{x}^{\prime}}
\newcommand{\outputMatPostEst}{\widehat{\mathbf{Y}}^{\prime}}
\newcommand{\outputVecPostEst}{\widehat{\mathbf{y}}^{\prime}}
\newcommand{\outputVecSumEst}{\widehat{\mathbf{y}}^{\prime\prime}}
\newcommand{\outputMatSumEst}{\widehat{\mathbf{Y}}^{\prime\prime}}
\newcommand{\unitNum}{C}
\newcommand{\unitIdx}{c}

\newcommand{\postcoderFlat}{\Tilde{P}_{\textrm{post}}}
\newcommand{\channelInvFlat}{\Tilde{H}^{-1}}
\newcommand{\channelBlock}{\bar{H}}

\newcommand{\precoderInit}{P_{\textrm{pre}}^{(0)}}
\newcommand{\precoderOpt}{P_{\textrm{pre}}^{\star}}
\newcommand{\postcoderInit}{P_{\textrm{post}}^{(0)}}
\newcommand{\postcoderOpt}{P_{\textrm{post}}^{\star}}
\newcommand{\postcoderFlatInit}{\Tilde{P}_{\textrm{post}}^{(0)}}

\newcommand{\airtimeMvm}{T}
\newcommand{\airtimeMac}{\tau}

\newcommand{\energyMvm}{E}
\newcommand{\energyMvmPre}{E_{\textrm{pre}}}
\newcommand{\energyMvmIFFT}{E_{\textrm{ifft}}}
\newcommand{\energyMvmDAC}{E_{\textrm{dac}}}
\newcommand{\energyMvmADC}{E_{\textrm{adc}}}
\newcommand{\energyMvmFFT}{E_{\textrm{fft}}}
\newcommand{\energyMvmPost}{E_{\textrm{post}}}
\newcommand{\energyMvmSum}{E_{\textrm{sum}}}

\newcommand{\outputVecEst}{\hat{\mathbf{y}}}

\newcommand{\userNum}{U}

\newcommand{\noise}{\sigma}

\definecolor{colorYes}{HTML}{2CA02C}
\definecolor{colorNo}{HTML}{D62728}
\newcommand{\cmark}{\textcolor{colorYes}{\ding{51}}}
\newcommand{\xmark}{\textcolor{colorNo}{\ding{55}}}

\title[{\name}]{{\namebf}: Broadcasting \texorpdfstring{\underline{AI}}{AI} Weights over \texorpdfstring{\underline{R}}{R}adio for Memory-Free Edge LLM Inference via RF Computing}

\author{Zhihui Gao}
\affiliation{%
  \institution{Massachusetts Institute of Technology}
  \city{Cambridge, MA}
  \country{USA}}

\author{Tingjun Chen}
\affiliation{%
  \institution{Duke University}
  \city{Durham, NC}
  \country{USA}}

\author{Dirk Englund}
\affiliation{%
  \institution{Massachusetts Institute of Technology}
  \city{Cambridge, MA}
  \country{USA}}

\renewcommand{\shortauthors}{Gao et al.}

\begin{abstract}
Next-generation large language models (LLMs) are expanding from the cloud to ubiquitous edge devices.
However, edge devices typically either lack the memory to store increasingly large LLM weights or, even with enough memory, spend unaffordable energy on loading the weights.
This raises our question: can an edge device run an LLM without storing or loading its weights, but receive them over the air and consume them on the fly?
Inspired by wireless broadcasting, we present {\name}, an LLM inference architecture for edge devices, which is composed of:
(\emph{i}) a central radio (e.g., 5G base stations) that broadcasts the LLM weights into the air, and (\emph{ii}) the edge user that receives the weights and completes the general matrix-vector multiplication (GEMV) of LLM inference directly in the radio frequency (RF) domain using RF mixers.
To further shorten the airtime, {\name} exploits MIMO spatial multiplexing and proposes an energy-efficient precoder--postcoder pair on the edge to calibrate its own wireless channel.
Since the central radio stays user-unaware, {\name} is user-scalable so that one broadcast serves unlimited users within its coverage.
We implement {\name} on the NVIDIA Sionna ray-traced channels of two real-world urban scenes and the profiling of a real RF mixer.
With a WikiText-2 perplexity degradation of {4.0\%} on LLaMA-3.1-8B, {\name} saves the energy by {157.7}$\times$/{40.4}$\times$ against the FP16 and weight-only quantization baselines; with {20} users, its airtime is {104.1}$\times$/{26.0}$\times$ shorter, respectively.
\end{abstract}

\keywords{LLM inference, edge computing, RF computing, RF mixer, weight broadcasting, MIMO communication}

\maketitle

\section{Introduction}

Trillions of edge devices today are embracing emerging machine learning (ML) techniques for a wide range of applications, such as continuous health monitoring~\cite{abbaspourazad2024biosignals, pan2025cgmm}, on-device speech recognition~\cite{radford2023whisper, yazdani2016asr}, and cooperative autonomous driving~\cite{hu2022where2comm, zhang2021emp}.
Among these, the emerging large language models (LLMs) provide a unique capability, allowing understanding, reasoning, and generating natural language~\cite{grattafiori2024llama3, touvron2023llama2, jiang2023mistral, gemmateam2026gemma4}.
However, the usage of LLMs on the edge mostly relies on cloud computing, where the locally generated prompts are uploaded to a cloud server, and the responses are downloaded afterwards.
This again exposes two typical concerns of cloud computing: the privacy of the sensitive prompts, as well as the round-trip latency, especially when the number of edge users is scaled up~\cite{kang2017neurosurgeon, xu2025llmnpu}.

The key challenge that hinders such an on-device deployment lies in the daunting model size of today's LLMs.
For example, LLaMA-3.1-8B~\cite{grattafiori2024llama3} occupies {16}\thinspace{GB} in FP16, and still {4}\thinspace{GB} after 4-bit weight-only quantization (WoQ)~\cite{frantar2023gptq, lin2024awq}.
Such a model size stresses an edge device in two stages.
First, its memory is typically too small to hold the whole model, so a local deployment has to keep downloading the weights and discarding them afterwards.
Second, even if an edge device could store such an LLM entirely, the energy of loading the weights from memory to the computing unit is unaffordable.
This is because LLM inference is memory-bounded, especially during the decoding phase, where fetching one weight feeds only two floating-point operations (FLOPs)~\cite{guo2025survey}.
The weight movement therefore dominates the energy, much more than the arithmetic it feeds~\cite{antepara2025energy, nvidia2024blackwell}.

To overcome this, various computing paradigms have been proposed recently.
For example, in-memory computing completes the computing in analog within the memory itself, on a ReRAM~\cite{shafiee2016isaac, chi2016prime, wan2022neurram}, PCM~\cite{khaddamaljameh2022hermes, legallo2023multicore}, or SRAM~\cite{liu2026ouroboros} crossbar, so that the weights are never fetched.
Over-the-air computing~\cite{nazer2007computation, zhu2020broadband}, in turn, superposes the signals of many transmitters over the wireless channel, so that the summation is completed by the channel itself.
However, in both cases the size of the general matrix-vector multiplication (GEMV) is frozen and limited by the hardware complexity, e.g., a crossbar of {128}$\times${128} cells~\cite{shafiee2016isaac} and an array of a few dozen antennas~\cite{reus2023airfc}.
Such a limited GEMV size is naturally incompatible with the LLM whose GEMV sizes keep increasing.

\begin{figure}[!t]
    \centering
    \includegraphics[width=0.99\columnwidth]{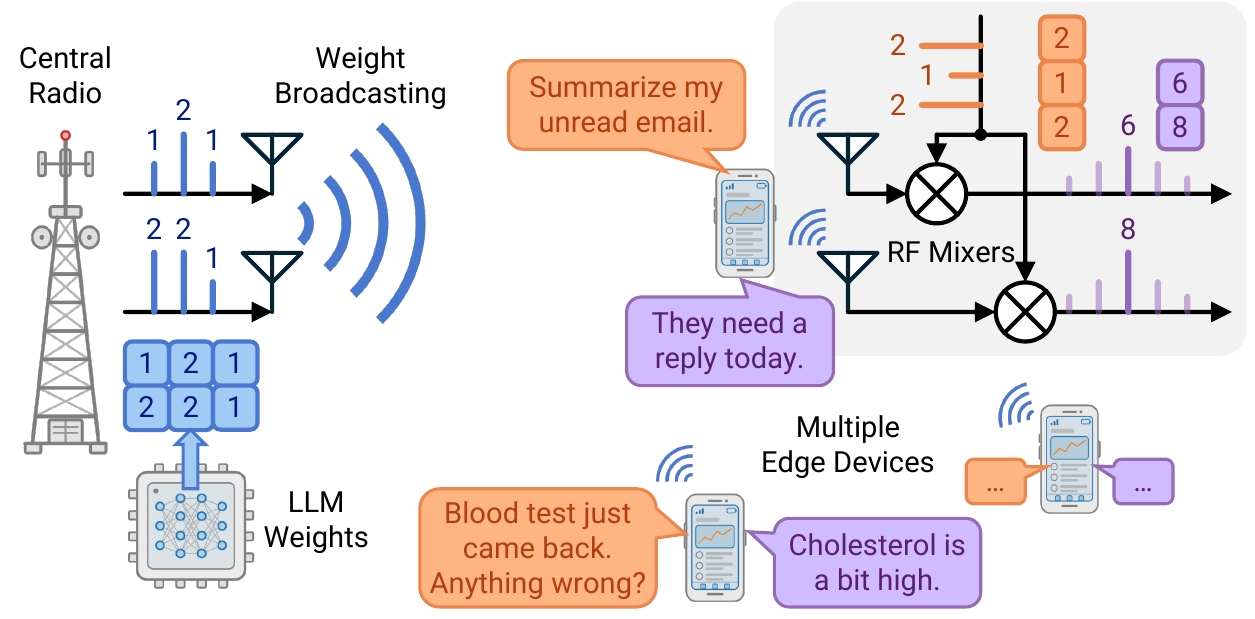}
    \caption{System overview of {\name}: a central radio frequency-encodes the LLM weights onto an OFDM waveform and broadcasts it over the air to serve any edge user within the coverage; the edge user drives the LO with its private activations, and the RF mixer's output directly carries the GEMV outputs. The LLM inference is completed once the weight waveform arrives, without ever storing it in memory.}
    \vspace{-12pt}
    \label{fig:system-introduction}
\end{figure}

\begin{table*}[t]
    \centering
    \caption{The comparison of {\name} and existing edge computing paradigms.}
    \label{tab:comparison}
    \small
    \setlength{\tabcolsep}{4pt}
    \renewcommand{\arraystretch}{1.15}
    \vspace{-6pt}
    \begin{tabular}{lcccccc}
    \toprule
    \textbf{Computing Paradigm} & \shortstack{\textbf{Memory-}\\\textbf{free}} & \shortstack{\textbf{Energy-}\\\textbf{efficient}} & \shortstack{\textbf{Prompt-}\\\textbf{private}} & \shortstack{\textbf{Wireless}\\\textbf{(untethered)}} & \shortstack{\textbf{User-}\\\textbf{scalable}} & \shortstack{\textbf{LLM-scale}\\\textbf{GEMV}} \\
    \midrule
    Cloud computing~\cite{kwon2023vllm, stojkovic2025dynamollm, kang2017neurosurgeon} & \cmark & \cmark & \xmark & \cmark & \xmark & \cmark \\
    Edge computing~\cite{frantar2023gptq, lin2024awq, song2024powerinfer, alizadeh2024llmflash} & \xmark & \xmark & \cmark & \cmark & \cmark & \cmark \\
    In-memory computing~\cite{shafiee2016isaac, wan2022neurram, legallo2023multicore} & \cmark & \cmark & \cmark & \xmark & \cmark & \xmark \\
    Photonic computing~\cite{shen2017deep, zhong2023lightning, sludds2022delocalized} & \xmark & \cmark & \cmark & \xmark & \cmark & \xmark \\
    Over-the-air computing~\cite{nazer2007computation, zhu2020broadband, reus2023airfc} & \xmark & \xmark & \cmark & \cmark & \xmark & \xmark \\
    \midrule
    \textbf{{\namebf}} & \cmark & \cmark & \cmark & \cmark & \cmark & \cmark \\
    \bottomrule
    \end{tabular}
    \vspace{-8pt}
\end{table*}

Our motivation lies in the wireless network these edge users already live in.
Taking 5G networks as an example, a 5G base station already establishes wireless links to each edge user under its coverage.
For these users, accessing the LLM weights is read-only, and thus a purely downlink process; once they run the same LLM, it is exactly what broadcasting was invented for: one transmission simultaneously serving every user in the coverage.
Moreover, we employ radio-frequency (RF) computing~\cite{gao2026disaggregated, gao2026radio}, which completes the computing in RF via the passive RF mixer that every receiver already carries.
The computing is thereby finished in analog once the broadcast signal arrives, without the weights ever being digitized into the local memory.

As shown in Fig.~\ref{fig:system-introduction}, we present {\name}, an LLM inference architecture with (\emph{i}) a central radio (e.g., a 5G base station) that frequency-encodes the LLM weights onto the subcarriers of an orthogonal frequency-division multiplexing (OFDM) waveform and broadcasts it, and (\emph{ii}) any edge device in the coverage receives that waveform and computes on it inside its own RF front-end.
The key computing component is the RF mixer every wireless device already carries: it multiplies its two inputs in the time domain, corresponding to a linear convolution in the frequency domain.
Specifically, driving its local oscillator (LO) with a waveform encoding the activations therefore convolves the arriving weight spectrum with the activation spectrum, and the very middle subcarrier of this convolution is exactly the inner product of the weight and activation vectors.
A comparison of existing computing paradigms is presented in Tab.~\ref{tab:comparison}.

On top of this principle, {\name} makes two designs.
First, it exploits $\antNum \times \antNum$ multiple-input multiple-output (MIMO) spatial multiplexing, which delivers $\antNum$ weight streams at a time and thus shortens the airtime by $\antNum\times$.
Second, it calibrates the MIMO channel with $\unitNum$ calibration units, each pairing a per-subcarrier precoder and a per-antenna postcoder, which compensates for the channel at a cost far below an explicit inversion, keeping the energy low.

Delivering the weights this way also fits broadcasting better than conventional communication in two aspects.
First, {\name} needs no handshake with users, as the computation is inherently tolerant to the errors introduced by thermal noise.
In addition, its weight waveform stays identical regardless of the users' SNRs, so no per-user modulation and coding scheme (MCS) is needed either.

We implement {\name} on ray-traced channels of two urban scenes, Paris and Munich, and profile a real RF mixer, running nine LLMs from {1B} to {70B} parameters end-to-end with every weight-activation GEMV in RF computing.
On LLaMA-3.1-8B and a selected user in the scene of Paris, {\name} degrades the WikiText-2 perplexity (PPL) by only {4.0\%}, while cutting the per-token energy by {157.7}$\times$/{40.4}$\times$ against FP16/4-bit weight baselines with remote weight storage, and by {51.5}$\times$/{13.9}$\times$ against fetching them from local memory; with {20} users, its per-token airtime is {104.1}$\times$/{26.0}$\times$ shorter than FP16 and 4-bit WoQ.

We summarize the contribution of {\name} as below.
\begin{itemize}[leftmargin=*, topsep=2pt, itemsep=2pt]
    \item We propose {\name}, a new computing paradigm for LLM inference on the edge, which is \emph{memory-access-free}, \emph{user-scalable}, and \emph{energy-efficient}.
    \item We design an RF computing engine that shortens the airtime with MIMO spatial multiplexing, while maintaining the computing accuracy and energy efficiency.
    \item We evaluate {\name} on Sionna ray-tracing and real RF mixer profiling, showing its near-lossless PPL on LLM tasks with orders-of-magnitude less energy and airtime.
\end{itemize}

\section{Preliminaries}

\subsection{OFDM and MIMO Channel Model}

\myparatight{OFDM signal structure}
The signals are based on orthogonal frequency-division multiplexing (OFDM), characterized by the FFT size $\fftSize$, i.e., the number of subcarriers in the frequency domain, and the occupied bandwidth $\band$. The resulting subcarrier spacing is $\freqSub = \band / \fftSize$.

\myparatight{MIMO signal notation}
We consider an $\antTxNum \times \antRxNum$ MIMO system, where $\antTxNum$ denotes the number of transmitting antennas on the central radio and $\antRxNum$ denotes the number of receiving antennas on each edge device.
On the central radio side, the transmitted signals in the frequency domain are denoted by a complex-valued matrix $\signalSpecTx \in \mathbb{C}^{\antTxNum \times \fftSize}$, whose $\antTxIdx$-th row $\signalSpecTx[\antTxIdx, :]$ gives the signal transmitted by the $\antTxIdx$-th antenna.
The received signals in the frequency domain are similarly denoted by $\signalSpecRx \in \mathbb{C}^{\antRxNum \times \fftSize}$, whose $\antRxIdx$-th row $\signalSpecRx[\antRxIdx, :]$ gives the signal received by the $\antRxIdx$-th antenna.

\myparatight{Wireless channel and received signal}
The wireless channel is defined as a three-dimensional complex-valued matrix $\channel \in \mathbb{C}^{\antTxNum \times \antRxNum \times \fftSize}$, where $\channel[\antTxIdx, \antRxIdx, :]$ is the channel response between the $\antTxIdx$-th transmitting antenna and the $\antRxIdx$-th receiving antenna across all subcarriers.
The received signal is then given by
\begin{align}
    \signalSpecRx[\antRxIdx, \specIdx] = \sum_{\antTxIdx=0}^{\antTxNum-1} \channel[\antTxIdx, \antRxIdx, \specIdx] \cdot \signalSpecTx[\antTxIdx, \specIdx] + \frac{\noise}{\sqrt{\fftSize}} \cdot \myGauss,
    \label{eq: preliminaries mimo channel}
\end{align}
where $\noise^2$ is the sum power of the additive white Gaussian noise (AWGN) across all the $\fftSize$ subcarriers, and $\myGauss$ is the i.i.d. complex-valued standard Gaussian distribution. Note that we assume that within a single edge device, the noise power is identical for all the antennas and subcarriers.
Conventionally, the channel $\channel$ can either be pre-calibrated on the central radio side or post-calibrated on the edge device side.
Specifically for the edge-side post-calibration, given the estimated channel $\widehat{\channel}$, a per-subcarrier (pseudo-)inverse matrix is performed for each $\widehat{\channel}[:, :, \specIdx]$ to recover the transmitted signal by
\begin{align}
    \widehat{\signalSpecTx}[:, \specIdx] = \left(\widehat{\channel}[:, :, \specIdx]\right)^{-1} \cdot \signalSpecRx[:, \specIdx].
    \label{eq: preliminaries mimo calibration}
\end{align}
This recovery process requires $8\antTxNum \antRxNum \fftSize$ floating-point operations (FLOPs) on the edge device.

\subsection{Analog GEMV via RF Mixers}
\label{sec:analog-gemv}

An RF mixer is a fundamental component on edge devices for up/down conversion. Essentially, it performs analog multiplication in the time domain, corresponding to the linear convolution in the frequency domain, where OFDM signals reside.
Such a convolution operation can be leveraged for GEMV in ML inference.

\myparatight{Complex-valued linear convolution}
The complex-valued nature of wireless signals allows an RF mixer to perform, in analog, the complex-valued linear convolution (denoted by $\ast$) between two input signals of equal subcarrier spacing $\freqSub$.
To simplify, we assume they have the same FFT size ($\fftSize$) so that we denote the frequency-domain input activation signal by $\inputSpec \in \mathbb{C}^{\fftSize}$, and the weight signal by $\weightSpec \in \mathbb{C}^{\fftSize}$.
This process yields the output activation signal $\outputSpec \in \mathbb{C}^{2\fftSize-1}$ as
\begin{align}
    \outputSpec[\specIdx] = \left(\weightSpec \ast \inputSpec\right)[\specIdx] = \sum_{\specIdx^{\prime}} \weightSpec[\specIdx^{\prime}] \cdot \inputSpec[\specIdx - \specIdx^{\prime}].
    \label{eq: preliminaries linear convolution}
\end{align}

\myparatight{Leveraging linear convolution for GEMVs/IPs}
We examine the real-valued GEMV that ML inference is built upon, formulated by $\outputFC = \weightFC \cdot \inputFC$ with $\weightFC \in \mathbb{R}^{\outputSize \times \inputSize}$, $\inputFC \in \mathbb{R}^{\inputSize}$, and $\outputFC \in \mathbb{R}^{\outputSize}$.
This GEMV can be broken down into $\outputSize$ independent $\inputSize$-element inner products (IPs), each of which takes $2\inputSize-1$ FLOPs.
Since the linear convolution in {\eqref{eq: preliminaries linear convolution}} is complex-valued whereas the GEMV is real-valued, a single pair of input and weight signals carries \emph{two} such IPs.
Specifically, for the IP pair given by the $\outputIdx$-th and the $\outputIdx^{\prime}$-th rows of $\weightFC$, we let the FFT size be $\fftSize = \inputSize$, and let the input signal directly carry $\inputFC$, whereas the weight signal carries these two rows in its real and imaginary parts, both in a reversed subcarrier order, i.e.,
\begin{align}
    \inputSpec[\specIdx] &= \inputFC[\specIdx], \notag \\
    \weightSpec_{\outputIdx, \outputIdx^{\prime}}[\specIdx] &= \weightFC[\outputIdx, \inputSize-1-\specIdx] + \iu \cdot \weightFC[\outputIdx^{\prime}, \inputSize-1-\specIdx].
    \label{eq: preliminaries gemv mapping}
\end{align}
Substituting {\eqref{eq: preliminaries gemv mapping}} into {\eqref{eq: preliminaries linear convolution}}, the middle subcarrier of the output signal, $\outputSpec[\inputSize-1]$, is exactly the one that accumulates all the $\inputSize$ element-wise products, i.e.,
\begin{align}
    \outputSpec_{\outputIdx, \outputIdx^{\prime}}[\inputSize-1] &= \sum_{\specIdx=0}^{\inputSize-1} \left( \weightFC[\outputIdx, \specIdx] + \iu \cdot \weightFC[\outputIdx^{\prime}, \specIdx] \right) \cdot \inputFC[\specIdx] \notag \\
    &= \outputFC[\outputIdx] + \iu \cdot \outputFC[\outputIdx^{\prime}],
    \label{eq: preliminaries gemv output}
\end{align}
so that the two IPs are read off the real and the imaginary parts of this same subcarrier without interfering with each other, while the remaining $2\inputSize-2$ subcarriers are discarded.
Therefore, the output can be extracted by a low-pass filter (LPF) of a single-subcarrier bandwidth $\freqSub$ so that it can be digitized with far fewer samples to save the energy consumption on the ADCs.
This process is repeated until all the $\outputSize$ IPs are covered, taking $\myCeil{\outputSize/2}$ weight signals $\weightSpec_{\outputIdx, \outputIdx^{\prime}}$ in total, all sharing the same input signal $\inputSpec$.

Beyond GEMV, this linear convolution maps even more naturally onto convolutional layers (Appendix~\ref{app:conv-mapping})~\cite{gao2026radio}.
\section{RF Computing Engine}

\begin{figure*}[!t]
    \centering
    \includegraphics[width=1.98\columnwidth]{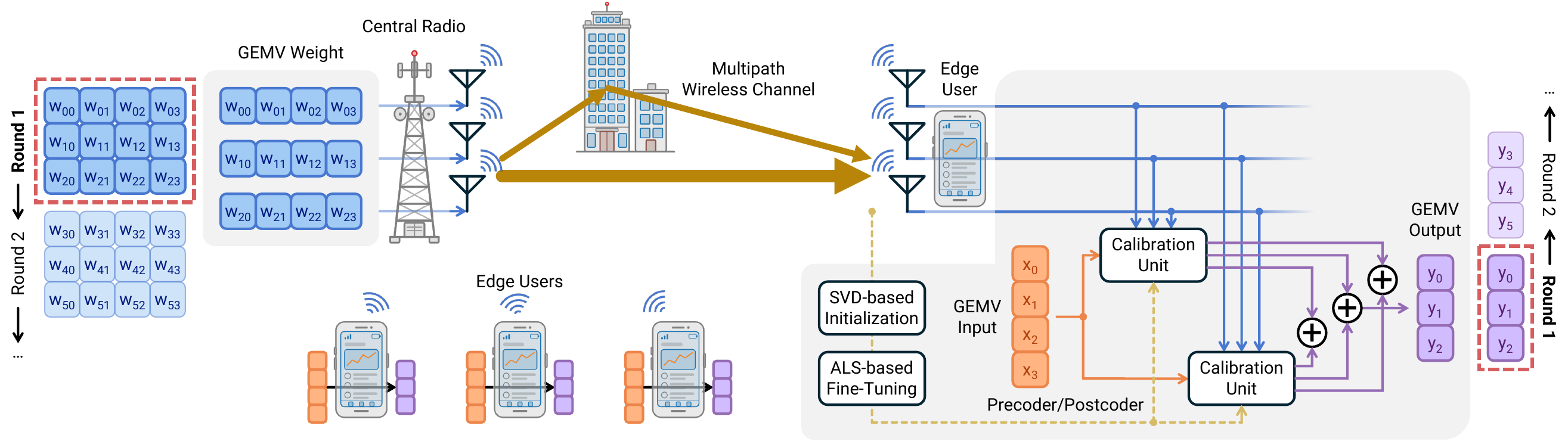}
    \caption{The overview of {\name}'s RF computing engine, which completes $\antNum$ IPs of the GEMV per round (or $2\antNum$ if real-valued).
    The central radio broadcasts the weights of these IPs over its $\antNum$ antennas, without user-specific precoding.
    Each edge user then feeds the $\antNum$ received weight waveforms, together with its own GEMV input, into $\unitNum$ calibration units, whose precoders and postcoders are derived from that user's own channel by an SVD-based initialization and an ALS-based fine-tuning.
    Summing the outputs of the $\unitNum$ units gives the GEMV output of this round.}
    \vspace{-12pt}
    \label{fig:system-overview}
\end{figure*}

In this section, we present the RF computing engine of {\name} on the edge device based on MIMO, as shown in Fig.~\ref{fig:system-overview}, where {\name} calibrates the channel of the broadcast weights with a pair of precoder and postcoder that are compatible with the RF mixer-driven analog GEMV.

\subsection{Problem Formulation}

We first introduce the basic configuration of the MIMO channel modeling, and then formulate the problem of designing the edge device hardware.

\myparatight{GEMV decoupling and IP parallelization}
In {\name}, we consider an $\antNum \times \antNum$ MIMO system for any central radio and edge device pair.
This supports up to $\antNum$ parallel spatial streams, each carrying an IP pair of {\eqref{eq: preliminaries gemv mapping}}, so that one round of {\name} completes $2\antNum$ IPs and a GEMV takes $\myCeil{\outputSize/(2\antNum)}$ rounds.
For these $2\antNum$ IPs per round, we denote the weight signals by $\weightMat = \{ \weightVec_{\antIdx} \}$ and the shared input signal by $\inputVec$, all spanning the $\fftSize = \inputSize$ subcarriers, i.e., $\weightVec_{\antIdx}, \inputVec \in \mathbb{C}^{\fftSize}$; their outputs are denoted by $\outputMat = \{ \outputVec_{\antIdx} \}$ with $\outputVec_{\antIdx} \in \mathbb{C}$.

\myparatight{MIMO channel modeling}
In {\name}, we let the central radio broadcast signals $\weightMat$ with $\weightVec_{\antIdx}$ by the $\antIdx$-th antenna without precoding.
For a specific edge device with a channel $\channel \in \mathbb{C}^{\antNum \times \antNum \times \fftSize}$, the received signals are denoted by $\weightMatRx = \{ \weightVecRx_{\antIdx} \}$ with $\weightVecRx_{\antIdx} \in \mathbb{C}^{\fftSize}$ received by the $\antIdx$-th antenna.
This signal transmission process directly follows the MIMO channel model in {\eqref{eq: preliminaries mimo channel}}, with $\antTxNum = \antRxNum = \antNum$.

\myparatight{Channel calibration challenges}
The design of the central radio is edge device-unaware, i.e., does not incorporate any information specific to a particular edge device (e.g., its channel), so that its broadcasting can be generalized for any edge device within its coverage.
In this case, the channel effect on $\weightMat$ should be entirely calibrated at the edge device side.
However, the conventional channel calibration in {\eqref{eq: preliminaries mimo calibration}} is not feasible for {\name} for two reasons.
First, the channel effect applies to all the $\fftSize$ subcarriers of $\weightMat$, whereas the outputs $\outputMat$ are digitized only after the LPF has collapsed them into a single subcarrier, where the $\fftSize$ per-subcarrier channel matrices are already mixed together and no longer separable.
Second, even if they were separable, applying {\eqref{eq: preliminaries mimo calibration}} over the $\fftSize$ subcarriers would take $8\antNum^2\fftSize$ FLOPs, approximately $2\antNum\times$ the cost of the $2\antNum$ IPs that it enables, overshadowing the energy savings that motivate {\name}.

\subsection{Edge Calibration Unit Design}
\label{sec: design-calibration-unit}

\begin{figure}[!t]
  \centering
  \includegraphics[width=0.99\columnwidth]{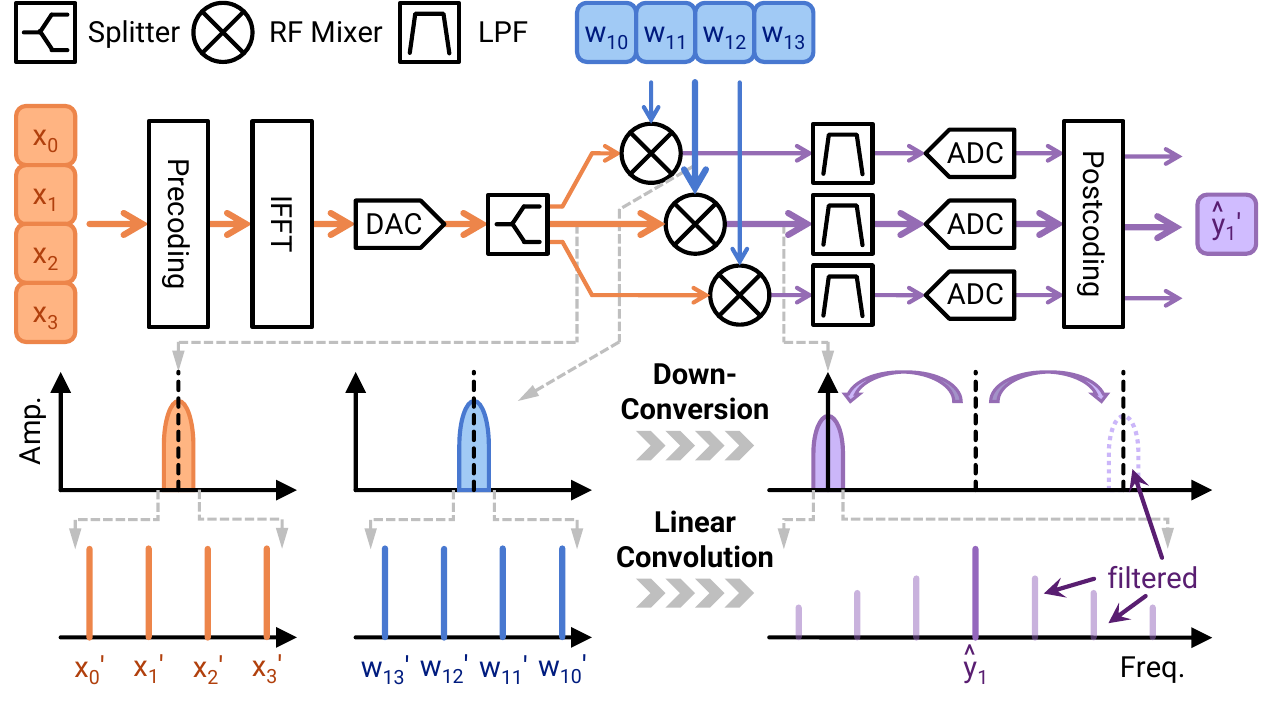}
  \caption{The detailed calibration unit architecture and its handling of an inner product between the received weight vector $\weightVecRx_{\antIdx}$ and the precoded input vector $\inputVecPre$.}
  \vspace{-12pt}
  \label{fig:hardware-computing-unit}
\end{figure}

We design the $\unitNum$ calibration units on the edge device to calibrate the channel effect, as shown in Fig.~\ref{fig:hardware-computing-unit}.

\myparatight{Calibration unit architecture}
Generally, a calibration unit takes the digital signal $\inputVec$ as input, which is (\emph{i}) precoded and IFFTed in digital, (\emph{ii}) split and duplicated into $\antNum$ ways in analog, (\emph{iii}) respectively mixed with the analog received signals $\weightMatRx$ across the $\antNum$ antennas and low-pass filtered, and (\emph{iv}) postcoded in digital.
The key design of this calibration unit is the precoding on the $\inputVec$ and the postcoding on the $\outputMat$.

\myparatight{Precoding for per-subcarrier calibration}
The precoding is performed on the input signal $\inputVec$ to calibrate the non-flat frequency response of the channel.
This is because $\inputVec$ spans all the $\fftSize$ subcarriers, yet lacks any antenna-dimension information.
Hereby, we define the precoder $\precoder \in \mathbb{C}^{\fftSize}$.
Denoting the precoded signal by $\inputVecPre \in \mathbb{C}^{\fftSize}$, the precoding process can be expressed as a per-subcarrier multiplication between $\inputVec$ and $\precoder$:
\begin{align}
    \inputVecPre[\specIdx] = \inputVec[\specIdx] \cdot \precoder[\specIdx], \quad \forall \specIdx = 0, 1, \ldots, \fftSize-1.
    \label{eq: design precoding}
\end{align}

\myparatight{Precoded linear convolution in analog}
After the precoding, the input signal $\inputVecPre$ is split into $\antNum$ ways and fed to the $\antNum$ RF mixers, which perform the $\antNum$ complex-valued convolutions in parallel against the respective received signals $\weightVecRx_{\antIdx}$, i.e.,
\begin{align}
    \outputVecEst_{\antIdx} = \myTrans{\weightVecRx_{\antIdx}} \cdot \inputVecPre, \quad \forall \antIdx = 0, 1, \ldots, \antNum-1,
    \label{eq: design convolution}
\end{align}
where $\outputVecEst_{\antIdx} \in \mathbb{C}$ is the unpostcoded output of antenna $\antIdx$.

\myparatight{Postcoding for per-antenna calibration}
The postcoding is performed on the output signal $\outputMat$ to calibrate the inter-antenna interference.
This is because $\outputMat$ retains the full antenna dimension, while its frequency dimension is already down-sampled and aliased.
Specifically, we define the postcoder $\postcoder \in \mathbb{C}^{\antNum \times \antNum}$, which is applied identically across all subcarriers; also, we denote the postcoded output signals by $\outputMatPostEst = \{ \outputVecPostEst_{\antIdx} \}$ with $\outputVecPostEst_{\antIdx} \in \mathbb{C}$.
The postcoding process can be written as:
\begin{align}
    \outputVecPostEst_{\antIdx} = \sum_{\antIdx^{\prime}=0}^{\antNum-1} \postcoder[\antIdx, \antIdx^{\prime}] \cdot \outputVecEst_{\antIdx^{\prime}}.
    \label{eq: design postcoding}
\end{align}

\myparatight{Cooperation over multiple calibration units}
We place $\unitNum$ such calibration units on the edge device, as shown in Fig.~\ref{fig:hardware-computing-unit}.
Each calibration unit receives the same input signal $\inputVec$ and the same received signals $\weightMatRx$ from the $\antNum$ antennas, while applying a different precoder and postcoder, jointly denoted by $\precoder \in \mathbb{C}^{\unitNum \times \fftSize}$ and $\postcoder \in \mathbb{C}^{\unitNum \times \antNum \times \antNum}$.
Correspondingly, the precoded input signals, unpostcoded outputs, and postcoded outputs are denoted by $\inputVecPre_{\unitIdx}$, $\outputVecEst_{\unitIdx, \antIdx}$, and $\outputVecPostEst_{\unitIdx, \antIdx}$, respectively.
Finally, the outputs from all the calibration units are summed together into the final output, denoted by $\outputMatSumEst = \{ \outputVecSumEst_{\antIdx} \}$ with $\outputVecSumEst_{\antIdx} \in \mathbb{C}$, i.e.,
\begin{align}
    \outputVecSumEst_{\antIdx} = \sum_{\unitIdx=0}^{\unitNum-1} \outputVecPostEst_{\unitIdx, \antIdx}, \quad \forall \antIdx = 0, 1, \dots, \antNum-1,
    \label{eq: design calibration unit sum}
\end{align}

\subsection{Precoder and Postcoder Optimization}
\label{sec:precoder-postcoder}

Putting {\eqref{eq: design precoding}}, {\eqref{eq: design convolution}}, {\eqref{eq: design postcoding}}, and {\eqref{eq: design calibration unit sum}} together, the ultimate goal of {\name} is to minimize the mean squared error (MSE) between the final output signal $\outputVecSumEst$ and the ground truth output signal $\outputVec$ by optimizing the precoder $\precoder$ and postcoder $\postcoder$ jointly across all calibration units.
We formulate this as the optimization problem in {\OptOrig}.
\begin{tcolorbox}[
    boxsep=0mm,left=1mm,right=1mm,top=-1mm,bottom=1mm,
    toptitle=1mm,bottomtitle=1mm,lefttitle=2mm,
    title={{\OptOrig}: Original precoder \& postcoder optimization},
    boxrule=1pt,sharp corners=all]
  \small
  \begin{align}
    \precoderOpt, \postcoderOpt = \arg\min_{\precoder, \postcoder} \ \mathbb{E}_{\inputVec, \weightMat}\left[ \sum_{\antIdx=0}^{\antNum-1} \big\| \outputVecSumEst_{\antIdx} - \outputVec_{\antIdx} \big\|_2^2 \right]
    \label{eq: pre-post-coder opt}
  \end{align}
\end{tcolorbox}
\noindent
Directly solving {\OptOrig} requires the statistics of $\weightMat$ and $\inputVec$.
In practice, however, $\weightMat$ and $\inputVec$ are not known a priori at the edge device, and obtaining them would require an online training phase that feeds known instances of $\weightMat$ and $\inputVec$. The overhead of such training is unaffordable, far exceeding the rate at which the channel $\channel$ itself varies over time.

\myparatight{Waiving the need of $\weightMat$ and $\inputVec$}
Fortunately, we observe that $\outputVecSumEst_{\antIdx} - \outputVec_{\antIdx}$ is bilinear in $(\weightMat, \inputVec)$, so the expectation objective in {\OptOrig} depends only on the first/second-order statistics of $\weightMat$ and $\inputVec$.
Hence, we assume that the entries of $\weightMat$ and $\inputVec$ are zero-mean and mutually independent, and i.i.d.\ within each; we also assume a high SNR region that ignores the noise term in \eqref{eq: preliminaries mimo channel}.
Under these two assumptions, the second-order statistics of $\weightMat$ and $\inputVec$ reduce to scaled identity covariance matrices, so the expectation term in {\OptOrig} collapses to a positive multiple of a channel-dependent-only objective, which becomes {\OptChan} below.
\begin{tcolorbox}[
    boxsep=0mm,left=1mm,right=1mm,top=-1mm,bottom=1mm,
    toptitle=1mm,bottomtitle=1mm,lefttitle=2mm,
    title={{\OptChan}: Simplified channel-dependent optimization},
    boxrule=1pt,sharp corners=all]
  \small
  \begin{align}
    &\precoderOpt, \postcoderOpt = \arg\min_{\precoder, \postcoder} \\
    &\sum_{\specIdx=0}^{\fftSize-1} \left\| \sum_{\unitIdx=0}^{\unitNum-1} \precoder[\unitIdx, \specIdx] \cdot \postcoder[\unitIdx, :, :] \cdot \myTrans{\channel[:, :, \specIdx]} - \mathbf{I}_{\antNum} \right\|^2,
    \label{eq: channel dependent opt}
  \end{align}
\end{tcolorbox}
\noindent
where $\mathbf{I}_{\antNum}$ is the $\antNum \times \antNum$ identity matrix.
Intuitively, {\OptChan} requires the $\unitNum$ precoder-postcoder pairs to jointly zero-force every per-subcarrier channel matrix $\channel[:, :, \specIdx]$ back to the identity matrix, independently of the weights $\weightMat$ or inputs $\inputVec$.
In the following, {\name} solves {\OptChan} in two steps: initialization and fine-tuning.

\myparatight{Step 1: closed-form approximation for initialization}
Generally, the objective in {\OptChan} is bilinear in $\precoder$ and $\postcoder$, which jointly have to reproduce the inverse of the channel $\channel$ across all the $\fftSize$ subcarriers.
\begin{tcolorbox}[
    boxsep=0mm,left=1mm,right=1mm,top=-1mm,bottom=1mm,
    toptitle=1mm,bottomtitle=1mm,lefttitle=2mm,
    title={{\OptApprox}: Approximation with closed-form solution},
    boxrule=1pt,sharp corners=all]
  \small
  \begin{align}
    \precoderInit, \postcoderFlatInit = \arg\min_{\postcoderFlat, \precoder} \left\| \postcoderFlat \cdot \precoder - \channelInvFlat \right\|^2.
    \label{eq: matrix operation opt}
  \end{align}
\end{tcolorbox}
\noindent
Relative to {\OptChan}, {\OptApprox} introduces a single approximation: it drops the inherent per-subcarrier weighting by the channel magnitudes, as the residual in {\OptChan} is exactly that of {\OptApprox} right-multiplied by $\myTrans{\channel[:, :, \specIdx]}$.
Specifically, $(\Tilde{\cdot})$ denotes merging the two antenna dimensions into one, so that {\OptApprox} introduces the flattened postcoder $\postcoderFlat \in \mathbb{C}^{\antNum^2 \times \unitNum}$ and the flattened channel inverse $\channelInvFlat \in \mathbb{C}^{\antNum^2 \times \fftSize}$, the latter inverted per subcarrier.
By the Eckart--Young--Mirsky theorem~\cite{mirsky1960symmetric}, {\OptApprox} can be solved in closed form by taking the singular value decomposition (SVD) of $\channelInvFlat$ and keeping its $\unitNum$ largest singular values.
Unflattening the resulting $\postcoderFlatInit$ back into the two antenna dimensions provides the initialization of {\OptChan}, denoted by $\precoderInit$ and $\postcoderInit$.

\myparatight{Step 2: alternating least squares (ALS) for fine-tuning}
Starting from $\precoderInit$ and $\postcoderInit$, we refine towards the objective in {\OptChan} by alternating between two convex and closed-form least-squares updates until convergence:
\begin{itemize}[leftmargin=*, topsep=2pt, itemsep=1pt]
    \item \textit{Precoder step}: fixing $\postcoder$, we optimize $\precoder$ to balance the zero-forcing residual across the frequency dimension, i.e., the $\fftSize$ subcarriers, which reduces to a linear least-squares problem solved independently per subcarrier.
    \item \textit{Postcoder step}: fixing $\precoder$, we optimize $\postcoder$ to cancel the inter-antenna interference respectively for each subcarrier, which also reduces to a linear least-squares problem, solved independently per output antenna.
\end{itemize}
Both steps admit closed-form solutions, so each ALS iteration is cheap; given the Step~1 initialization, convergence typically occurs within a handful of iterations.
A detailed optimization process of these two steps is shown in Appendix~\ref{app:opt-derivation} and~\ref{app:opt-convergence}.

\subsection{System Efficiency Analysis}

We analyze the efficiency of {\name} along two dimensions: the airtime it occupies the wireless channel, and the energy it consumes on the edge device.

\myparatight{Airtime analysis}
The airtime of {\name} is inversely proportional to the number of antennas $\antNum$ and the occupied bandwidth $\band$.
Interestingly, this form exactly follows a conventional communication link with spatial multiplexing on $\antNum$ and frequency multiplexing on $\band$.
On the other hand, {\name}'s airtime is independent of the SNR, which instead determines the computing accuracy; {\name} therefore does not customize the transmitted signals for each user, and is naturally fit to broadcasting.
Leveraged by broadcasting, one transmission of the weights serves arbitrarily many edge users at a constant airtime, so the airtime amortized per user decreases as $1/\userNum$.
The complete analysis is given in Appendix~\ref{sec:throughput-analysis}.

\myparatight{Energy analysis}
Overall, {\name} achieves good energy efficiency on RF computing, whose energy per FLOP is further averaged down as the GEMV scales up to LLM.
This is because the energy consumption of almost every hardware component in {\name} is amortized by either the input size $\inputSize$ or the output size $\outputSize$, leaving the DAC, whose per-FLOP energy is set by the antenna number $\antNum$ and the unit number $\unitNum$, as the floor that a large GEMV converges to.
Such a scaling law is contributed by two key designs.
First, the RF mixer and the LPF collapse the whole input dimension into a single subcarrier with a downsampling ratio of $\inputSize$, so the ADC takes only one sample per inner product, which usually dominates the energy budget of an analog computing system; the postcoding runs on that same single sample per antenna, so its energy is amortized by $\inputSize$ as well.
Second, the precoding and the IFFT are paid only once per GEMV, as the same precoded input signal $\inputVecPre$ is reused across all the $\myCeil{\outputSize/(2\antNum)}$ rounds, so their energy is amortized by the output size $\outputSize$.
The per-component analysis is in Appendix~\ref{sec:energy-analysis}.

\myparatight{Tradeoff on computing accuracy, airtime, and energy}
The calibration unit number $\unitNum$ is where these three metrics meet.
On the accuracy side, {\OptChan} calibrates the channel with a rank-$\unitNum$ approximation, so a larger $\unitNum$ leaves a smaller residual and thus a more accurate channel calibration; and the more antennas $\antNum$ are deployed, the larger $\unitNum$ this accuracy requires.
On the energy side, however, every energy term grows linearly with $\unitNum$.
Scaling up $\antNum$ therefore shortens the airtime but pays for its accuracy in energy, so $\unitNum$ should be the smallest that meets the target task's accuracy.

\section{Implementation}
\label{sec:implementation}

We implement {\name} on the wireless channels of real-world urban scenes by a ray-tracing simulator, and on the distortion of a real RF mixer, measured on a wired testbed.

\subsection{Channel Modeling by Ray Tracing}
\label{sec:impl-raytracing}

\begin{figure}[t]
    \centering
    \begin{minipage}[b]{0.28\linewidth}
        \centering
        \includegraphics[width=\linewidth]{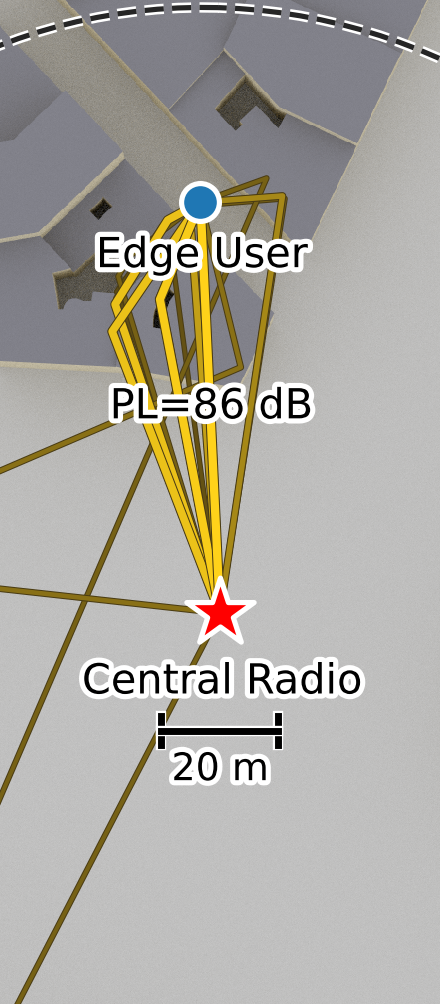}
        \subcaption{Paris}
        \vspace{-7pt}
    \end{minipage}
    \hfill
    \begin{minipage}[b]{0.28\linewidth}
        \centering
        \includegraphics[width=\linewidth]{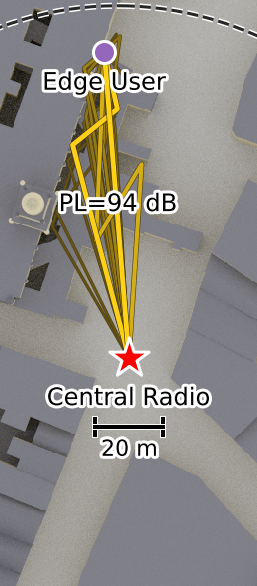}
        \subcaption{Munich}
        \vspace{-7pt}
    \end{minipage}
    \hfill
    \begin{minipage}[b]{0.40\linewidth}
        \centering
        \includegraphics[width=\linewidth]{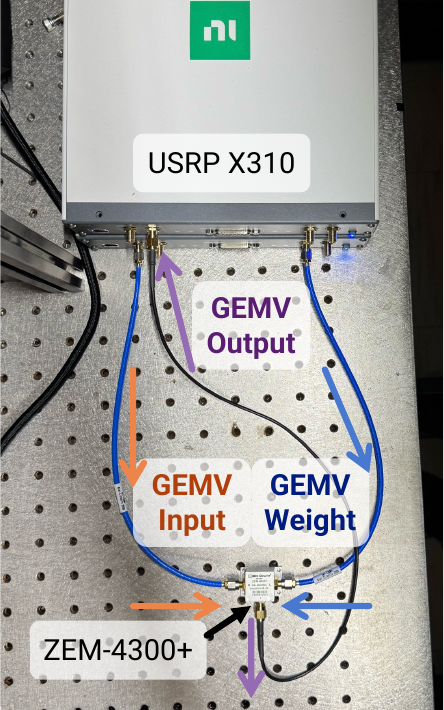}
        \subcaption{Testbed}
        \vspace{-7pt}
    \end{minipage}
    \caption{The two Sionna ray-tracing scenes in Paris (a) and Munich (b), and the testbed for RF mixer profiling.}
    \vspace{-12pt}
    \label{fig:scene-map}
\end{figure}

\myparatight{Sionna ray-tracing simulator}
We build the wireless channel between the central radio and the edge devices on top of NVIDIA Sionna's ray-tracing engine (v2.0.1)~\cite{hoydis2022sionna, hoydis2023sionnart}.
Specifically, Sionna simulates the wireless signal propagation process by shooting $10^7$ rays into a digital twin of a real-world city scene and resolving their interactions (e.g., reflection, refraction, and diffraction) with the buildings, grounds, etc.
Assembling these propagation paths, it provides the realistic channel $\channel$ that the weight signal $\weightMat$ propagates through to each of the edge devices.
As shown in Fig.~\ref{fig:scene-map}(a) and (b), we evaluate {\name} in two contrasting urban scenes: (\emph{i}) Place Charles de Gaulle in Paris, open enough that a line-of-sight (LOS) path always holds, and (\emph{ii}) the city center of Munich, densely built so that multipath is rich while the LOS path is often blocked.

\myparatight{Central radio configuration}
The central radio sits at the center of each scene, {10}\thinspace{m} above the ground.
It carries $\antNum$ antennas ($\antNum=4$ by default) on $\myCeil{\antNum/2}$ physical elements, each of two orthogonal polarizations, spaced {1.7}\thinspace{m} apart.
It broadcasts the weight signal on the CBRS band~\cite{sohul2015cbrs} at a {3.5}\thinspace{GHz} carrier frequency over {100}\thinspace{MHz} of bandwidth.
The default broadcast power per antenna is set to {50}\thinspace{dBm}, and the coverage radius is {100}\thinspace{m}.

\myparatight{Edge device configuration}
We randomly drop {100} edge devices at {1.5}\thinspace{m} height within the central radio's coverage zone.
Each edge device carries the same cross-polarized $\antNum$-port array with {4.3}\thinspace{cm} spacing (half-wavelength), and $\unitNum=8$ calibration units, whose coders are derived from a least-squares per-antenna pilot estimate $\widehat{\channel}$.
We assume the devices' noise figure of {4}\thinspace{dB}, corresponding to a noise floor of {-90}\thinspace{dBm} across the {100}\thinspace{MHz} bandwidth; their path loss (PL) from the central radio is estimated by its own channel as $-10\log_{10}(\mathbb{E}[|\channel|^2])$.

\subsection{Hardware Profiling on Real RF Mixer}
\label{sec:impl-mixer}

RF mixers are not built for analog multiplication, and hence distort the desired product.
We therefore measure the distortion of a real RF mixer, ZEM-4300+~\cite{minicircuits2024zem4300}, driven by USRP X310 radios~\cite{ettus2024x310} as the transceivers, as shown in Fig.~\ref{fig:scene-map}(c).
The profiling is wired, so the power entering the mixer stays far above the noise floor; the link is thus considered noiseless, and the computing errors are fully attributed to the mixer's own distortion.
Specifically, for each power at the mixer's RF port, which receives the weight signal over the air in a deployment, we sweep the power at its LO port and keep the lowest distortion error, i.e., the best effective number of bits (ENOB), that the mixer reaches.
The measured distortion ENOB curve over the RF power is shown in Fig.~\ref{fig:usrp-rf-power} of Appendix~\ref{app:setup-measured}.
In our evaluation, we read this curve at each edge device's received weight signal power and plug in the corresponding distortion error.
\section{Evaluation}

\subsection{Benchmarking on GEMV}
\label{sec:eval-gemm}

\begin{figure}[t]
    \centering
    \begin{minipage}{0.49\linewidth}
        \centering
        \includegraphics[width=\linewidth]{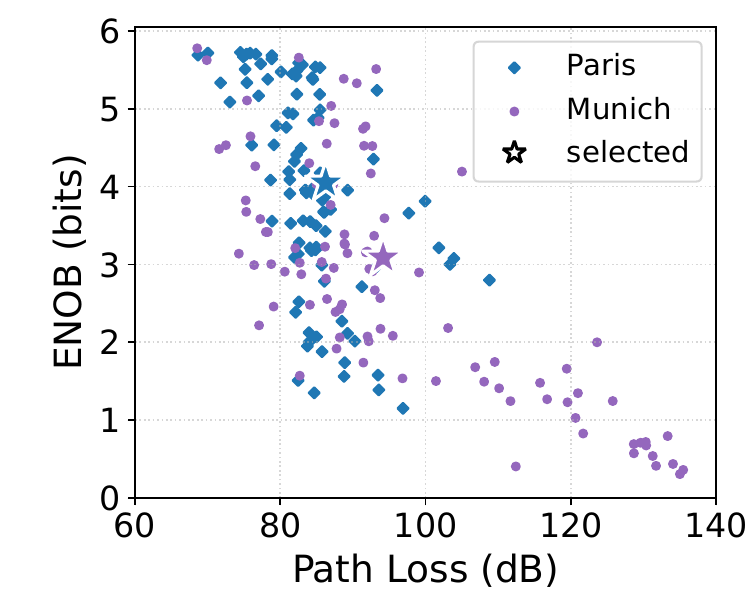}
        \par\vspace{-4pt}
        \subcaption{ENOB vs.\ user PL}
        \label{fig:layer-pathloss}
        \vspace{-7pt}
    \end{minipage}
    \hfill
    \begin{minipage}{0.49\linewidth}
        \centering
        \includegraphics[width=\linewidth]{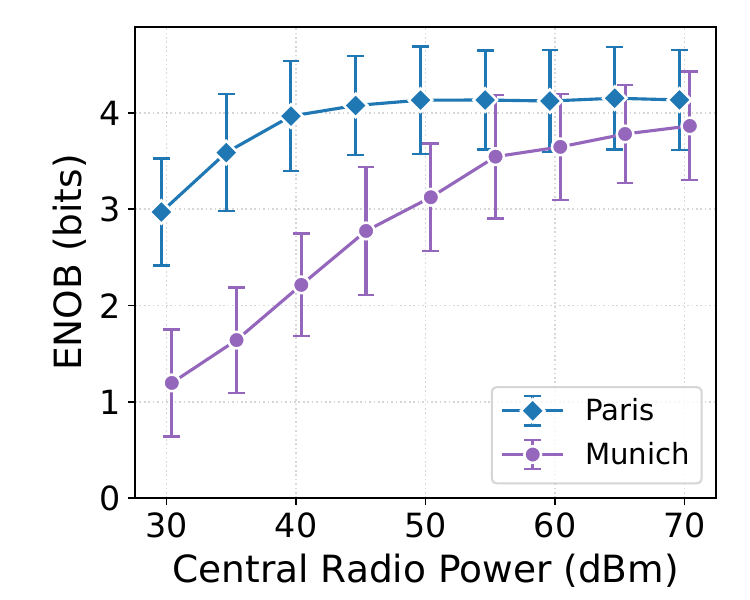}
        \par\vspace{-4pt}
        \subcaption{ENOB vs.\ central radio power}
        \label{fig:layer-centpower-all}
        \vspace{-7pt}
    \end{minipage}
    \caption{In both Paris and Munich, the ENOB of a randomized GEMV over user PL (a) and broadcast power (b).}
    \vspace{-12pt}
    \label{fig:layer-pathloss-centpower}
\end{figure}

We first benchmark {\name} on a single GEMV with randomized weights and input activations.
The computing accuracy is measured by ENOB defined in Appendix~\ref{app:setup-metrics}.

\myparatight{GEMV accuracy over the user PL}
Following the default setup in \S\ref{sec:impl-raytracing}, Fig.~\ref{fig:layer-pathloss} scatters the ENOB of a randomized GEMV by each of the {100} candidate users in Paris and Munich against their PLs.
The ENOB generally decreases with the PL, as a higher PL lowers the power, and thus the SNR, of the weight signal arriving at the user.
The open scene of Paris, where an LOS path is more likely and the PL is therefore lower (median {84.0}\thinspace{dB}), reaches a median ENOB of {3.97}\thinspace{bit} with {5/95}-th percentiles of {1.58/5.69}\thinspace{bit}, whereas the users of Munich, mostly without an LOS path (median {91.4}\thinspace{dB}), reach only {2.91}\thinspace{bit} with {0.53/5.33}\thinspace{bit}.
In addition, the PL does not determine the ENOB, which further depends on how the received power is spread over the multi-path components: within a PL of {89--91}\thinspace{dB} alone, the ENOB still ranges over {2.01--5.33}\thinspace{bit}.
We empirically select the user with approximately {4.00}\thinspace{bit} at {86.3}\thinspace{dB} in Paris and that of approximately {3.00}\thinspace{bit} at {94.1}\thinspace{dB} in Munich, starred in the figure, as the default for the experiments below.

\myparatight{GEMV accuracy over the central radio power}
Fig.~\ref{fig:layer-centpower-all} sweeps the central radio transmit power from {30}\thinspace{dBm} to {70}\thinspace{dBm} on the two selected users.
Both scenes follow the same two-stage trend: the ENOB first grows by roughly {1}\thinspace{bit} per {10}\thinspace{dB} of transmit power, and then saturates.
The Munich user, for example, climbs from {1.20}\thinspace{bit} at {30}\thinspace{dBm} to {3.55}\thinspace{bit} at {55}\thinspace{dBm}, but to only {3.87}\thinspace{bit} when further increased to {70}\thinspace{dBm} power.
In the first stage, the GEMV accuracy is limited by the received weight signal power against the noise floor; in the second, by the imperfect channel compensation of the $\unitNum$ units and the RF mixer's imperfect multiplication.

\begin{figure}[t]
    \centering
    \begin{minipage}{0.49\linewidth}
        \centering
        \includegraphics[width=\linewidth]{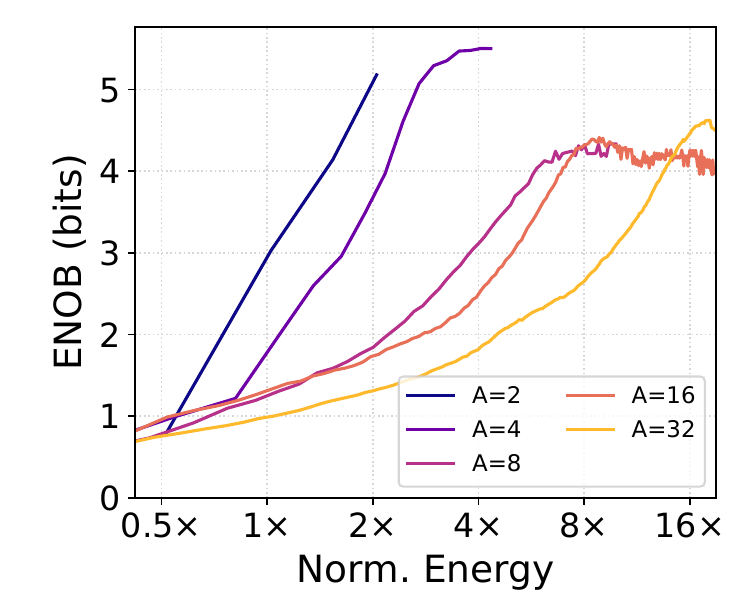}
        \par\vspace{-4pt}
        \subcaption{ENOB vs.\ energy}
        \label{fig:layer-unit-pareto-etoile}
        \vspace{-7pt}
    \end{minipage}
    \hfill
    \begin{minipage}{0.49\linewidth}
        \centering
        \includegraphics[width=\linewidth]{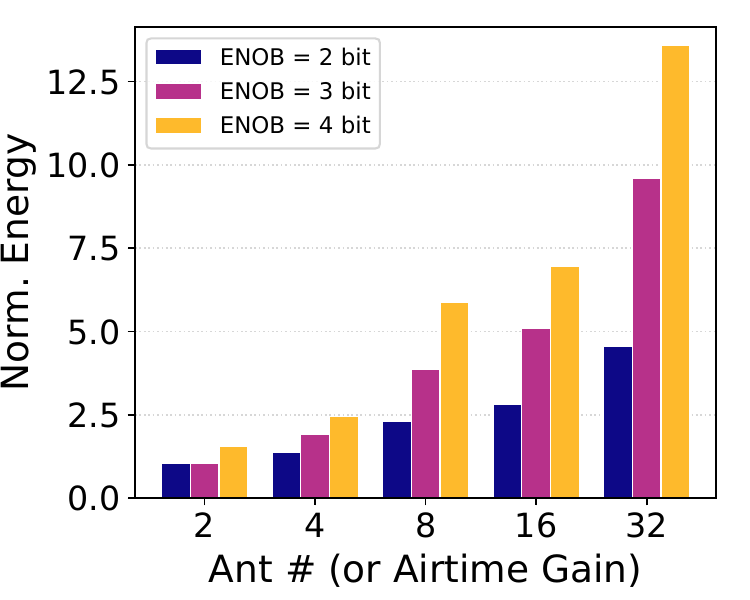}
        \par\vspace{-4pt}
        \subcaption{Energy per ENOB level}
        \label{fig:layer-unit-group-etoile}
        \vspace{-7pt}
    \end{minipage}
    \caption{With $\antNum =\left\{2^1, \dots, 2^5\right\}$, the ENOB over $\unitNum$ or energy (a), and the minimum energy to achieve 2--4 ENOBs in Paris.}
    \vspace{-12pt}
    \label{fig:layer-unit-pareto}
\end{figure}

\begin{figure}[!t]
    \centering
    \begin{minipage}{0.49\linewidth}
        \centering
        \includegraphics[width=\linewidth]{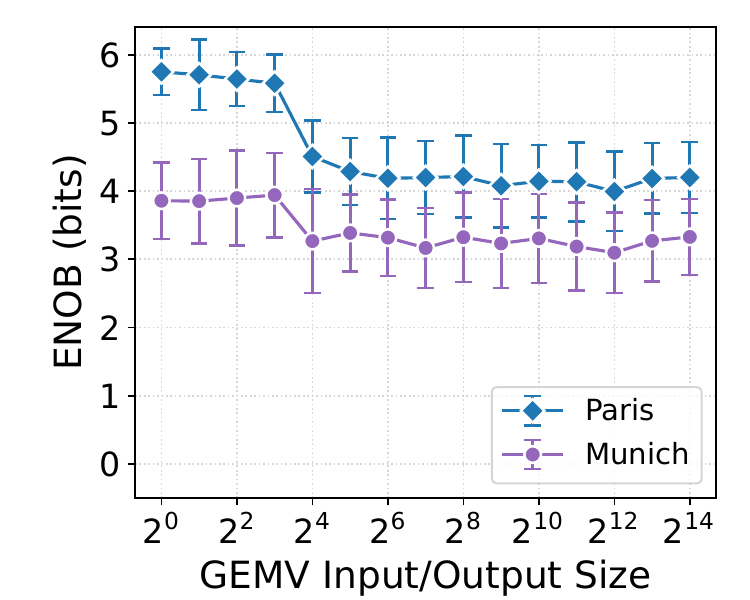}
        \par\vspace{-4pt}
        \subcaption{ENOB}
        \label{fig:layer-scalability-enob}
        \vspace{-7pt}
    \end{minipage}
    \hfill
    \begin{minipage}{0.49\linewidth}
        \centering
        \includegraphics[width=\linewidth]{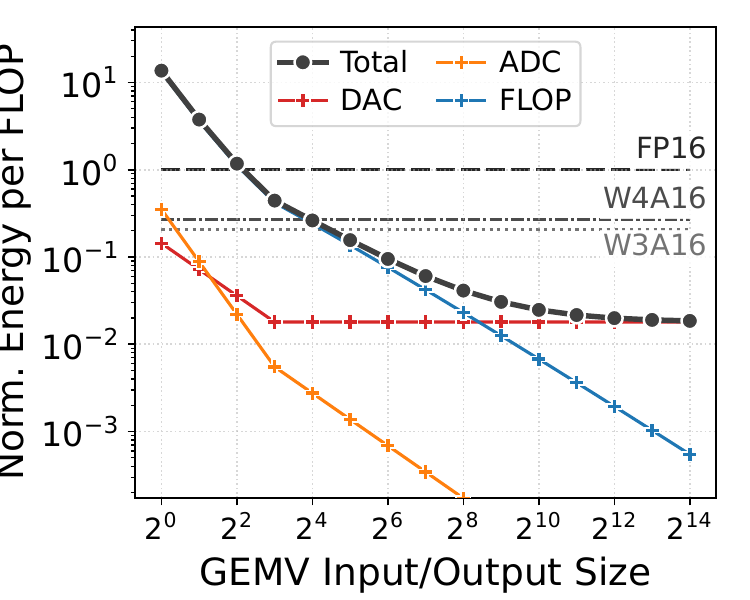}
        \par\vspace{-4pt}
        \subcaption{Norm. energy per FLOP}
        \label{fig:layer-scalability-energy}
        \vspace{-7pt}
    \end{minipage}
    \caption{When scaling up the GEMV sizes, the ENOB (a) and the energy per FLOP breakdown of {\name} (b).}
    \vspace{-12pt}
    \label{fig:layer-scalability}
\end{figure}

\begin{table*}[h]
    \centering
    \caption{Over the nine LLMs, the WikiText-2 PPL of {\name} on the selected users of Paris and Munich vs. the FP16, WoQ, and analog computing baselines, at the sequence length of {2,048}.}
    \label{tab:wikitext2_ppl}
    \small
    \setlength{\tabcolsep}{5pt}
    \vspace{-6pt}
    \begin{tabular}{c|c|cccc|cc|c|cc}
    \toprule
    \multirow{2}{*}{\textbf{Precision}} & \multirow{2}{*}{\textbf{Method}} & \multicolumn{4}{c|}{\textbf{LLaMA-3}} & \multicolumn{2}{c|}{\textbf{LLaMA-2}} & \textbf{Mistral} & \multicolumn{2}{c}{\textbf{Gemma-4}} \\
    \cline{3-6}\cline{7-8}\cline{9-9}\cline{10-11}
    & & 3.2-1B & 3.2-3B & 3.1-8B & 3.1-70B$^{\dagger}$ & 7B & 13B & 7B & E2B & E4B \\
    \midrule
    \textbf{FP16} & - & 9.77 & 7.82 & 6.24 & 2.81 & 5.47 & 4.88 & 5.25 & 8.24 & 7.07 \\
    \midrule
    \multirow{3}{*}{\textbf{W4A16}} & RTN & 11.71 & 8.49 & 6.83 & 3.35 & 5.72 & 4.98 & 5.42 & 10.60 & 193.64 \\
    & GPTQ & 10.63 & 8.78 & 6.65 & 3.37 & 5.62 & 4.99 & 5.38 & 8.95 & 11.51 \\
    & AWQ & 10.95 & 8.29 & 6.66 & 3.27 & 5.60 & 4.97 & 5.37 & 9.57 & 11.72 \\
    \midrule
    \multirow{3}{*}{\textbf{ENOB 4}} & AWGN & 12.28 & 8.75 & 7.11 & 4.30 & 40.94 & 5.06 & 16.31 & 9.51 & 11.60 \\
    & HyFlexPIM & 10.45 & 8.11 & 6.48 & 3.04 & 5.56 & 4.93 & 5.32 & 8.82 & 15.25 \\
    & \methodEtoile & \cellEtoile{10.41} & \cellEtoile{8.08} & \cellEtoile{6.49} & \cellEtoile{3.21} & \cellEtoile{5.82} & \cellEtoile{4.96} & \cellEtoile{5.45} & \cellEtoile{8.77} & \cellEtoile{8.49} \\
    \midrule
    \multirow{3}{*}{\textbf{W3A16}} & RTN & 34.82 & 13.25 & 10.86 & 5.24 & 6.66 & 5.52 & 6.15 & 264.35 & $1.8{\times}10^{7}$ \\
    & GPTQ & 16.46 & 15.21 & 8.24 & 5.33 & 6.42 & 5.44 & 5.89 & 11.79 & 41.82 \\
    & AWQ & 18.38 & 10.62 & 8.28 & 4.80 & 6.27 & 5.35 & 5.93 & 17.92 & $1.5{\times}10^{3}$ \\
    \midrule
    \multirow{3}{*}{\textbf{ENOB 3}} & AWGN & 72.87 & 18.76 & 26.51 & 24.40 & 515.84 & 6.31 & 332.53 & 13.11 & 18.76 \\
    & HyFlexPIM & 13.45 & 9.35 & 7.25 & 3.90 & 5.92 & 5.14 & 5.59 & 11.55 & $1.2{\times}10^{3}$ \\
    & \methodMunich & \cellMunich{12.20} & \cellMunich{8.92} & \cellMunich{7.37} & \cellMunich{4.99} & \cellMunich{6.47} & \cellMunich{5.12} & \cellMunich{5.78} & \cellMunich{10.17} & \cellMunich{16.37} \\
    \bottomrule
    \end{tabular}
    \par\vspace{2pt}
    {\footnotesize $^{\dagger}$keep the first block's value projection ({0.012\%} of the weights) in FP16, to alleviate its massive outliers.}
    \vspace{-8pt}
\end{table*}

\myparatight{Trading off computing accuracy, energy, and airtime}
Compared to WISE~\cite{gao2026disaggregated}, the special case of {\name} with $\antNum=\unitNum=1$, {\name} involves MIMO to save the airtime of a GEMV: the airtime is proportional to the inverse antenna number $1/\antNum$. On the other hand, the corresponding wireless channel takes more calibration units $\unitNum$ to compensate, and the energy consumption is proportional to $\unitNum$.
Fig.~\ref{fig:layer-unit-pareto-etoile} illustrates this trade-off in Paris on a single-round GEMV of $\outputSize = 2\antNum$ and $\inputSize = {4,096}$, sweeping $\unitNum = \left\{1, 2, \dots, \antNum^2\right\}$ at $\antNum = \left\{2, 4, 8, 16, 32\right\}$, i.e., an airtime gain of {2/4/8/16/32}$\times$ over WISE, with each array at the central radio power of {40/50/60/70/80}\thinspace{dBm}.
Given $\antNum$, a larger $\unitNum$ compensates for the channel more accurately and is saturated at some point depending on the channel's complexity.
At $\antNum={4}$, for example, the ENOB climbs from {0.55}\thinspace{bit} at $\unitNum={1}$ to {3.97} at {8} units and {5.07} at {10}; five further units then add only {0.43}\thinspace{bit} while costing {1.5}$\times$ the energy.
Fig.~\ref{fig:layer-unit-group-etoile} extracts the minimum energy reaching each ENOB level, normalized to that by WISE: {3}\thinspace{bit} costs {1.02/1.90/3.86/5.09/9.60}$\times$ and {4}\thinspace{bit} costs {1.54/2.44/5.86/6.96/13.59}$\times$ at $\antNum={2/4/8/16/32}$.
This suggests that, to reduce the airtime with a larger $\antNum$, the same ENOB requires more energy consumption, because an $\antNum^2$-entry per-subcarrier matrix takes more units to invert.
The same trends hold in Munich, whose results are shown in Appendix~\ref{app:gemv}.

\myparatight{Scaling up the GEMV size}
In Fig.~\ref{fig:layer-scalability-enob}, we scale up the GEMV's input and output sizes together, over $\inputSize=\outputSize = \left\{2^0, 2^1, \dots, 2^{14}\right\}$.
Up to $\inputSize=\outputSize = 2^3$, where a GEMV is completed within a single round, the computing accuracy stays at {5.58--5.75}\thinspace{bit} in Paris and {3.85--3.94}\thinspace{bit} in Munich.
Beyond that point, where multiple rounds are needed, the ENOB drops slightly to {3.99--4.51} and {3.10--3.39}\thinspace{bit}, respectively, which is maintained as the size keeps scaling up.
{\name} therefore supports good computing accuracy on the large-scale GEMVs that are widely used in LLMs.
Fig.~\ref{fig:layer-scalability-energy} further illustrates the energy per FLOP over the same sweep of GEMV scales, normalized to that of the FP16 baseline.
That of FP16 and of the WoQ baselines, counting both the data movement and the digital computing, is constant over the sweep, with W4A16/W3A16 {3.72/4.80}$\times$ more energy efficient than FP16.
In contrast, {\name}'s energy per FLOP scales down as the GEMV scales up.
A small GEMV is bottlenecked by the digital arithmetic of the precoding, the IFFT, and the postcoding (e.g., only {3.81}$\times$ more energy efficient than FP16 at $\inputSize=2^{4}$), after which the bottleneck moves to the DAC, whose per-FLOP cost is constant over the GEMV size (e.g., {50.2}$\times$ more energy efficient than FP16 at $\inputSize=2^{12}$).
The ADC energy, which dominates conventional analog computing, is nearly free because of {\name}'s downsampling ratio of $\inputSize$.
In applications whose GEMV input is itself an analog waveform, e.g., a received wireless signal or a biomedical sensor readout, this bottlenecked DAC energy is waived as well.
This is exactly the regime an LLM runs in, whose inference is dominated by large GEMVs, and where {\name} therefore costs the edge device the least.

\subsection{Benchmarking on LLM}
\label{sec:eval-llm-baseline}

We run whole LLMs with all weight-activation GEMVs computed in analog by {\name}, and the remaining in digital.
The LLM tasks and the WoQ/analog computing baselines are described in Appendix~\ref{app:setup-baselines} and~\ref{app:setup-metrics}.

\myparatight{PPL on WikiText-2}
In Tab.~\ref{tab:wikitext2_ppl}, we present the WikiText-2 PPL at the sequence length of {2,048} over nine LLMs of four families: LLaMA-3.2 (1B and 3B), LLaMA-3.1 (8B and 70B), LLaMA-2 (7B and 13B), Mistral-7B, and Gemma-4 (E2B and E4B).
The selected user in Paris, resolving about {4}\thinspace{bit}, is read against the W4A16 baselines (RTN, GPTQ~\cite{frantar2023gptq}, and AWQ~\cite{lin2024awq}) and the ENOB-4 analog computing baselines (AWGN and HyFlexPIM~\cite{song2025hybrid}); the user in Munich, at about {3}\thinspace{bit}, against their W3A16 and ENOB-3 counterparts.
In Paris, specifically, {\name} beats the best of the WoQ baselines on {7} out of the {9} LLMs, and the best of the analog computing baselines on {4} of the {9}, staying within {5.6\%} of it on the other five.
In Munich, where every baseline loses more at {3}\thinspace{bit}, it beats the best of the W3A16 and of the ENOB-3 baselines on {7} and {5} of the {9}, respectively.
Such comparable or slightly lower PPLs show that the errors of {\name}, including the thermal noise, the hardware distortion, and the imperfect channel compensation, still leave enough computing accuracy for an LLM inference.

\begin{figure}[t]
    \centering
    \includegraphics[width=\linewidth]{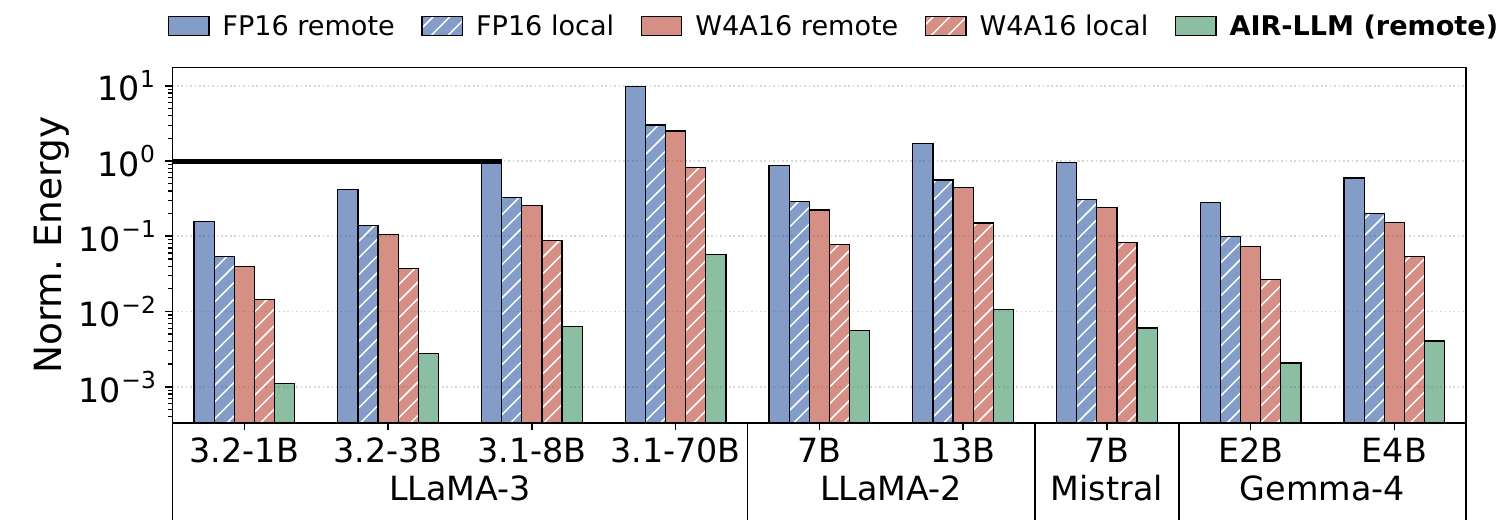}
    \caption{The normalized energy per token of {\name} vs. four baselines over the nine LLMs.}
    \vspace{-12pt}
    \label{fig:llm-energy}
\end{figure}

\setcounter{topnumber}{1}
\myparatight{Energy consumption per token}
Fig.~\ref{fig:llm-energy} illustrates the per-token energy the edge device spends on the nine LLMs at the selected user in Paris, against four baselines that combine the remote/local weight storage with the FP16/W4A16 precision.
All these energy consumptions are normalized to that of the remote FP16 on LLaMA-3.1-8B.
On the same LLaMA-3.1-8B, {\name} costs {157.7}$\times$ and {40.4}$\times$ less energy than the remote FP16 and W4A16, respectively, and against the local FP16 and W4A16, whose weights are more easily accessible from the device's own memory, {\name} still wins by {51.5}$\times$ and {13.9}$\times$.
Such an energy gain holds over all nine LLMs.
Specifically for the two W4A16 baselines, whose PPL is comparable to {\name}, {\name} gains {35.4--44.0}$\times$ against the remote one, owing to the downsampling ratio of $\inputSize$ at {\name}'s ADC, as well as the much lighter digital pipeline behind it.
{\name} even achieves {12.9--14.3}$\times$ against the local one, whose weights are easier to access than {\name}'s, thanks to the waiving of the energy-expensive data movement.
The Munich device draws the same conclusion against its own W3A16 baselines, as reported in Fig.~\ref{fig:llm-energy-munich}.

\begin{figure}[t]
    \centering
    \includegraphics[width=\linewidth]{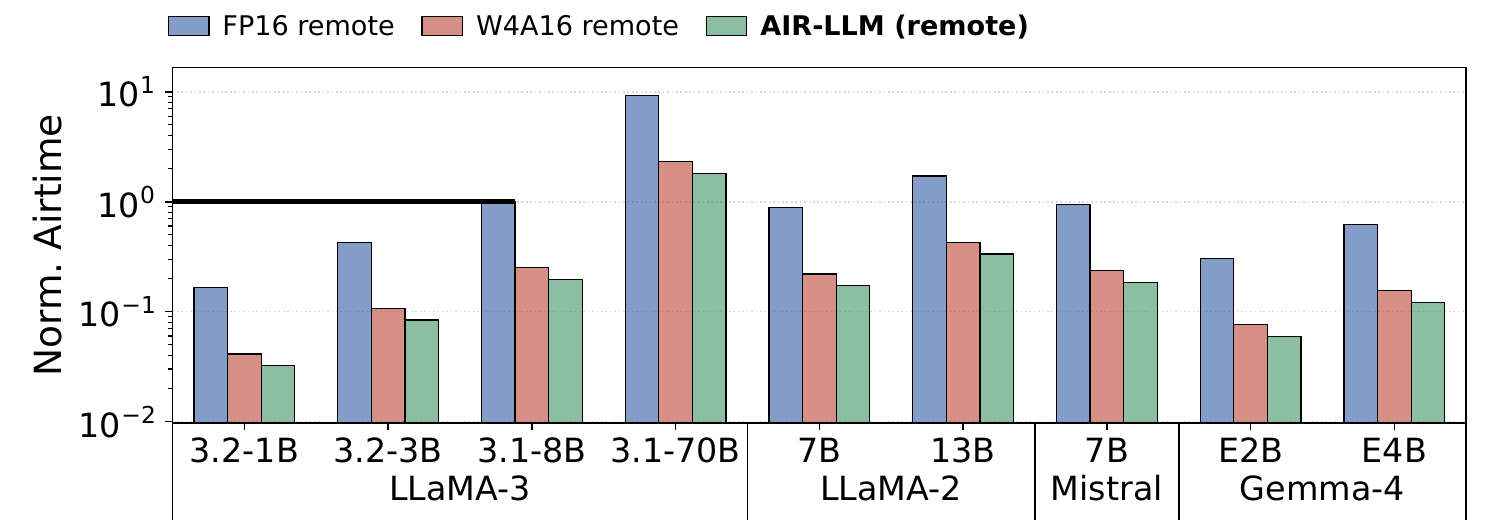}
    \caption{The normalized airtime per token of {\name} vs. two remote baselines over the nine LLMs.}
    \vspace{-12pt}
    \label{fig:llm-airtime}
\end{figure}

\myparatight{Airtime per token}
In Paris, Fig.~\ref{fig:llm-airtime} presents the per-token airtime occupying the wireless channel, including {\name} and the two remote baselines on the same nine LLMs.
Similarly, the airtime is normalized to that of the remote FP16 on LLaMA-3.1-8B; the two local baselines occupy no channel and are therefore skipped.
On LLaMA-3.1-8B, {\name} takes {5.11}$\times$ less airtime than the remote FP16 baseline, and slightly less airtime of {1.28}$\times$ compared to the W4A16 baseline.
Both ratios stay constant at {5.11}$\times$ and {1.28}$\times$ across all nine LLMs, as the airtime of {\name} and that of the remote baselines are proportional to the same weight count.
In this single-user case, the close airtime to the W4A16 baseline comes from the physical limit of the wireless channel {\eqref{eq: gemv airtime per flop}}: the airtime is scaled down linearly with the bandwidth $\band$ and the antenna number $\antNum$ for both the communication and {\name}'s computing, and their remaining difference in the SNR is canceled out once the baseline is quantized to the specific weight width giving the close PPL performance to {\name}.
Similar ratios are also observed in Munich in Fig.~\ref{fig:llm-airtime-munich} of Appendix~\ref{app:supp-llm}.
In a dense-user scenario, however, {\name}'s broadcasting capability significantly outperforms these WoQ baselines regarding the airtime, which is later shown in \S\ref{sec:eval-llm-user}.

\setcounter{topnumber}{2}

\begin{table}[h]
    \centering
    \caption{In the prefilling phase, the PPL of {\name} in Paris and Munich over different sequence lengths.}
    \label{tab:ppl_seqlen}
    \small
    \setlength{\tabcolsep}{3pt}
    \vspace{-6pt}
    \begin{tabular}{c|ccccccc}
    \toprule
    \multirow{2}{*}{\textbf{\shortstack{Precision/\\User}}} & \multicolumn{7}{c}{\textbf{Sequence length}} \\
    \cline{2-8}
    & 256 & 512 & 1,024 & 2,048 & 4,096 & 8,192 & 16,384 \\
    \midrule
    \textbf{FP16} & 10.37 & 8.18 & 6.96 & 6.24 & 5.85 & 5.62 & 5.48 \\
    \midrule
    \cellEtoile{Paris} & \cellEtoile{11.05} & \cellEtoile{8.57} & \cellEtoile{7.26} & \cellEtoile{6.49} & \cellEtoile{6.08} & \cellEtoile{5.84} & \cellEtoile{5.70} \\
    \cellMunich{Munich} & \cellMunich{12.99} & \cellMunich{9.88} & \cellMunich{8.30} & \cellMunich{7.37} & \cellMunich{6.87} & \cellMunich{6.58} & \cellMunich{6.43} \\
    \bottomrule
    \end{tabular}
    \vspace{-8pt}
\end{table}

\myparatight{PPL over the sequence length}
In Tab.~\ref{tab:ppl_seqlen}, we sweep the sequence length from {256} to {16,384} tokens to evaluate {\name}'s robustness over the context length of the prefilling phase.
The PPLs of the FP16 baseline and of {\name} in Paris/Munich fall as the context grows, and {\name}'s relative gap to the FP16 reference narrows and then flattens, spanning {3.9--6.6\%} and {17.1--25.3\%} over the sweep while settling at about {4.0\%} and {17.3\%} beyond {2,048} tokens.
This suggests that {\name}'s computing error does not accumulate along the context, and that the same robustness would potentially extend to the much longer contexts demanded by future LLM applications.
Note that {\name}'s per-token energy and airtime are constant along this axis, since both are dominated by streaming the weights of each layer once per token, independent of the token number.

\subsection{Impact of the System Factors}
\label{sec:eval-llm-factor}

We examine more deployment factors, such as transmit power, antenna/calibration unit number, and frequency band.

\begin{table}[h]
    \centering
    \caption{The PPL of {\name} in Paris and Munich over different central radio broadcasting power.}
    \label{tab:ppl_txpower}
    \small
    \setlength{\tabcolsep}{3pt}
    \vspace{-6pt}
    \begin{tabular}{c|cccccc}
    \toprule
    \multirow{2}{*}{\textbf{User}} & \multicolumn{6}{c}{\textbf{Central radio transmit power (dBm)}} \\
    \cline{2-7}
    & 35 & 40 & 45 & 50 & 55 & 60 \\
    \midrule
    \cellEtoile{Paris} & \cellEtoile{6.81} & \cellEtoile{6.59} & \cellEtoile{6.51} & \cellEtoile{6.49} & \cellEtoile{6.49} & \cellEtoile{6.48} \\
    \cellMunich{Munich} & \cellMunich{$3.2{\times}10^{3}$} & \cellMunich{270.42} & \cellMunich{15.76} & \cellMunich{7.37} & \cellMunich{6.95} & \cellMunich{6.79} \\
    \bottomrule
    \end{tabular}
    \vspace{-8pt}
\end{table}

\myparatight{Impact of the central radio transmit power}
We evaluate the impact of the transmit power under the same LLaMA-3.1-8B setup in Paris and Munich, which is presented in Tab.~\ref{tab:ppl_txpower}.
The Paris user is already saturated at {35}\thinspace{dBm} with {9.1\%} above the FP16 reference of {6.24}, and raising the power to {60}\thinspace{dBm} narrows that gap only to {3.8\%}.
This hints that the computing error of this user comes from the imprecise channel calibration and the RF mixer's imperfection.
In comparison, the Munich user is still power-limited: its PPL stays orders of magnitude above the FP16 reference below {50}\thinspace{dBm}, and then improves steeply to {8.8\%} at {60}\thinspace{dBm}, meaning that its computing accuracy is mainly determined by the thermal noise.
Both observations match the two-stage trend of Fig.~\ref{fig:layer-centpower-all}.
In {\name}, users under different channel conditions are limited by different sources of the computing error, which in turn decides whether more transmit power pays off.

\begin{table}[h]
    \centering
    \caption{The PPL of {\name} in Paris and Munich at $\antNum = \left\{4, 8\right\}$ with a varying number of calibration units, $\unitNum$.}
    \label{tab:ppl_unit}
    \small
    \setlength{\tabcolsep}{4pt}
    \renewcommand{\arraystretch}{1.1}
    \vspace{-6pt}
    \begin{tabular}{c@{\hspace{6pt}}cc|c@{\hspace{6pt}}cc}
    \toprule
    \multicolumn{3}{c|}{\textbf{Ant \# = {4} @ {50}\thinspace{dBm}}} & \multicolumn{3}{c}{\textbf{Ant \# = {8} @ {60}\thinspace{dBm}}} \\
    \cmidrule(r){1-3} \cmidrule(l){4-6}
    \textbf{Unit \#} & \cellEtoile{Paris} & \cellMunich{Munich} & \textbf{Unit \#} & \cellEtoile{Paris} & \cellMunich{Munich} \\
    \midrule
      2 & \cellEtoile{$2.2{\times}10^{5}$} & \cellMunich{$6.7{\times}10^{5}$} &   8 & \cellEtoile{$4.3{\times}10^{4}$} & \cellMunich{$1.2{\times}10^{4}$} \\
      4 & \cellEtoile{64.48} & \cellMunich{140.60} &  16 & \cellEtoile{14.65} & \cellMunich{24.58} \\
      6 & \cellEtoile{7.46} & \cellMunich{9.26} &  24 & \cellEtoile{7.68} & \cellMunich{8.37} \\
      8 & \cellEtoile{6.49} & \cellMunich{7.37} &  32 & \cellEtoile{6.74} & \cellMunich{6.95} \\
     12 & \cellEtoile{6.33} & \cellMunich{7.41} &  48 & \cellEtoile{6.50} & \cellMunich{6.59} \\
     16 & \cellEtoile{6.32} & \cellMunich{7.46} &  64 & \cellEtoile{6.50} & \cellMunich{6.59} \\
    \bottomrule
    \end{tabular}
    \vspace{-8pt}
\end{table}

\begin{table}[!b]
    \centering
    \caption{The PPL of {\name} at ISM ({0.915}\thinspace{GHz}) and mmWave ({28}\thinspace{GHz}) bands with different $(\antNum, \unitNum)$ combinations.}
    \label{tab:ppl_band_main}
    \small
    \setlength{\tabcolsep}{2.5pt}
    \renewcommand{\arraystretch}{1.1}
    \vspace{-6pt}
    \begin{tabular}{c|c|c|c|cc|cc}
    \toprule
    \textbf{Freq.} & \textbf{BW} & \textbf{Power} & \multirow{2}{*}{\textbf{User}} & \multicolumn{2}{c|}{\textbf{$\antNum={4}$}} & \multicolumn{2}{c}{\textbf{$\antNum={8}$}} \\
    \cmidrule(lr){5-6} \cmidrule(l){7-8}
    \textbf{(GHz)} & \textbf{(MHz)} & \textbf{(dBm)} & & \textbf{$\unitNum={8}$} & \textbf{$\unitNum={16}$} & \textbf{$\unitNum={32}$} & \textbf{$\unitNum={64}$} \\
    \midrule
    \multirow{2}{*}{\textbf{0.915}} & \multirow{2}{*}{25} & \multirow{2}{*}{30/40} & \cellEtoile{Paris} & \cellEtoile{6.73} & \cellEtoile{6.35} & \cellEtoile{6.32} & \cellEtoile{6.32} \\
    & & & \cellMunich{Mun.} & \cellMunich{7.32} & \cellMunich{7.35} & \cellMunich{6.62} & \cellMunich{6.72} \\
    \midrule
    \multirow{2}{*}{\textbf{3.5}} & \multirow{2}{*}{100} & \multirow{2}{*}{50/60} & \cellEtoile{Paris} & \cellEtoile{6.49} & \cellEtoile{6.32} & \cellEtoile{6.74} & \cellEtoile{6.50} \\
    & & & \cellMunich{Mun.} & \cellMunich{7.37} & \cellMunich{7.46} & \cellMunich{6.95} & \cellMunich{6.59} \\
    \midrule
    \multirow{2}{*}{\textbf{28}} & \multirow{2}{*}{400} & \multirow{2}{*}{80/90} & \cellEtoile{Paris} & \cellEtoile{7.07} & \cellEtoile{6.35} & \cellEtoile{8.59} & \cellEtoile{6.78} \\
    & & & \cellMunich{Mun.} & \cellMunich{198.95} & \cellMunich{$1.4{\times}10^{3}$} & \cellMunich{402.40} & \cellMunich{$4.1{\times}10^{3}$} \\
    \bottomrule
    \end{tabular}
\end{table}

\myparatight{Impact of the antenna and calibration unit number}
The joint impact of $\antNum$ and $\unitNum$ on LLaMA-3.1-8B's PPL is shown in Tab.~\ref{tab:ppl_unit}, where we consider $\antNum=\left\{4, 8\right\}$ with different $\unitNum$.
Given $\antNum={4}$, the PPL falls from above {$10^{5}$} at {2} units to {6.49/7.37} at {8} units in Paris/Munich and then saturates; as for $\antNum={8}$, the saturation point moves to around {32} units, with {6.74/6.95} PPLs, respectively.
Theoretically, a perfect inversion takes $\unitNum=\antNum^2$ units in {\OptApprox}; in practice, fortunately, approximately $\antNum^2/2$ units already achieve a good PPL on LLM applications, saving approximately {1/2} energy consumption.
This is because a realistic channel comprises a limited number of multi-path components, whose singular values decay rapidly.

\myparatight{Performance in other bands}
We repeat the sweeps above in the {0.915}\thinspace{GHz} ISM band with {25}\thinspace{MHz} bandwidth, and the {28}\thinspace{GHz} mmWave band with {400}\thinspace{MHz} bandwidth, on the same two users in Paris/Munich, as summarized in Tab.~\ref{tab:ppl_band_main}.
Moving the frequency changes both the PL and the richness of the multipath; the former is compensated by setting its own broadcast power.
The PPLs at {0.915}\thinspace{GHz} are accordingly comparable to those at {3.5}\thinspace{GHz}, within {0.42} of each other across the eight $(\antNum, \unitNum)$ combinations, whereas the sparser multipath at {28}\thinspace{GHz} leaves the Munich user unusable, with PPLs always $>10^2$ even at {90}\thinspace{dBm}.
The full results can be found in Appendix~\ref{app:supp-factors}.

\subsection{Broadcasting to Massive Edge Users}
\label{sec:eval-llm-user}

Finally, we move to the {20}-edge-user scenes served by a single central radio's broadcasting LLM weights.

\begin{figure}[t]
    \centering
    \begin{minipage}{0.49\linewidth}
        \centering
        \includegraphics[width=\linewidth]{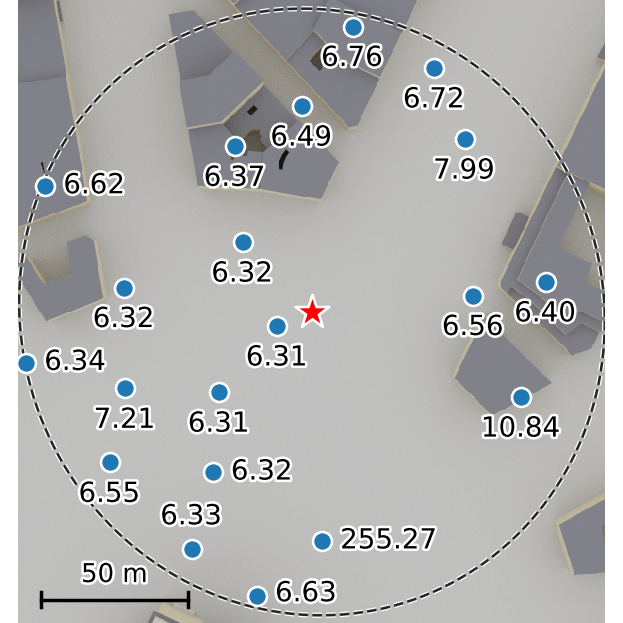}
        \subcaption{Paris}
        \label{fig:llm-user-etoile}
        \vspace{-7pt}
    \end{minipage}
    \hfill
    \begin{minipage}{0.49\linewidth}
        \centering
        \includegraphics[width=\linewidth]{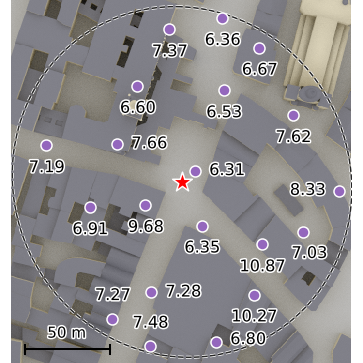}
        \subcaption{Munich}
        \label{fig:llm-user-munich}
        \vspace{-7pt}
    \end{minipage}
    \caption{The broadcasting of LLaMA-3.1-8B to the {20} users with WikiText-2 PPL labeled in Paris (a) and Munich (b).}
    \vspace{-12pt}
    \label{fig:llm-user}
\end{figure}
\setcounter{topnumber}{1}

\begin{figure}[t]
    \centering
    \begin{minipage}{0.49\linewidth}
        \centering
        \includegraphics[width=\linewidth]{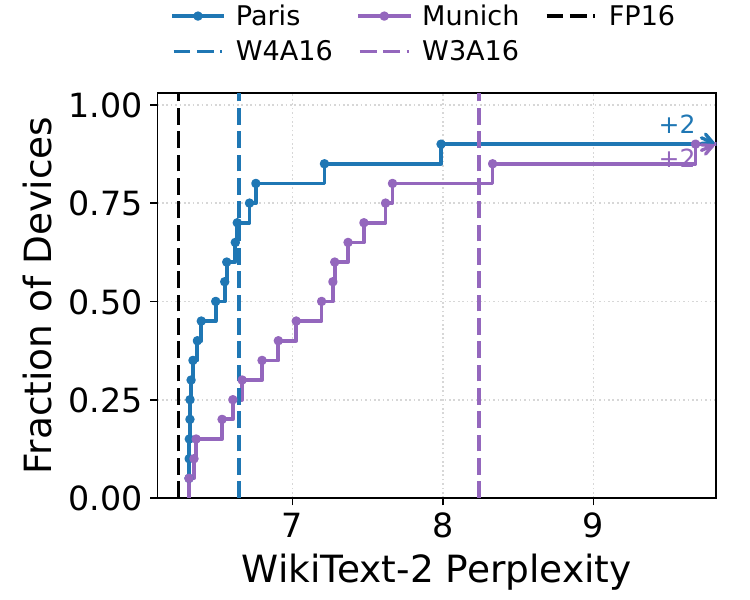}
        \par\vspace{-4pt}
        \subcaption{PPL}
        \label{fig:llm-user-cdf-ppl}
        \vspace{-7pt}
    \end{minipage}
    \hfill
    \begin{minipage}{0.49\linewidth}
        \centering
        \includegraphics[width=\linewidth]{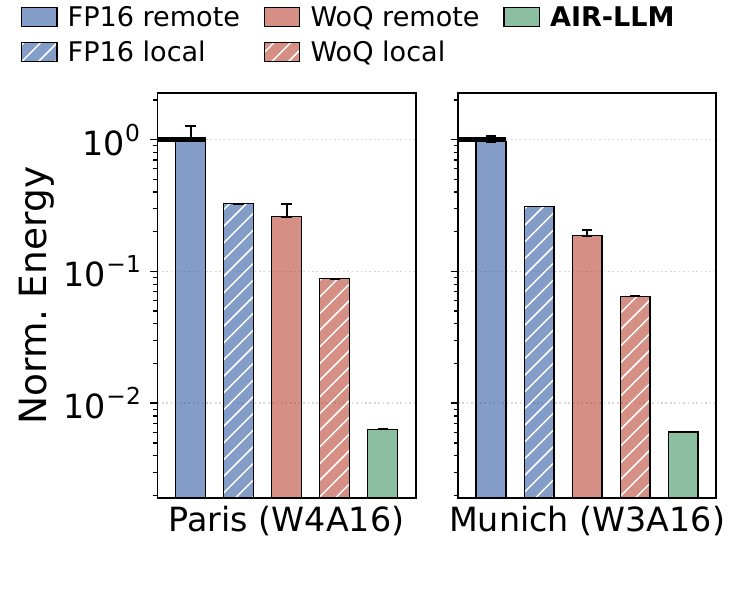}
        \par\vspace{-4pt}
        \subcaption{Energy per token}
        \label{fig:llm-user-cdf-energy}
        \vspace{-7pt}
    \end{minipage}
    \caption{In both scenes, the PPL CDFs vs. those by FP/WoQ baselines (a), and the energy per token comparison (b).}
    \vspace{-12pt}
    \label{fig:llm-user-cdf}
\end{figure}

\myparatight{PPL over the edge users}
Fig.~\ref{fig:llm-user} marks the positions of the {20} edge users per scene, spanning PLs of {70.0--99.9}\thinspace{dB} in Paris and {68.6--105.0}\thinspace{dB} in Munich.
We also label each user with its respective PPL on LLaMA-3.1-8B using {\name}, and the CDFs are further given in Fig.~\ref{fig:llm-user-cdf-ppl}.
Specifically for Paris, the median PPL is {6.52}, better than the W4A16 baseline of {6.65}, and in Munich it is {7.23}, better than the W3A16 baseline of {8.24}; {70\%} and {80\%} of the users, respectively, fall below these two baselines.
Note that one extreme user in Paris suffers an outlier PPL of {255.27}.
Despite its PL of {84.0}\thinspace{dB}, its received power concentrates on too few multi-path components to support spatial multiplexing, which leaves its channel matrix $\channel$ nearly singular.
Such a degradation also appears for the communication-based baselines whose channel capacity collapses on the same link; for {\name}, it can be alleviated by falling back to fewer antennas $\antNum$, which trades the airtime for a better-conditioned $\channel$.

\myparatight{Energy over the edge users}
The averaged per-token energy of {\name} over the same set of {20} users against the four baselines of each scene is shown in Fig.~\ref{fig:llm-user-cdf-energy}.
In Paris, {\name} consumes {160.5}$\times$/{41.1}$\times$ less mean energy than the remote FP16 and W4A16 baselines, and {51.5}$\times$/{13.9}$\times$ less than the two local ones; similar numbers of {160.6}$\times$/{31.2}$\times$ and {51.5}$\times$/{10.7}$\times$ can also be observed in Munich comparing to the W3A16 baselines.
{\name}'s energy consumption is constant regardless of users' positions and wireless channels, and is stably better than both the communication-based remote baselines, whose energy follows each user's own link, and the data movement-based local ones.

\begin{figure}[t]
    \centering
    \includegraphics[width=\linewidth]{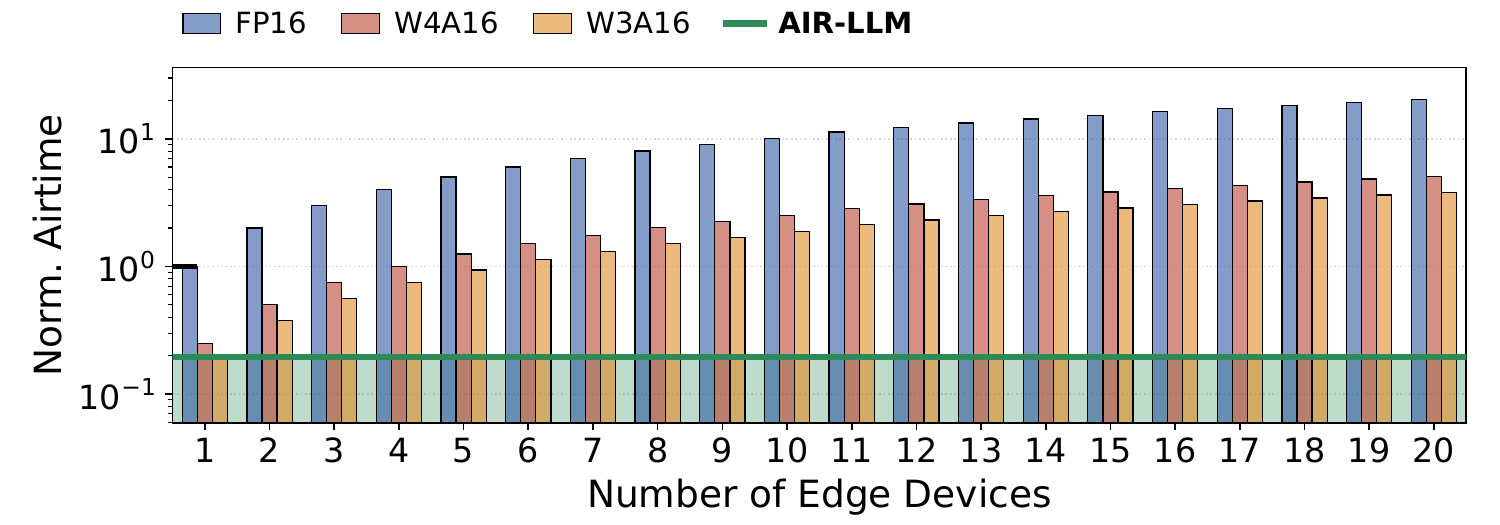}
    \caption{The cumulative airtime over $\userNum = \left\{1 \dots 20\right\}$ by {\name} and the FP16/WoQ baselines in Paris.}
    \vspace{-12pt}
    \label{fig:llm-user-airtime}
\end{figure}

\myparatight{Airtime over the edge users}
Fig.~\ref{fig:llm-user-airtime} plots the total airtime that one token takes while the first $\userNum$ users are served.
The communication baselines are based on unicasting, where every user is served separately at its own MCS customized to its channel quality.
Thus, serving $\userNum$ users costs $\userNum \times$ the airtime of a single user, and the total grows linearly.
In comparison, {\name} broadcasts the weights for the {20} users simultaneously, because the central radio is user-unaware and one uniform transmission is fed to every user inside the coverage.
Therefore, the airtime of {\name} is constant over $\userNum$, which is {104.1}$\times$/{26.0}$\times$ less than that by the FP16/W4A16 baselines with {20} users.
Munich shows the same in Fig.~\ref{fig:llm-user-airtime-munich} of Appendix~\ref{app:supp-users}.
This broadcast- and user-scalable feature lets {\name} serve the thriving number of future edge users.

\section{Related Works}

\myparatight{LLM accelerators}
Serving an LLM is memory-bounded by the weight movement rather than arithmetic~\cite{guo2025survey}.
One category of works shrinks the bits per weight and co-designs the compute units around the resulting mixed-format GEMVs~\cite{frantar2023gptq, lin2024awq, xiao2023smoothquant, tseng2024quip}, with the hardware absorbing the outliers~\cite{guo2023olive, lee2024tender} or the dequantization~\cite{park2025figlut, mo2025lut, wu2026oasis}; a second attacks the movement rather than the format, placing the compute next to the memory that holds the weights and the KV cache~\cite{park2024attacc, heo2024neupims, yun2024duplex}.
All share one premise: the weights are already stored in the device's memory, and only the cost of fetching and multiplying a resident weight is optimized.
{\name} removes that premise: no weight is stored on the edge device; each arrival of the broadcast weight waveform is consumed and computed directly in the RF/analog domain.

\myparatight{Analog computing for ML inference}
Analog computing performs the FLOPs in physics rather than in digital, escaping the energy floor of digital arithmetic: in-memory computing completes a GEMV on a ReRAM, PCM, or SRAM crossbar through Ohm's and Kirchhoff's laws~\cite{shafiee2016isaac, wan2022neurram, legallo2023multicore}, while photonic computing multiplies operands encoded onto light~\cite{shen2017deep, zhu2024lightening, zhong2023lightning}.
However, such designs are usually limited by their analog error and readout converters~\cite{zhang2025aim, khaddamaljameh2022hermes}, yet already carry real models at a handful of effective bits~\cite{ambrogio2023analogai, zhou2025hundred, wu2024bwq}; {\name} likewise reports every result at the effective bits its link resolves~\cite{dettmers2023case}.
Closest in spirit is the work that lets operand \emph{delivery} and computation happen in the same physical layer, over fiber~\cite{sludds2022delocalized} or over the air~\cite{reus2023airfc, gao2026disaggregated, gao2026radio}, the latter inside a commodity receiver's RF mixer; unlike over-the-air computation, which superposes many transmitters into a \emph{sum}~\cite{nazer2007computation, zhu2020broadband}, {\name}'s mixer forms a \emph{product} of an arriving weight waveform with a local input.
In addition, {\name} generalizes the single-antenna and single-weight-stream into an $\antNum \times \antNum$ MIMO engine to largely reduce the airtime, whose calibration units zero-force a real, frequency-selective channel.

\myparatight{LLM deployment on the edge}
On-device LLM serving is a capacity problem before a compute one, and the systems response keeps as few weights resident as possible: pruning~\cite{niu2020patdnn}, activation sparsity~\cite{song2024powerinfer, wang2024coreinfer}, flash streaming~\cite{alizadeh2024llmflash}, or accelerator--model co-design~\cite{yu2024cambriconllm}.
Yet even a quantized resident model stays memory-bound on commodity phones, where energy and thermals make continuous generation untenable~\cite{laskaridis2024melt}.
The complementary response moves the model off the device~\cite{kwon2023vllm, stojkovic2025dynamollm, kang2017neurosurgeon}, trading the memory wall for a per-query network cost and for handing the prompt to the server.
Both leave each device paying, privately and per token, for the same weights that every other device in the cell also pays for~\cite{qu2024trimcaching}.
{\name} instead disaggregates the weight storage to the central radio while keeping both the computation and the prompt on the edge, and because its transmission is edge-device-unaware, one broadcast is mixed by every device in the coverage at once, amortizing the delivery cost over the devices served.
Such one-to-many analog delivery has precedent in SoftCast~\cite{jakubczak2011softcast}, which broadcasts video over raw OFDM at a fidelity set by each receiver's own channel.

\section{Conclusion}

In this paper, we present {\name}, an LLM inference architecture for edge users that is \emph{memory-access-free}, \emph{user-scalable} and \emph{energy-efficient}.
Specifically, {\name} exploits MIMO spatial multiplexing to deliver multiple weight streams at a time and thereby shorten the airtime, and the design of the precoder--postcoder pair on the edge user energy-efficiently calibrates its own wireless channel.
Exhaustive evaluations on the NVIDIA Sionna ray-traced channels and the profiling of a real RF mixer reveal that, with a slight PPL gap to the FP16 baseline, {\name} cuts the per-token energy and airtime by orders of magnitude.
We believe {\name} opens up a new path for the next generation of edge intelligence.

\begin{acks}
This work was supported in part by the NSF AI Institute for Edge Computing Leveraging Next Generation Networks (Athena) under award \#2112562, by the Army Research Office (ARO) under grant W911NF-25-1-0241, and by the Air Force Research Laboratory (AFRL) under Cooperative Agreement No. FA8750-25-2-0500.
\end{acks}

\bibliographystyle{ACM-Reference-Format}
\bibliography{reference}

\clearpage
\appendix
\setcounter{topnumber}{2}
\setcounter{figure}{0}
\setcounter{table}{0}
\renewcommand{\thefigure}{A\arabic{figure}}
\renewcommand{\thetable}{A\arabic{table}}
\newcommand{\notationHead}{%
  \toprule
  Symbol & Dimension & Description \\
  \midrule
}

\newcommand{\notationRowsMain}{%
  $\antIdx$ & $0,\ldots,\antNum-1$ & Antenna index \\
  $\antTxIdx$ & $0,\ldots,\antNum-1$ & Transmit-antenna index \\
  $\antRxIdx$ & $0,\ldots,\antNum-1$ & Receive-antenna index \\
  $\outputIdx$ & $0,\ldots,\outputSize-1$ & Output index of an ML layer \\
  $\specIdx$ & $0,\ldots,\fftSize-1$ & Subcarrier index \\
  $\unitIdx$ & $0,\ldots,\unitNum-1$ & Calibration unit index \\
  \midrule
  $\band$ & $\mathbb{R}$ & Accessible bandwidth \\
  $\fftSize$ & $\mathbb{Z}$ & FFT size, $\fftSize = \inputSize$ \\
  $\freqSub$ & $\mathbb{R}$ & Subcarrier spacing \\
  $\antNum$ & $\mathbb{Z}$ & Number of antennas (Tx and Rx) \\
  $\antTxNum$ & $\mathbb{Z}$ & Number of transmit antennas \\
  $\antRxNum$ & $\mathbb{Z}$ & Number of receive antennas \\
  $\inputSize$ & $\mathbb{Z}$ & Number of inputs per ML layer \\
  $\outputSize$ & $\mathbb{Z}$ & Number of outputs per ML layer \\
  $\unitNum$ & $\mathbb{Z}$ & Number of calibration units \\
  $\userNum$ & $\mathbb{Z}$ & Number of edge users served by one broadcast \\
  $\noise^2$ & $\mathbb{R}$ & Total noise power across subcarriers \\
  $\myGauss$ & -- & I.i.d. standard complex Gaussian distribution~\eqref{eq: preliminaries mimo channel} \\
  \midrule
  $\weightFC$ & $\mathbb{R}^{\outputSize\times\inputSize}$ & Weight matrix of a GEMV layer \\
  $\inputFC$ & $\mathbb{R}^{\inputSize}$ & Input vector of a GEMV layer \\
  $\outputFC$ & $\mathbb{R}^{\outputSize}$ & Output vector of a GEMV layer \\
  \midrule
  $\signalSpecTx$ & $\mathbb{C}^{\antTxNum\times\fftSize}$ & Transmitted frequency-domain signals~\eqref{eq: preliminaries mimo channel} \\
  $\signalSpecRx$ & $\mathbb{C}^{\antRxNum\times\fftSize}$ & Received frequency-domain signals~\eqref{eq: preliminaries mimo channel} \\
  $\weightSpec$ & $\mathbb{C}^{\fftSize}$ & Weight signal on the subcarriers~\eqref{eq: preliminaries linear convolution} \\
  $\inputSpec$ & $\mathbb{C}^{\fftSize}$ & Input signal on the subcarriers~\eqref{eq: preliminaries linear convolution} \\
  $\outputSpec$ & $\mathbb{C}^{2\fftSize-1}$ & Mixer output signal~\eqref{eq: preliminaries linear convolution} \\
  \midrule
  $\channel$ & $\mathbb{C}^{\antNum\times\antNum\times\fftSize}$ & MIMO channel tensor \\
  $\weightMat$ & $\mathbb{C}^{\antNum\times\fftSize}$ & Transmitted weight signals~\eqref{eq: preliminaries mimo channel} \\
  $\weightVec_{\antIdx}$ & $\mathbb{C}^{\fftSize}$ & Weight signal transmitted by antenna $\antIdx$ \\
  $\inputVec$ & $\mathbb{C}^{\fftSize}$ & Shared input signal \\
  $\outputMat$ & $\mathbb{C}^{\antNum}$ & Ground-truth output signals \\
  $\outputVec_{\antIdx}$ & $\mathbb{C}$ & Ground-truth output signal at antenna $\antIdx$ \\
  $\weightMatRx$ & $\mathbb{C}^{\antNum\times\fftSize}$ & Received weight signals~\eqref{eq: preliminaries mimo channel} \\
  $\weightVecRx_{\antIdx}$ & $\mathbb{C}^{\fftSize}$ & Weight signal received at antenna $\antIdx$ \\
  \midrule
  $\precoder$ & $\mathbb{C}^{\unitNum\times\fftSize}$ & Precoder coefficients, jointly across units \\
  $\postcoder$ & $\mathbb{C}^{\unitNum\times\antNum\times\antNum}$ & Postcoder coefficients, jointly across units \\
  $\inputVecPre_{\unitIdx}$ & $\mathbb{C}^{\fftSize}$ & Precoded input signal at unit $\unitIdx$~\eqref{eq: design precoding} \\
  $\outputVecEst_{\unitIdx,\antIdx}$ & $\mathbb{C}$ & Pre-postcoding output estimate~\eqref{eq: design convolution} \\
  $\outputVecPostEst_{\unitIdx,\antIdx}$ & $\mathbb{C}$ & Postcoded output estimate~\eqref{eq: design postcoding} \\
  $\outputVecSumEst_{\antIdx}$ & $\mathbb{C}$ & Combined output estimate across units~\eqref{eq: design calibration unit sum} \\
  $\outputMatPostEst$ & $\mathbb{C}^{\antNum}$ & Postcoded output estimates across antennas \\
  $\outputMatSumEst$ & $\mathbb{C}^{\antNum}$ & Combined output estimates across antennas \\
  $\postcoderFlat$ & $\mathbb{C}^{\antNum^2\times\unitNum}$ & Flattened postcoder~\eqref{eq: matrix operation opt} \\
  $\channelInvFlat$ & $\mathbb{C}^{\antNum^2\times\fftSize}$ & Flattened per-subcarrier channel inverse~\eqref{eq: matrix operation opt} \\
  $\precoderInit$ & $\mathbb{C}^{\unitNum\times\fftSize}$ & SVD-based initialization of $\precoder$ \\
  $\postcoderInit$ & $\mathbb{C}^{\unitNum\times\antNum\times\antNum}$ & SVD-based initialization of $\postcoder$ \\
  $\postcoderFlatInit$ & $\mathbb{C}^{\antNum^2\times\unitNum}$ & Flattened form of $\postcoderInit$ \\
  $\precoderOpt$ & $\mathbb{C}^{\unitNum\times\fftSize}$ & Optimized precoder \\
  $\postcoderOpt$ & $\mathbb{C}^{\unitNum\times\antNum\times\antNum}$ & Optimized postcoder \\
}

\newcommand{\notationRowsApp}{%
  $\blockIdx$ & $0,\ldots,\blockNum-1$ & Block index \\
  $\blockNum$ & $\mathbb{Z}$ & Number of blocks \\
  $\blockSizeOutput$ & $\mathbb{Z}$ & Output block size (subcarriers per block) \\
  $\blockSizeInput$ & $\mathbb{Z}$ & Input block size \\
  $\blockSizeWeight$ & $\mathbb{Z}$ & Weight block size \\
  $\channelBlock$ & $\mathbb{C}^{\antNum\times\antNum\times\blockNum}$ & Per-block averaged channel \\
  \midrule
  $\fftSize$ & $\mathbb{Z}$ & FFT size, $\fftSize = \blockNum\blockSizeOutput$ \\
  $\weightVec$ & $\mathbb{C}^{\blockNum\times\blockSizeWeight}$ & Weight operand of a BCO~\eqref{eq: preliminaries bco} \\
  $\inputVec$ & $\mathbb{C}^{\blockNum\times\blockSizeInput}$ & Input operand of a BCO~\eqref{eq: preliminaries bco} \\
  $\outputVec$ & $\mathbb{C}^{\blockSizeOutput}$ & Output block of a BCO~\eqref{eq: preliminaries bco} \\
  $\weightSpec, \inputSpec$ & $\mathbb{C}^{\blockNum\blockSizeOutput}$ & Zero-padded weight and input signals~\eqref{eq: preliminaries zero padding} \\
  $\outputSpec$ & $\mathbb{C}^{\blockSizeOutput}$ & Mixer output signal after the LPF \\
  $\precoder$ & $\mathbb{C}^{\unitNum\times\blockNum}$ & Per-block precoder coefficients \\
  $\channelInvFlat$ & $\mathbb{C}^{\antNum^2\times\blockNum}$ & Flattened per-block channel inverse~\eqref{eq: channel inverse flat} \\
  \midrule
  $\kernelSize$ & $\mathbb{Z}$ & Conv2D kernel size, $\kernelSize \times \kernelSize$ \\
  $\imageSize$ & $\mathbb{Z}$ & Conv2D image size, $\imageSize \times \imageSize$ \\
  $\convRatio$ & $\mathbb{R}$ & Zero-padding overhead of Conv2D~\eqref{eq: cnn padding ratio} \\
  $\weightCONV$ & $\mathbb{R}^{\inputSize\times\outputSize\times\kernelSize\times\kernelSize}$ & Kernels of a Conv2D layer \\
  $\inputCONV$ & $\mathbb{R}^{\inputSize\times\imageSize\times\imageSize}$ & Input feature map of a Conv2D layer \\
  $\outputCONV$ & $\mathbb{R}^{\outputSize\times\imageSize\times\imageSize}$ & Output feature map of a Conv2D layer \\
  \midrule
  $\airtimeMvm$ & $\mathbb{R}$ & Airtime of one round~\eqref{eq: bco airtime round} \\
  $\airtimeMac$ & $\mathbb{R}$ & Airtime per FLOP~\eqref{eq: bco airtime per flop} \\
  $\energyMvm$ & $\mathbb{R}$ & Energy consumption of one module~\eqref{eq: bco energy breakdown} \\
  $\energyMvmPre$ & $\mathbb{R}$ & Precoding energy~\eqref{eq: bco energy breakdown} \\
  $\energyMvmIFFT$ & $\mathbb{R}$ & IFFT energy~\eqref{eq: bco energy breakdown} \\
  $\energyMvmDAC$ & $\mathbb{R}$ & DAC energy~\eqref{eq: bco energy breakdown} \\
  $\energyMvmADC$ & $\mathbb{R}$ & ADC energy~\eqref{eq: bco energy breakdown} \\
  $\energyMvmFFT$ & $\mathbb{R}$ & FFT energy~\eqref{eq: bco energy breakdown} \\
  $\energyMvmPost$ & $\mathbb{R}$ & Postcoding energy~\eqref{eq: bco energy breakdown} \\
  $\energyMvmSum$ & $\mathbb{R}$ & Cross-unit summation energy~\eqref{eq: bco energy breakdown} \\
}

\newcommand{\notationTableMain}{%
  \captionof{table}{Summary of the notation used in the main body.}
  \label{tab:notation}
  \begin{tabular}{@{}cll@{}}
  \notationHead
  \notationRowsMain
  \bottomrule
  \end{tabular}%
}

\newcommand{\notationTableApp}{%
  \captionof{table}{Additional notation used only in the appendices.}
  \label{tab:notation-appendix}
  \begin{tabular}{@{}cll@{}}
  \notationHead
  \notationRowsApp
  \bottomrule
  \end{tabular}%
}

\makeatletter
\if@twocolumn
\twocolumn[{%
\section{Table of Notations}
\label{app:notation}

\begin{center}
  \footnotesize
  \setlength{\tabcolsep}{4pt}
  \begin{minipage}[t]{0.50\textwidth}
    \centering
    \notationTableMain
  \end{minipage}
  \hfill
  \begin{minipage}[t]{0.47\textwidth}
    \centering
    \notationTableApp
  \end{minipage}
\end{center}
}]
\else
\clearpage
\section{Table of Notations}
\label{app:notation}

\begin{center}
  \footnotesize
  \notationTableMain
\end{center}
\clearpage
\begin{center}
  \footnotesize
  \notationTableApp
\end{center}
\fi
\makeatother

\clearpage

\section{Blocked Correlation Operation}
\label{app:bco}
\label{app:opt-approx}

We generalize the linear-convolution mapping into a \emph{blocked correlation operation} (BCO), where the mixer retains a whole output \emph{block} of subcarriers rather than a single one.

\subsection{Definition and Realization}
\label{app:bco-def}

\myparatight{Definition of BCO}
We define the BCO by $\star$.
Its two operands, weights and inputs, are both complex valued, given by $\weightVec \in \mathbb{C}^{\blockNum \times \blockSizeWeight}$ and $\inputVec \in \mathbb{C}^{\blockNum \times \blockSizeInput}$, each consisting of $\blockNum$ blocks.
These $\blockNum$ blocks are correlated in parallel, and their correlation outputs are accumulated into a single output block $\outputVec \in \mathbb{C}^{\blockSizeOutput}$, defined as
\begin{align}
    \outputVec[\specIdx] &= \left(\weightVec \star \inputVec\right)[\specIdx] \notag \\
    &= \sum_{\blockIdx=0}^{\blockNum-1} \sum_{\specIdx^{\prime}=\max(0,\, \specIdx-\blockSizeInput+1)}^{\min(\specIdx,\, \blockSizeWeight-1)} \weightVec[\blockIdx, \specIdx^{\prime}] \cdot \inputVec[\blockIdx, \specIdx^{\prime} - \specIdx + \blockSizeInput - 1],
    \label{eq: preliminaries bco}
\end{align}
where the block sizes satisfy $\blockSizeOutput = \blockSizeWeight + \blockSizeInput - 1$.
A BCO thus costs $\blockNum\blockSizeWeight\blockSizeInput$ additions and as many multiplications, i.e., $2\blockNum\blockSizeWeight\blockSizeInput$ FLOPs.
The GEMV mapping of {\eqref{eq: preliminaries gemv mapping}} is the degenerate case $\blockNum = \inputSize$ and $\blockSizeWeight = \blockSizeInput = \blockSizeOutput = 1$, in which every block holds one scalar and {\eqref{eq: preliminaries bco}} reduces to the $\inputSize$-element inner product of {\eqref{eq: preliminaries gemv output}}.

\myparatight{Realizing BCO with an RF mixer}
In the frequency domain, an RF mixer convolves two analog signals, $\weightSpec$ and $\inputSpec$, into $\outputSpec$, which is the correlation of {\eqref{eq: preliminaries bco}} once one operand is reversed in subcarrier order, as in {\eqref{eq: preliminaries gemv mapping}}.
To realize a BCO, we let the FFT size be $\fftSize = \blockNum \cdot \blockSizeOutput$, i.e., one length-$\blockSizeOutput$ subcarrier slot per block.
The frequency-domain weight signal $\weightSpec \in \mathbb{C}^{\blockNum \blockSizeOutput}$ is obtained from $\weightVec$ by zero-padding each block $\weightVec[\blockIdx, :]$ into the length-$\blockSizeOutput$ segment $\weightSpec[\blockIdx \blockSizeOutput : (\blockIdx+1)\blockSizeOutput - 1]$, and the input signal $\inputSpec \in \mathbb{C}^{\blockNum \blockSizeOutput}$ is obtained from $\inputVec$ in the same way, i.e.,
\begin{align}
    \weightSpec[\blockIdx \blockSizeOutput + \specIdx] &=
        \begin{cases}
            \weightVec[\blockIdx, \specIdx], & \specIdx < \blockSizeWeight, \\
            0, & \specIdx \geq \blockSizeWeight,
        \end{cases}
    \label{eq: preliminaries zero padding} \\
    \inputSpec[\blockIdx \blockSizeOutput + \specIdx] &=
        \begin{cases}
            \inputVec[\blockIdx, \specIdx], & \specIdx < \blockSizeInput, \\
            0, & \specIdx \geq \blockSizeInput,
        \end{cases}
        \notag
\end{align}
for all $\blockIdx = 0, 1, \dots, \blockNum-1$ and $\specIdx = 0, 1, \dots, \blockSizeOutput-1$.
The zero padding guarantees that the correlation of one block never leaks into the subcarrier slot of the next.
Correlating $\weightSpec$ with $\inputSpec$ spans $2\blockNum-1$ slots, and every matching block pair $(\blockIdx, \blockIdx)$ lands in the \emph{same}, middle slot of index $\blockNum-1$, which therefore accumulates the correlation outputs of all $\blockNum$ blocks, exactly as {\eqref{eq: preliminaries bco}} prescribes.
This middle slot is extracted by the LPF, whose output signal $\outputSpec \in \mathbb{C}^{\blockSizeOutput}$ satisfies $\outputSpec[\specIdx] = \outputVec[\specIdx]$ for all $\specIdx = 0, 1, \dots, \blockSizeOutput-1$.
Since the subcarrier spacing $\freqSub$ is shared by all three signals, this output occupies a bandwidth of only $\band / \blockNum$ and is digitized with $\blockNum\times$ fewer ADC samples than the full-band signal, which is where {\name}'s ADC energy saving comes from.

\myparatight{Implication on the calibration units}
The calibration unit of \S\ref{sec:precoder-postcoder} carries over unchanged, with one structural caveat on the precoder.
Under BCO, the $\blockSizeOutput$ subcarriers within a block jointly carry one correlation, so they must be scaled \emph{identically}: a per-subcarrier precoder would reweight the individual correlation taps and destroy the block structure of {\eqref{eq: preliminaries bco}}.
The precoder is therefore constant within each block, i.e., $\precoder \in \mathbb{C}^{\unitNum \times \blockNum}$ with the subcarrier $\specIdx$ driven by $\precoder[\unitIdx, \myFloor{\specIdx/\blockSizeOutput}]$, which is precisely the form optimized in {\eqref{eq: bco channel dependent opt}} of \S\ref{app:opt-derivation}.
The postcoder, in contrast, acts across antennas only and is applied identically to all subcarriers, so it is unaffected.
For a GEMV, $\blockSizeOutput = 1$ makes the per-block and the per-subcarrier precoder identical, recovering the main-body design.

\subsection{Pre-/Post-coder Optimization}
\label{app:opt-derivation}

Following the optimization derivation in \S\ref{sec:precoder-postcoder}, the BCO version of the optimization can be written as
\begin{tcolorbox}[
    boxsep=0mm,left=1mm,right=1mm,top=-1mm,bottom=1mm,
    toptitle=1mm,bottomtitle=1mm,lefttitle=2mm,
    title={{\OptOrig} under BCO},
    boxrule=1pt,sharp corners=all]
  \small
  \begin{align}
    \min_{\precoder, \postcoder} \ \mathbb{E}_{\inputVec, \weightMat}\left[ \sum_{\antIdx=0}^{\antNum-1} \big\| \outputVecSumEst_{\antIdx} - \outputVec_{\antIdx} \big\|_2^2 \right]
    \label{eq: bco pre-post-coder opt}
  \end{align}
\end{tcolorbox}
\myparatight{Waiving the need of $\weightMat$ and $\inputVec$}
The argument of \S\ref{sec:precoder-postcoder} carries over verbatim, as the BCO changes neither the objective nor its bilinearity in $(\weightMat, \inputVec)$.
Assuming zero-mean i.i.d.\ entries and a high SNR region, the expectation in \eqref{eq: bco pre-post-coder opt} again collapses to a positive multiple of a channel-dependent-only objective, which becomes
\begin{tcolorbox}[
    boxsep=0mm,left=1mm,right=1mm,top=-1mm,bottom=1mm,
    toptitle=1mm,bottomtitle=1mm,lefttitle=2mm,
    title={{\OptChan} under BCO},
    boxrule=1pt,sharp corners=all]
  \small
  \begin{align}
    \min_{\precoder, \postcoder} \sum_{\specIdx=0}^{\fftSize-1} \left\| \sum_{\unitIdx=0}^{\unitNum-1} \precoder\!\left[\unitIdx, \myFloor{\frac{\specIdx}{\blockSizeOutput}}\right] \cdot \postcoder[\unitIdx, :, :] \cdot \myTrans{\channel[:, :, \specIdx]} - \mathbf{I}_{\antNum} \right\|^2,
    \label{eq: bco channel dependent opt}
  \end{align}
\end{tcolorbox}
\noindent
where $\mathbf{I}_{\antNum}$ is the $\antNum \times \antNum$ identity matrix.
The only difference from {\OptChan} is the precoder index: the floor $\myFloor{\specIdx / \blockSizeOutput}$ ties all $\blockSizeOutput$ subcarriers of a block to one coefficient, as \S\ref{app:bco-def} requires.
Its two-step solution below follows \S\ref{sec:precoder-postcoder}: a closed-form initialization, then ALS fine-tuning.

\begin{figure*}[t]
    \centering
    \begin{minipage}{0.49\linewidth}
        \centering
        \includegraphics[width=\linewidth]{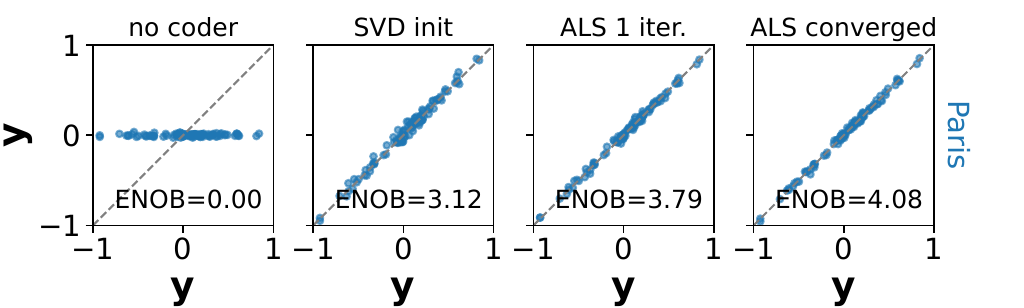}
        \subcaption{Paris: $\outputVecEst$ against $\outputVec$ at each stage of the solve}
        \label{fig:als-convergence-scatter-etoile}
    \end{minipage}
    \hfill
    \begin{minipage}{0.49\linewidth}
        \centering
        \includegraphics[width=\linewidth]{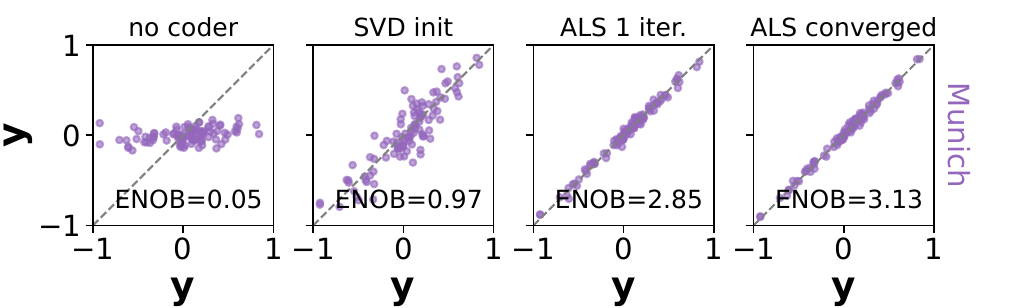}
        \subcaption{Munich: $\outputVecEst$ against $\outputVec$ at each stage of the solve}
        \label{fig:als-convergence-scatter-munich}
    \end{minipage}
    \begin{minipage}{0.49\linewidth}
        \centering
        \includegraphics[width=\linewidth]{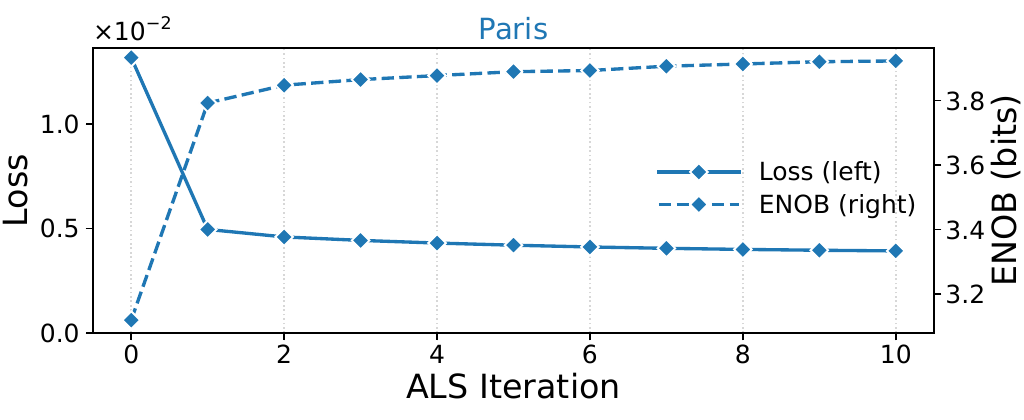}
        \subcaption{Paris: {\OptChan} loss and the ENOB per ALS iteration}
        \label{fig:als-convergence-rate-etoile}
    \end{minipage}
    \hfill
    \begin{minipage}{0.49\linewidth}
        \centering
        \includegraphics[width=\linewidth]{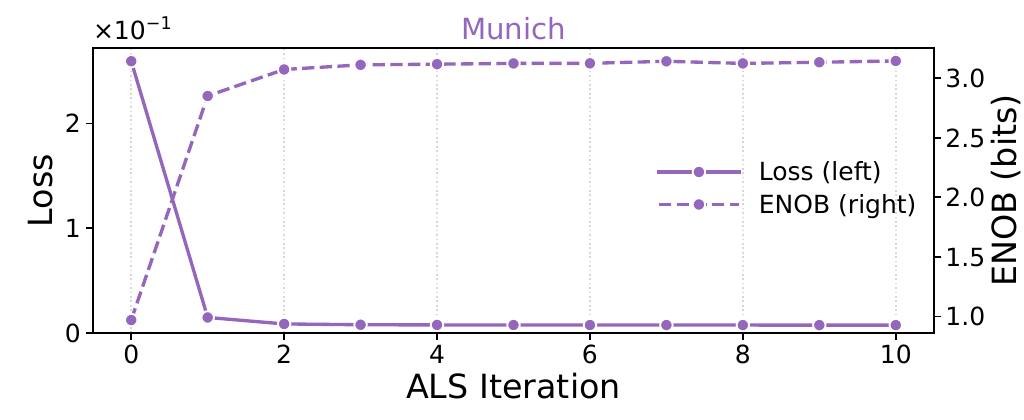}
        \subcaption{Munich: {\OptChan} loss and the ENOB per ALS iteration}
        \label{fig:als-convergence-rate-munich}
    \end{minipage}
    \caption{What the two steps of {\OptChan} buy on the selected users at $\antNum = {4}$ and $\unitNum = {8}$: $\outputVecEst$ against $\outputVec$ at four stages of the solve (a, b), and the {\eqref{eq: bco channel dependent opt}} loss and its ENOB per ALS iteration (c, d).}
    \vspace{-12pt}
    \label{fig:als-convergence}
\end{figure*}

\myparatight{Step 1: closed-form initialization}
Since one precoder coefficient now serves a whole block, we average the channel within each block into $\channelBlock \in \mathbb{C}^{\antNum \times \antNum \times \blockNum}$, which is accurate as long as the frequency response varies smoothly across the $\blockSizeOutput$ subcarriers of a block:
\begin{align}
    \channelBlock[\antTxIdx, \antRxIdx, \blockIdx] = \frac{1}{\blockSizeOutput} \sum_{\specIdx=0}^{\blockSizeOutput-1} \channel[\antTxIdx, \antRxIdx, \blockIdx\blockSizeOutput+\specIdx].
\end{align}
The optimization then collapses from $\fftSize$ subcarriers to $\blockNum$ blocks, i.e.,
\begin{tcolorbox}[
    boxsep=0mm,left=1mm,right=1mm,top=-1mm,bottom=1mm,
    toptitle=1mm,bottomtitle=1mm,lefttitle=2mm,
    title={{\OptChan} under BCO with per-block channel},
    boxrule=1pt,sharp corners=all]
  \small
  \begin{align}
    \min_{\precoder, \postcoder} \sum_{\blockIdx=0}^{\blockNum-1} \left\| \left( \sum_{\unitIdx=0}^{\unitNum-1} \precoder[\unitIdx, \blockIdx] \cdot \postcoder[\unitIdx, :, :] \right) \cdot \myTrans{\channelBlock[:, :, \blockIdx]} - \mathbf{I}_{\antNum} \right\|^2
    \label{eq: channel block opt}
  \end{align}
\end{tcolorbox}

As in {\OptApprox}, we then flatten the two antenna dimensions of the postcoder into $\postcoderFlat \in \mathbb{C}^{\antNum^2 \times \unitNum}$, i.e.,
\begin{align}
    \postcoderFlat[\antTxIdx \antNum + \antRxIdx, \unitIdx] = \postcoder[\unitIdx, \antTxIdx, \antRxIdx].
\end{align}
The flattened channel inverse $\channelInvFlat \in \mathbb{C}^{\antNum^2 \times \blockNum}$ is now inverted per block rather than per subcarrier, so its $\blockIdx$-th column is the zero-forcing solution of block $\blockIdx$ alone:
\begin{align}
    \channelInvFlat[\antTxIdx \antNum + \antRxIdx, \blockIdx] = \left[ \left( \myTrans{\channelBlock[:, :, \blockIdx]} \right)^{-1} \right]_{\antTxIdx, \antRxIdx}.
    \label{eq: channel inverse flat}
\end{align}
This turns \eqref{eq: channel block opt} into a standard rank-$\unitNum$ matrix approximation problem.
\begin{tcolorbox}[
    boxsep=0mm,left=1mm,right=1mm,top=-1mm,bottom=1mm,
    toptitle=1mm,bottomtitle=1mm,lefttitle=2mm,
    title={{\OptApprox} under BCO},
    boxrule=1pt,sharp corners=all]
  \small
  \begin{align}
    \min_{\postcoderFlat, \precoder} \left\| \postcoderFlat \cdot \precoder - \channelInvFlat \right\|^2.
  \end{align}
\end{tcolorbox}
\noindent
By the Eckart--Young--Mirsky theorem~\cite{mirsky1960symmetric}, this is solved in closed form by the SVD of $\channelInvFlat$ truncated to its $\unitNum$ largest singular values, whose unflattened result $\precoderInit$ and $\postcoderInit$ initializes \eqref{eq: bco channel dependent opt}.
Compared with the main body, Step~1 under BCO carries one extra approximation, the per-block channel of \eqref{eq: channel block opt}, on top of dropping the weighting by the channel magnitudes.

\myparatight{Step 2: alternating least squares (ALS) for fine-tuning}
Starting from $\precoderInit$ and $\postcoderInit$, we alternate two closed-form least-squares updates on \eqref{eq: bco channel dependent opt} until convergence, both of which reinstate the per-subcarrier channel that Step~1 averages away:
\begin{itemize}[leftmargin=*, topsep=2pt, itemsep=1pt]
    \item \textit{Precoder step}: fixing $\postcoder$, \eqref{eq: bco channel dependent opt} decouples across blocks, since $\precoder[\unitIdx, \blockIdx]$ only affects the $\blockSizeOutput$ subcarriers of block $\blockIdx$.
    Per block, the $\unitNum$ coefficients $\{\precoder[\unitIdx, \blockIdx]\}_{\unitIdx=0}^{\unitNum-1}$ fit the $\unitNum$ fixed bases $\{\postcoder[\unitIdx, :, :] \cdot \myTrans{\channel[:, :, \blockIdx\blockSizeOutput+\specIdx]}\}_{\unitIdx=0}^{\unitNum-1}$ to $\mathbf{I}_{\antNum}$, stacked over the $\blockSizeOutput$ subcarriers within that block.
    \item \textit{Postcoder step}: fixing $\precoder$, the postcoder is solved from one least-squares problem stacked over all $\fftSize$ subcarriers, each contributing its own channel $\myTrans{\channel[:, :, \specIdx]}$ and precoder coefficient $\precoder\!\left[\unitIdx, \myFloor{\specIdx/\blockSizeOutput}\right]$; it further decouples across the $\antNum$ output antennas.
\end{itemize}
Both steps are closed-form via the normal equations, so each iteration is cheap and, from the Step~1 initialization, only a few of them are needed.

\subsection{Convergence of the Optimization}
\label{app:opt-convergence}

We study how the two-step solve converges on the GEMV, i.e., the degenerate case $\blockSizeOutput = 1$ in which \eqref{eq: bco channel dependent opt} reduces to the {\OptChan} of \S\ref{sec:precoder-postcoder}.
The two edge devices are the selected users of Paris and Munich of \S\ref{sec:eval-gemm}, resolving approximately {4.00} and {3.00}\thinspace{bit}, so the numbers below are directly comparable to the evaluation.

\myparatight{What each step buys}
Fig.~\ref{fig:als-convergence} traces the two steps on the selected edge device of each scene.
Before either of them, the raw link is unusable: fitting the single complex gain that an AGC and a carrier recovery already provide leaves {0.00}\thinspace{bit} in Paris and {0.05}\thinspace{bit} in Munich, i.e., the frequency selectivity of $\channel$ and its inter-antenna mixing alone destroy the GEMV.
The closed-form Step-1 initialization recovers most of it at once, cutting the loss of \eqref{eq: bco channel dependent opt} from {0.998} to {0.013} in Paris and from {0.947} to {0.259} in Munich, i.e., to {3.12} and {0.97}\thinspace{bit}.
The gap between the two scenes is the price of the per-block channel approximation of \eqref{eq: channel block opt}: the richer multipath of Munich makes the frequency response vary within a block, so the averaged $\channelBlock$ that Step~1 inverts sits further from the per-subcarrier channel that \eqref{eq: bco channel dependent opt} actually scores, and the SVD initialization alone is worth about one bit there against three in Paris.

\myparatight{How many ALS iterations}
Figs.~\ref{fig:als-convergence-rate-etoile} and~\ref{fig:als-convergence-rate-munich} answer how long Step~2 has to run.
A single ALS pass recovers most of what Step~1 leaves behind, dropping the loss by a further {2.7}$\times$ in Paris and {17.5}$\times$ in Munich to reach {3.79} and {2.85}\thinspace{bit}.
Everything after that pass is a slow creep: iterations {2} through {10} add only {0.13}\thinspace{bit} in Paris and {0.29}\thinspace{bit} in Munich, and the relative-tolerance test stops the two devices at {40} and {22} iterations, at {4.08} and {3.13}\thinspace{bit}.
Since one iteration is two closed-form least-squares solves and the optimization runs once per channel coherence time rather than per GEMV, we run it to that convergence throughout the evaluation, though stopping after the first pass would already sit within {0.29}\thinspace{bit} of it.

\section{Airtime and Energy Consumption}
\label{app:analysis}

This section analyzes the two efficiency metrics of {\name}, the airtime and the energy consumption per FLOP.
We derive them at the general block size of a BCO (\S\ref{app:bco-airtime} and \S\ref{app:bco-energy}), and then substitute the GEMV mapping of the main body, its degenerate case, to read off the airtime and the energy of a GEMV (\S\ref{sec:throughput-analysis} and \S\ref{sec:energy-analysis}).

\subsection{Airtime of a BCO}
\label{app:bco-airtime}

We analyze the \emph{airtime} of {\name}, i.e., how long the RF chain of an edge device is occupied to compute one FLOP.
It is bottlenecked by the duration of the signals, both the wireless one carrying $\weightMat$ from the central radio and the local one carrying $\inputVec$, which the bandwidth $\band$ sets.

\myparatight{Airtime per round}
Each round of {\name} receives one OFDM symbol carrying $\weightMat$ across the $\fftSize = \blockNum\blockSizeOutput$ subcarriers, feeding $\antNum$ antennas/RF mixers for $2\antNum$ parallel BCOs, two of them packed per antenna into the real and the imaginary parts; a layer of $\outputSize$ outputs, all sharing the same input $\inputVec$, thus takes $\myCeil{\outputSize/(2\antNum)}$ rounds.
The airtime per round then scales as
\begin{align}
    \airtimeMvm \propto \frac{\blockNum\blockSizeOutput}{\band}.
    \label{eq: bco airtime round}
\end{align}

\myparatight{Airtime per FLOP}
Normalizing {\eqref{eq: bco airtime round}} by the FLOPs that the $2\antNum$ parallel BCOs complete within the round gives the per-FLOP airtime of a general BCO,
\begin{align}
    \airtimeMac \propto \frac{\blockSizeOutput}{\antNum\band\blockSizeWeight\blockSizeInput}.
    \label{eq: bco airtime per flop}
\end{align}
The block number $\blockNum$ cancels out: splitting a layer into more, smaller blocks lengthens the round and adds FLOPs in the same proportion.
More generally, {\eqref{eq: bco airtime per flop}} rewards a block layout whose output block is short relative to the product of the two input block sizes, i.e., one whose correlation reuses each transmitted weight over many taps.

\subsection{Energy Consumption of a BCO}
\label{app:bco-energy}

We analyze the edge-device-side energy consumption of {\name} per FLOP by its complexity, i.e., the number of operations on three kinds of hardware, each costing a constant energy per operation: the DAC and the ADC, counted in samples, and the digital computing, counted in FLOPs.

\myparatight{Energy per module}
We account for the energy of one whole module, i.e., all $\myCeil{\outputSize/(2\antNum)}$ rounds of $2\antNum$ BCOs, and then normalize by the FLOPs it completes, breaking the module energy into seven components along the signal chain,
\begin{align}
    \energyMvm = \energyMvmPre + \energyMvmIFFT + \energyMvmDAC + \energyMvmADC + \energyMvmFFT + \energyMvmPost + \energyMvmSum,
    \label{eq: bco energy breakdown}
\end{align}
i.e., the precoding, the IFFT, the DAC, the ADC, the FFT, the postcoding, and the cross-unit summation.
Each of the $\unitNum$ calibration units precodes $\inputVec$ per block and takes an $\fftSize$-point IFFT.
Both are paid only \emph{once} for the whole module: $\precoder$ depends solely on the channel and every round mixes the same $\inputVec$ with a new set of broadcast weights, so the precoded waveform is buffered and replayed across the rounds.
Within each of the $\myCeil{\outputSize/(2\antNum)} \approx \outputSize/(2\antNum)$ rounds, every unit then converts the waveform with one DAC over $\fftSize = \blockNum\blockSizeOutput$ samples, digitizes the LPF output of $\blockSizeOutput$ samples per antenna, and decodes it with a $\blockSizeOutput$-point FFT plus postcoding, before the $\unitNum$ units are summed.
Accumulating the latter five over the rounds gives the ``Energy per module'' column of Tab.~\ref{tab:bco_energy_breakdown}, and dividing each term by the $2\outputSize\blockNum\blockSizeWeight\blockSizeInput$ FLOPs of the module gives its ``Energy per FLOP'' column.
Note that the ADC term is charged for $\blockSizeOutput$ samples rather than $\fftSize$, since the BCO output occupies only the reduced bandwidth $\band/\blockNum$.

\begin{table}[htbp]
    \centering
    \caption{The energy complexity of one module on {\name} under BCO and the seven terms per FLOP.}
    \label{tab:bco_energy_breakdown}
    \small
    \setlength{\tabcolsep}{4pt}
    \renewcommand{\arraystretch}{2.0}
    \vspace{-6pt}
    \begin{tabular}{ccc}
    \toprule
    \textbf{Term} & \textbf{Energy per module} & \textbf{Energy per FLOP} \\
    \midrule
    Precoding & $\mathcal{O}\left(\unitNum\blockNum\blockSizeInput\right)$ & $\mathcal{O}\left(\dfrac{\unitNum}{\outputSize\blockSizeWeight}\right)$ \\
    IFFT & $\mathcal{O}\left(\unitNum\blockNum\blockSizeOutput\log_2(\blockNum\blockSizeOutput)\right)$ & $\mathcal{O}\left(\dfrac{\unitNum\blockSizeOutput\log_2(\blockNum\blockSizeOutput)}{\outputSize\blockSizeWeight\blockSizeInput}\right)$ \\
    DAC & $\mathcal{O}\left(\dfrac{\unitNum\outputSize\blockNum\blockSizeOutput}{\antNum}\right)$ & $\mathcal{O}\left(\dfrac{\unitNum\blockSizeOutput}{\antNum\blockSizeWeight\blockSizeInput}\right)$ \\
    ADC & $\mathcal{O}\left(\unitNum\outputSize\blockSizeOutput\right)$ & $\mathcal{O}\left(\dfrac{\unitNum\blockSizeOutput}{\blockNum\blockSizeWeight\blockSizeInput}\right)$ \\
    FFT & $\mathcal{O}\left(\unitNum\outputSize\blockSizeOutput\log_2\blockSizeOutput\right)$ & $\mathcal{O}\left(\dfrac{\unitNum\blockSizeOutput\log_2\blockSizeOutput}{\blockNum\blockSizeWeight\blockSizeInput}\right)$ \\
    Postcoding & $\mathcal{O}\left(\unitNum\outputSize\antNum\blockSizeOutput\right)$ & $\mathcal{O}\left(\dfrac{\unitNum\antNum\blockSizeOutput}{\blockNum\blockSizeWeight\blockSizeInput}\right)$ \\
    Summation & $\mathcal{O}\left(\unitNum\outputSize\blockSizeOutput\right)$ & $\mathcal{O}\left(\dfrac{\unitNum\blockSizeOutput}{\blockNum\blockSizeWeight\blockSizeInput}\right)$ \\
    \bottomrule
    \end{tabular}
    \vspace{-8pt}
\end{table}

Unlike the airtime of {\eqref{eq: bco airtime per flop}}, the block number $\blockNum$ does not cancel here: it amortizes the receiver-side terms while it stretches the IFFT logarithmically, which is the trade-off a concrete block layout has to settle.

\subsection{Airtime of a GEMV}
\label{sec:throughput-analysis}

\myparatight{Airtime per GEMV}
Under the GEMV mapping, $\blockNum = \inputSize$ and $\blockSizeWeight = \blockSizeInput = \blockSizeOutput = 1$, so that $\fftSize = \inputSize$ and each round of {\eqref{eq: bco airtime round}} produces $2\antNum$ IPs.
A GEMV of $\outputSize$ IPs therefore takes $\myCeil{\outputSize/(2\antNum)}$ such rounds in serial, i.e., a signal duration per GEMV of
\begin{align}
    \airtimeMvm \propto \frac{\outputSize\inputSize}{\antNum\band}.
    \label{eq: design airtime mvm}
\end{align}

\myparatight{Airtime per FLOP}
Within this duration, {\name} completes the entire GEMV of $\outputSize(2\inputSize-1)$ FLOPs, so that
\begin{align}
    \airtimeMac \propto \frac{1}{\antNum\band},
    \label{eq: gemv airtime per flop}
\end{align}
\noindent
which is {\eqref{eq: bco airtime per flop}} at unit block sizes.

\myparatight{Airtime scaling law}
Notably, both layer sizes $\inputSize$ and $\outputSize$ cancel out of {\eqref{eq: gemv airtime per flop}}: a larger GEMV proportionally stretches the signal duration and the number of FLOPs completed within it, leaving the per-FLOP airtime unchanged.
What remains is governed solely by the bandwidth $\band$ and the antenna number $\antNum$, independent of the SNR.

\subsection{Energy Consumption of a GEMV}
\label{sec:energy-analysis}

\begin{table}[htbp]
    \centering
    \caption{The energy complexity of one GEMV on {\name} and the same seven terms per FLOP, along the signal chain.}
    \label{tab:energy_breakdown}
    \small
    \setlength{\tabcolsep}{4pt}
    \renewcommand{\arraystretch}{2.0}
    \vspace{-6pt}
    \begin{tabular}{ccc}
    \toprule
    \textbf{Term} & \textbf{Energy per GEMV} & \textbf{Energy per FLOP} \\
    \midrule
    Precoding & $\mathcal{O}\left(\unitNum\inputSize\right)$ & $\mathcal{O}\left(\dfrac{\unitNum}{\outputSize}\right)$ \\
    IFFT & $\mathcal{O}\left(\unitNum\inputSize\log_2\inputSize\right)$ & $\mathcal{O}\left(\dfrac{\unitNum\log_2\inputSize}{\outputSize}\right)$ \\
    DAC & $\mathcal{O}\left(\dfrac{\unitNum\inputSize\outputSize}{\antNum}\right)$ & $\mathcal{O}\left(\dfrac{\unitNum}{\antNum}\right)$ \\
    ADC & $\mathcal{O}\left(\unitNum\outputSize\right)$ & $\mathcal{O}\left(\dfrac{\unitNum}{\inputSize}\right)$ \\
    FFT & 0 & 0 \\
    Postcoding & $\mathcal{O}\left(\unitNum\outputSize\antNum\right)$ & $\mathcal{O}\left(\dfrac{\unitNum\antNum}{\inputSize}\right)$ \\
    Summation & $\mathcal{O}\left(\unitNum\outputSize\right)$ & $\mathcal{O}\left(\dfrac{\unitNum}{\inputSize}\right)$ \\
    \bottomrule
    \end{tabular}
    \vspace{-8pt}
\end{table}

\myparatight{Reading the seven terms at unit block sizes}
Substituting $\blockNum = \inputSize$ and $\blockSizeWeight = \blockSizeInput = \blockSizeOutput = 1$ into the ``Energy per module'' column of Tab.~\ref{tab:bco_energy_breakdown} gives the ``Energy per GEMV'' column of Tab.~\ref{tab:energy_breakdown}, and dividing by the $2\outputSize\inputSize$ FLOPs of the GEMV gives its ``Energy per FLOP'' column.
Two terms are worth reading out at these block sizes.
The ADC is charged for a single sample: the RF-mixer output has been low-pass filtered into a single-subcarrier bandwidth of $\freqSub$, so one complex sample captures the outputs of the whole IP pair by its real and imaginary components.
The FFT term vanishes with $\log_2\blockSizeOutput$, as the FFT that follows the ADC is trivial for its single sample per antenna.

\myparatight{Energy consumption scaling law}
Summing the six remaining per-FLOP terms of Tab.~\ref{tab:energy_breakdown} gives the energy per FLOP of {\name}.
Every term of this per-FLOP energy consumption is amortized by one of the dimensions as the GEMV scales up in LLM:
\begin{itemize}[leftmargin=*, topsep=2pt, itemsep=1pt]
    \item \textit{Amortization over the input size $\inputSize$}: the ADC, postcoding, and summation terms scale as $1/\inputSize$, as one ADC sample squeezes the whole input dimension into a scalar, on which the postcoding and the summation are performed regardless of $\inputSize$.
    \item \textit{Amortization over the output size $\outputSize$}: the precoding and IFFT terms scale as $1/\outputSize$, as the precoded waveform is stored and reused across all the rounds of a GEMV.
    \item \textit{Amortization over the antenna number $\antNum$}: the DAC term scales as $1/\antNum$, as its single waveform is fanned out to all the $\antNum$ RF mixers.
\end{itemize}
Among the three, the DAC term is the bottleneck, since $\antNum \ll \inputSize, \outputSize$ in practice: it is amortized only by how many antennas an edge device can physically deploy, which massive MIMO keeps relieving as it scales up the array.

\section{{\namebf} in CNN}
\label{app:cnn}

Convolutional (CONV) layers, which dominate CNNs, do not fit the one-subcarrier-per-output mapping of the main body: reusing the same kernel across all $\imageSize^2$ positions of an image would require re-transmitting that kernel once per output pixel, so the broadcast weight traffic, the very resource {\name} is built to save, would grow by $\imageSize^2$.
The BCO of Appendix~\ref{app:bco} removes that restriction, as one mixing round now retains a whole output block instead of a single subcarrier, and a kernel broadcast once is correlated against every position of the image within it.
We map a Conv2D layer onto the BCO (\S\ref{app:conv-mapping}), and then specialize the BCO airtime of \S\ref{app:bco-airtime} and the BCO energy of \S\ref{app:bco-energy} to it (\S\ref{app:conv-airtime} and \S\ref{app:conv-energy}).

\subsection{Mapping Conv2D onto BCO}
\label{app:conv-mapping}

\myparatight{Layer formulation}
We examine a general Conv2D layer with a kernel size of $\kernelSize \times \kernelSize$, an image size of $\imageSize \times \imageSize$, $\inputSize$ input channels, and $\outputSize$ output channels, so that $\weightCONV \in \mathbb{R}^{\inputSize \times \outputSize \times \kernelSize \times \kernelSize}$, $\inputCONV \in \mathbb{R}^{\inputSize \times \imageSize \times \imageSize}$, and $\outputCONV \in \mathbb{R}^{\outputSize \times \imageSize \times \imageSize}$.
We assume $\kernelSize$ is odd and that the correlation is zero-padded to preserve the image size.
Just as a GEMV decomposes into $\outputSize$ inner products, this Conv2D decomposes into $\outputSize$ independent sub-convolutions, one per output channel, each of which becomes one BCO; as the layer is real-valued whereas the BCO is complex-valued, two output channels are packed into the real and the imaginary parts of a single BCO, as the two IPs of {\eqref{eq: preliminaries gemv mapping}} are.

\myparatight{Blocks as input channels, subcarriers as pixels}
For each sub-convolution we set $\blockNum = \inputSize$, letting one block correspond to one input channel, so that the block accumulation in {\eqref{eq: preliminaries bco}} performs exactly the summation over input channels.
To embed the 2D image correlation into the 1D BCO, we flatten the $\imageSize \times \imageSize$ image row by row and pad each row with $\kernelSize$ zeros, so that the correlation of one row does not leak into the next; this gives each row a slot of length $\imageSize + \kernelSize$, hence
\begin{align}
    \blockSizeInput = \imageSize (\imageSize + \kernelSize).
    \label{eq: cnn block size input}
\end{align}
The kernel is flattened the same way, row by row, but each of its rows is instead padded with $\imageSize$ zeros so that its row stride matches that of the flattened image, giving
\begin{align}
    \blockSizeWeight = \kernelSize (\imageSize + \kernelSize).
    \label{eq: cnn block size weight}
\end{align}
Following $\blockSizeOutput = \blockSizeWeight + \blockSizeInput - 1$ and approximating away the $-1$, the output block length is
\begin{align}
    \blockSizeOutput \approx (\imageSize + \kernelSize)^2,
    \label{eq: cnn block size output}
\end{align}
out of which the $\imageSize^2$ subcarriers sitting at the valid correlation lags carry $\outputCONV$ of this output channel, and the remainder are discarded together with the LPF-rejected slots.
Repeating this over all $\outputSize$ output channels sweeps $\weightVec_{\outputIdx}$ and $\outputVec_{\outputIdx}$ while sharing the same $\inputVec$, exactly the structure the central radio broadcasts and the $\antNum$ antennas consume in parallel, two output channels per antenna.
Notably, the kernel is broadcast \emph{once} for the whole image: the reuse of a kernel across positions is absorbed by the correlation itself rather than paid for in weight traffic.

\myparatight{Padding overhead}
The $\outputSize$ BCOs nominally complete $2\inputSize\outputSize\kernelSize\imageSize(\imageSize+\kernelSize)^2$ FLOPs, of which only $2\inputSize\outputSize\kernelSize^2\imageSize^2$ are useful to the CONV layer; the rest are correlation taps created by the zero padding.
We denote this inefficiency by the ratio of nominal to useful FLOPs,
\begin{align}
    \convRatio = \frac{(\imageSize+\kernelSize)^2}{\kernelSize\imageSize} \geq 1,
    \label{eq: cnn padding ratio}
\end{align}
which approaches $\imageSize/\kernelSize$ for $\imageSize \gg \kernelSize$.
All per-FLOP metrics below are reported per \emph{useful} FLOP, i.e., scaled by $\convRatio$, so that they are directly comparable with the GEMV results of the main body.

\subsection{Airtime of Conv2D}
\label{app:conv-airtime}

\myparatight{Airtime per FLOP}
Substituting {\eqref{eq: cnn block size input}}--{\eqref{eq: cnn block size output}} into the per-FLOP airtime of a general BCO in {\eqref{eq: bco airtime per flop}}, and scaling by the padding overhead $\convRatio$ of {\eqref{eq: cnn padding ratio}}, yields the airtime per useful FLOP of a Conv2D layer,
\begin{align}
    \airtimeMac_{\textrm{CONV}} \propto \convRatio \cdot \frac{\blockSizeOutput}{\antNum\band\blockSizeWeight\blockSizeInput} = \frac{1}{\antNum\band} \cdot \frac{(\imageSize+\kernelSize)^2}{\kernelSize^2\imageSize^2}.
    \label{eq: cnn airtime per flop conv}
\end{align}

\myparatight{Comparison against GEMV}
Against the GEMV airtime $\airtimeMac_{\textrm{GEMV}}$ of {\eqref{eq: gemv airtime per flop}}, which shares the same proportionality constant, we have
\begin{align}
    \frac{\airtimeMac_{\textrm{CONV}}}{\airtimeMac_{\textrm{GEMV}}} =\frac{(\imageSize+\kernelSize)^2}{\kernelSize^2\imageSize^2} \xrightarrow{~\imageSize \gg \kernelSize~} \frac{1}{\kernelSize^2}.
    \label{eq: cnn airtime ratio}
\end{align}
Interestingly, despite the padding inefficiency $\convRatio$, a Conv2D layer is \emph{faster} per useful FLOP than a GEMV, occupying up to $\kernelSize^2\times$ less airtime.
This is because the BCO's block-wise correlation structure is intrinsically matched to the convolution's reuse of one kernel across all positions of the image: a single OFDM symbol that carries $\kernelSize^2$ kernel taps produces $\imageSize^2$ output pixels, whereas under the GEMV mapping the same symbol duration would produce only one.

\subsection{Energy Consumption of Conv2D}
\label{app:conv-energy}

\begin{table}[htbp]
    \centering
    \caption{The energy complexity of one Conv2D layer on {\name} and the same seven terms per useful FLOP, along the signal chain, assuming $\imageSize \gg \kernelSize$.}
    \label{tab:cnn_energy_breakdown}
    \small
    \setlength{\tabcolsep}{4pt}
    \renewcommand{\arraystretch}{2.0}
    \vspace{-6pt}
    \begin{tabular}{ccc}
    \toprule
    \textbf{Term} & \textbf{Energy per module} & \textbf{Energy per useful FLOP} \\
    \midrule
    Precoding & $\mathcal{O}\left(\unitNum\inputSize\imageSize^2\right)$ & $\mathcal{O}\left(\dfrac{\unitNum}{\outputSize\kernelSize^2}\right)$ \\
    IFFT & $\mathcal{O}\left(\unitNum\inputSize\imageSize^2\log_2(\inputSize\imageSize)\right)$ & $\mathcal{O}\left(\dfrac{\unitNum\log_2(\inputSize\imageSize)}{\outputSize\kernelSize^2}\right)$ \\
    DAC & $\mathcal{O}\left(\dfrac{\unitNum\inputSize\outputSize\imageSize^2}{\antNum}\right)$ & $\mathcal{O}\left(\dfrac{\unitNum}{\antNum\kernelSize^2}\right)$ \\
    ADC & $\mathcal{O}\left(\unitNum\outputSize\imageSize^2\right)$ & $\mathcal{O}\left(\dfrac{\unitNum}{\inputSize\kernelSize^2}\right)$ \\
    FFT & $\mathcal{O}\left(\unitNum\outputSize\imageSize^2\log_2\imageSize\right)$ & $\mathcal{O}\left(\dfrac{\unitNum\log_2\imageSize}{\inputSize\kernelSize^2}\right)$ \\
    Postcoding & $\mathcal{O}\left(\unitNum\outputSize\antNum\imageSize^2\right)$ & $\mathcal{O}\left(\dfrac{\unitNum\antNum}{\inputSize\kernelSize^2}\right)$ \\
    Summation & $\mathcal{O}\left(\unitNum\outputSize\imageSize^2\right)$ & $\mathcal{O}\left(\dfrac{\unitNum}{\inputSize\kernelSize^2}\right)$ \\
    \bottomrule
    \end{tabular}
    \vspace{-8pt}
\end{table}

Assuming $\imageSize \gg \kernelSize$, so that $\blockSizeInput \approx \blockSizeOutput \approx \imageSize^2$ and $\blockSizeWeight \approx \kernelSize\imageSize$ in {\eqref{eq: cnn block size input}}--{\eqref{eq: cnn block size output}}, and substituting the Conv2D mapping ($\blockNum = \inputSize$) into the ``Energy per module'' column of Tab.~\ref{tab:bco_energy_breakdown} gives that of Tab.~\ref{tab:cnn_energy_breakdown}, and dividing by the $2\inputSize\outputSize\kernelSize^2\imageSize^2$ useful FLOPs of the layer gives its ``Energy per useful FLOP'' column.

\myparatight{Takeaways}
Two opposite trends are visible in Tab.~\ref{tab:cnn_energy_breakdown}.
On the one hand, the ADC and the cross-unit-summation terms both carry $1/\inputSize$ and vanish as the layer deepens, exactly as in the GEMV case of Tab.~\ref{tab:energy_breakdown}: one ADC sample and one cross-unit addition now serve an entire input channel of the image rather than a single weight element, which is the saving that the reduced output bandwidth $\band/\blockNum$ buys and that the remote baseline cannot obtain.
The DAC term likewise decays as $1/\kernelSize^2$, mirroring the airtime gain of {\eqref{eq: cnn airtime ratio}}.
On the other hand, the digital encoding and decoding terms do not cancel as cleanly.
Each of them does carry an amortization of its own, the transmitter-side precoding and IFFT over $\outputSize$ and the receiver-side FFT and postcoding over $\inputSize$, so all three dimensions of \S\ref{sec:energy-analysis} survive the Conv2D mapping intact.
What survives with them, however, is a logarithmic factor the GEMV mapping does not have: the IFFT over all $\inputSize\imageSize^2$ subcarriers and the receive FFT over the $\imageSize^2$ subcarriers of a block leave a residual $\log_2(\inputSize\imageSize)$ and $\log_2\imageSize$ that both grow with the image size.
Consequently, unlike the airtime of \S\ref{app:conv-airtime}, where a larger kernel or image is strictly beneficial, the energy of Conv2D trades the analog savings on the converters against the digital cost of the wider OFDM symbol, and the two must be weighed together when choosing {\name}'s operating point.

\section{Experimental Setup}
\label{sec:appendix-implementation}

\begin{figure}[t]
    \centering
    \includegraphics[width=0.6\columnwidth]{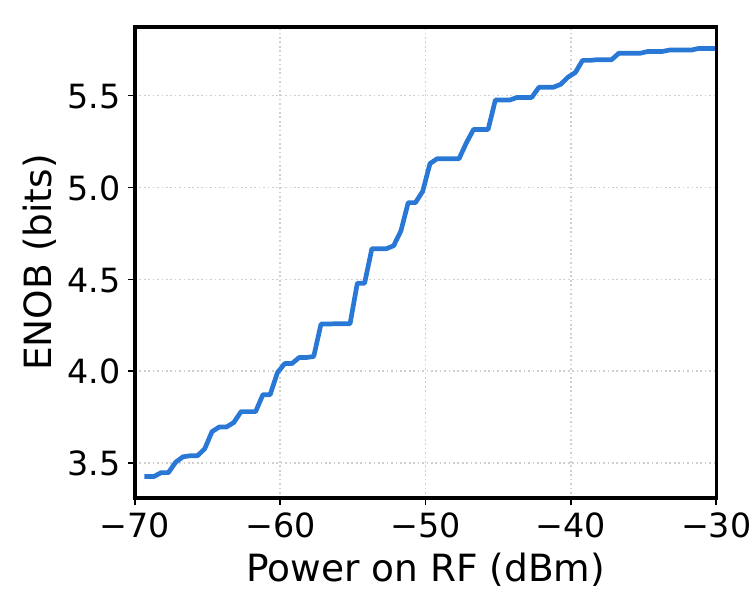}
    \caption{On the wired testbed of Fig.~\ref{fig:scene-map}(c), the best ENOB of the ZEM-4300+ mixer over its RF port power.}
    \vspace{-12pt}
    \label{fig:usrp-rf-power}
\end{figure}

\begin{figure}[t]
    \centering
    \begin{minipage}{0.49\linewidth}
        \centering
        \includegraphics[width=\linewidth]{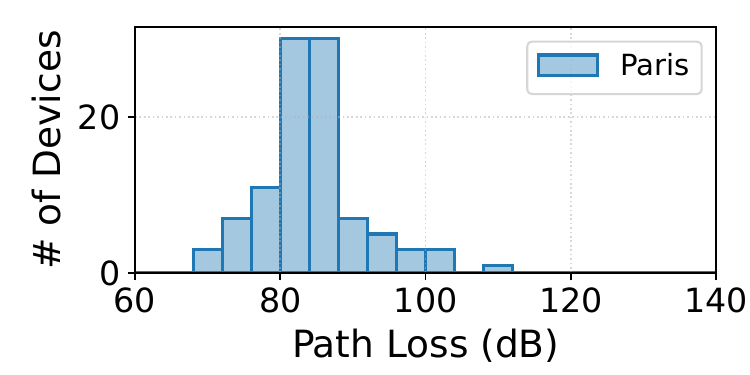}
        \par\vspace{-4pt}
        \subcaption{Paris}
        \vspace{-7pt}
    \end{minipage}
    \hfill
    \begin{minipage}{0.49\linewidth}
        \centering
        \includegraphics[width=\linewidth]{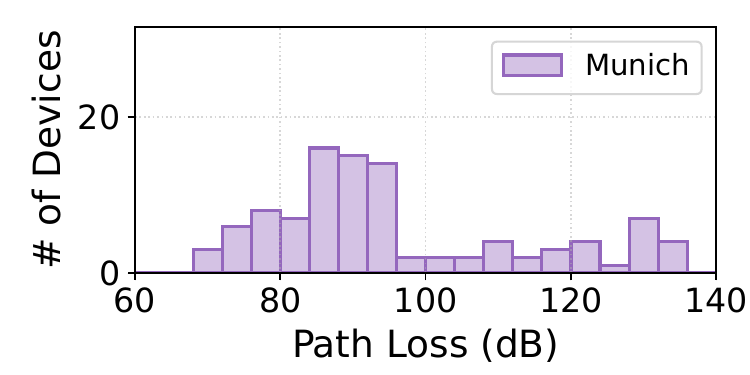}
        \par\vspace{-4pt}
        \subcaption{Munich}
        \vspace{-7pt}
    \end{minipage}
    \caption{Path-loss distribution across the candidate-device pool of each ray-tracing scene.}
    \vspace{-12pt}
    \label{fig:pathloss-hist}
\end{figure}

This section details the baselines, the tasks, and the metrics of the evaluation, together with the two measurements that the implementation of \S\ref{sec:implementation} rests on: the distortion profile of the RF mixer (Fig.~\ref{fig:usrp-rf-power}) and the path-loss distribution of each ray-traced scene (Fig.~\ref{fig:pathloss-hist}).

\subsection{Measured Distortion and Path Loss}
\label{app:setup-measured}

\myparatight{Distortion profile of the RF mixer}
Fig.~\ref{fig:usrp-rf-power} reports the wired sweep of \S\ref{sec:implementation}: at each power delivered to the mixer's RF port, the LO power is swept and the best ENOB is kept.
The ENOB climbs with the RF power and then flattens as the mixer's own nonlinearity takes over from the drive level; this knee is the distortion floor that the saturated users of \S\ref{sec:eval-llm-factor} sit on.

\myparatight{Path loss across the candidate devices}
Fig.~\ref{fig:pathloss-hist} shows the PL distribution of the {100} candidate devices dropped in each scene, from which the representative devices of \S\ref{sec:eval-gemm} are selected.
The open scene of Paris is tightly concentrated, a single mode of median {84.0}\thinspace{dB} with {5/95}-th percentiles of {74.9/97.8}\thinspace{dB}, since almost every drop keeps an LOS path to the central radio.
The dense scene of Munich is both worse and far wider, a median of {91.4}\thinspace{dB} and percentiles of {75.2/131.3}\thinspace{dB}, with a long shadowed tail of devices whose LOS path is blocked by a building.
Both scenes therefore span the range of link qualities from the noise-limited users of \S\ref{sec:eval-llm-factor} to the distortion-limited ones.

\subsection{Baselines}
\label{app:setup-baselines}

We benchmark {\name} against digital baselines along two dimensions: the \emph{precision} at which the weights are kept, and the \emph{weight storage} from which they reach the edge device.

\myparatight{Precision dimension}
We take FP16 as the exact reference, and further consider weight-only quantization (WoQ) at {4/3}\thinspace{bit}, i.e., W4A16 and W3A16, by RTN, GPTQ~\cite{frantar2023gptq}, and AWQ~\cite{lin2024awq}.

\myparatight{Analog computing baselines}
We further benchmark against two analog computing baselines at the same ENOB as {\name}'s device: (\emph{i}) \emph{AWGN}, additive Gaussian noise on each GEMV output, and (\emph{ii}) \emph{HyFlexPIM}, a multiplicative Gaussian error on each weight, i.e., the conductance spread of a memristive crossbar.
The former error is redrawn per GEMV, whereas the latter is frozen on the weights; {\name} sits in between, as its thermal noise is redrawn per transmission while its calibration residual and mixer distortion hold for as long as the channel does.

\myparatight{Weight storage dimension}
Under \emph{remote} storage, the weights are kept on the central radio and delivered to the edge device over the same wireless channel as {\name}'s by a standard communication link.
Under \emph{local} storage, they are instead pre-stored in the edge device's memory.
{\name} targets the remote storage, which is both more realistic for edge computing and more challenging than the local one.

\subsection{Evaluation Tasks and Metrics}
\label{app:setup-metrics}

\myparatight{Benchmarking GEMV with ENOB}
We first benchmark {\name} on a single GEMV $\outputVec = \weightMat \cdot \inputVec$, whose weights $\weightMat$ and inputs $\inputVec$ are randomly drawn from i.i.d.\ Gaussian distributions, and quantify its computing accuracy by the effective number of bits (ENOB), defined as
\begin{align}
    \textsf{ENOB} = -\frac{1}{2} \cdot \log_2\left(\mathbb{E}_{\inputVec, \weightMat}\left[\frac{1}{\outputSize}\sum_{\outputIdx=0}^{\outputSize-1}\left|\outputVec_\outputIdx-\outputVecEst_\outputIdx\right|^2\right]\right),
\end{align}
where both the ground truth $\outputVec$ and its estimate $\outputVecEst$ are normalized by the standard deviation of $\outputVec$.
A trivial estimate $\outputVecEst = \mathbf{0}$ thus sits at exactly {0}\thinspace{bit}.

\myparatight{PPL and zero-shot accuracy on LLM}
We then run whole LLMs over {\name}, including LLaMA-3.2 (1B and 3B)~\cite{meta2024llama32}, LLaMA-3.1 (8B and 70B)~\cite{grattafiori2024llama3}, LLaMA-2 (7B and 13B)~\cite{touvron2023llama2}, Mistral-7B~\cite{jiang2023mistral}, and Gemma-4 (E2B and E4B)~\cite{gemmateam2026gemma4}, where every weight-activation GEMV is performed by {\name}'s RF computing and the rest stays digital.
We report (\emph{i}) the PPL on WikiText-2~\cite{merity2017pointer} and (\emph{ii}) the zero-shot accuracy on PIQA~\cite{bisk2020piqa}, ARC-e and ARC-c~\cite{clark2018arc}, BoolQ~\cite{clark2019boolq}, HellaSwag~\cite{zellers2019hellaswag}, and WinoGrande~\cite{sakaguchi2020winogrande}.

\myparatight{Airtime and energy consumption}
We further report the airtime and the energy consumption per token.
The airtime is the time one token occupies the wireless link, the resource the central radio rations across the edge devices it serves; for the remote storage baselines, it is counted on the same ray-traced channel under a standard-compliant 5G NR link~\cite{3gpp2022ts38214, 3gpp2022ts38306}, while the local storage baselines occupy no airtime.
The energy consumption is based on the per-operation energies of state-of-the-art DACs~\cite{olieman2015dac}, ADCs~\cite{harpe2011adc, murmann2026adcsurvey}, digital computing~\cite{nvidia2024blackwell}, and data movement~\cite{antepara2025energy}.

\section{Supplementary Results}
\label{app:supp-results}

\subsection{Benchmarking on GEMV}
\label{app:gemv}

\begin{figure}[t]
    \centering
    \begin{minipage}{0.49\linewidth}
        \centering
        \includegraphics[width=\linewidth]{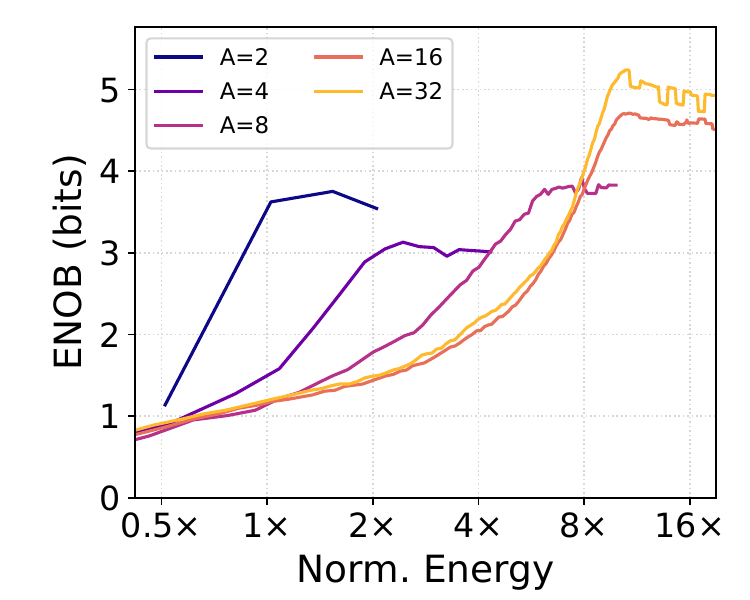}
        \par\vspace{-4pt}
        \subcaption{ENOB vs.\ energy}
        \label{fig:layer-unit-pareto-munich}
        \vspace{-7pt}
    \end{minipage}
    \hfill
    \begin{minipage}{0.49\linewidth}
        \centering
        \includegraphics[width=\linewidth]{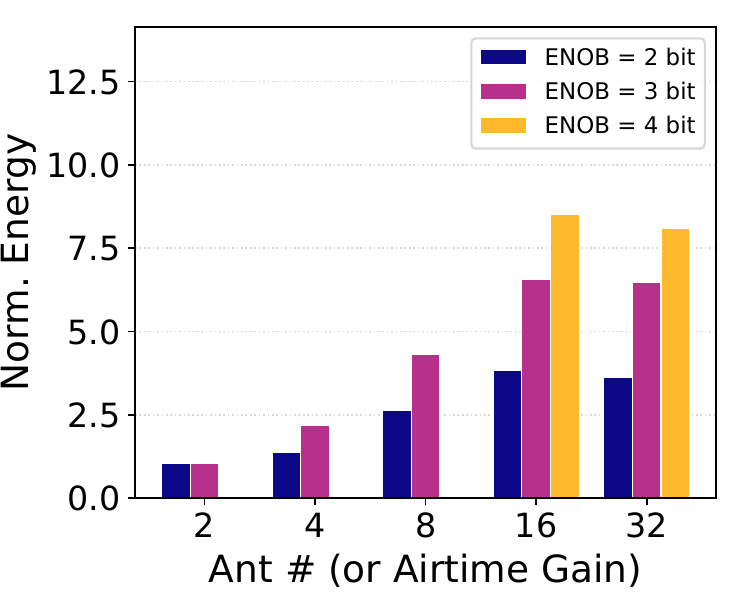}
        \par\vspace{-4pt}
        \subcaption{Energy per ENOB level}
        \label{fig:layer-unit-group-munich}
        \vspace{-7pt}
    \end{minipage}
    \caption{Fig.~\ref{fig:layer-unit-pareto} in Munich: the ENOB over energy (a), and the minimum energy to achieve 2--4 ENOBs (b).}
    \vspace{-12pt}
    \label{fig:layer-unit-munich}
\end{figure}

\myparatight{Trading off accuracy, energy, and airtime in Munich}
Fig.~\ref{fig:layer-unit-munich} repeats the sweep of Fig.~\ref{fig:layer-unit-pareto} on the selected device of Munich, over the same $\antNum = \left\{2, 4, 8, 16, 32\right\}$ at the same paired central radio powers.
The trade-off of \S\ref{sec:eval-gemm} survives the denser scene, with the front sitting lower throughout.
Reading the minimum energy per ENOB level off Fig.~\ref{fig:layer-unit-group-munich}, {2}\thinspace{bit} costs {1.02/1.36/2.62/3.85/3.61}$\times$ and {3}\thinspace{bit} costs {1.02/2.17/4.32/6.55/6.46}$\times$ at $\antNum={2/4/8/16/32}$, against the {1.02/1.90/3.86/5.09/9.60}$\times$ that {3}\thinspace{bit} costs in Paris, and {4}\thinspace{bit} is out of reach below $\antNum={16}$, where it costs {8.52}$\times$ and {8.08}$\times$.
What moves is the ENOB the sweep saturates at: at $\antNum={4}$ this device tops out at {3.13}\thinspace{bit} against {5.50} in Paris, and no number of calibration units lifts the smaller arrays over {4}\thinspace{bit}.
The ordering over $\antNum$ then inverts against Paris.
The saturation ENOB \emph{rises} with the array here, from {3.13}\thinspace{bit} at $\antNum={4}$ to {3.92}, {4.71}, and {5.24}\thinspace{bit} at $\antNum={8/16/32}$, so much so that $\antNum={32}$ reaches {4}\thinspace{bit} more cheaply than $\antNum={16}$ does; in Paris it instead falls from {5.50}\thinspace{bit} at $\antNum={4}$ to {4.33--4.62}\thinspace{bit} over the same three arrays.
The two scenes read one cause from opposite sides: spatial multiplexing needs multipath to multiplex over, so the open scene runs out of usable spatial dimensions as its array grows, while the dense one keeps finding more of them.

\subsection{Benchmarking on LLM}
\label{app:supp-llm}

\begin{figure}[t]
    \centering
    \begin{minipage}{\linewidth}
        \centering
        \includegraphics[width=\linewidth]{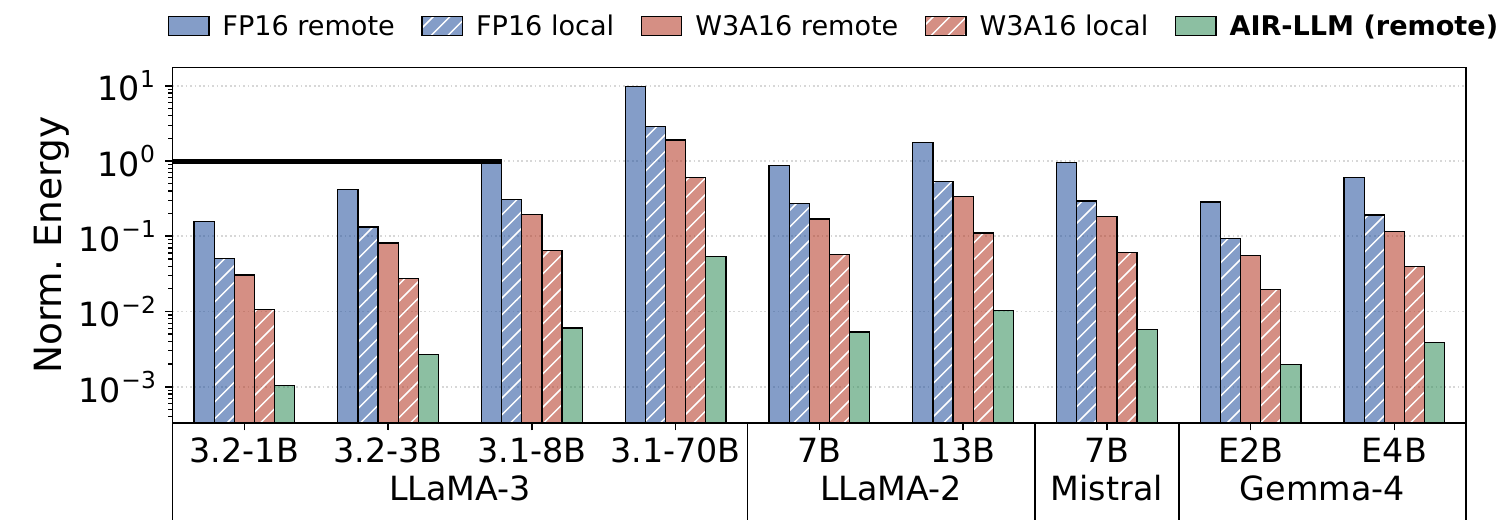}
        \par\vspace{-4pt}
        \subcaption{Energy per token}
        \label{fig:llm-energy-munich}
        \vspace{-7pt}
    \end{minipage}
    \par\vspace{2pt}
    \begin{minipage}{\linewidth}
        \centering
        \includegraphics[width=\linewidth]{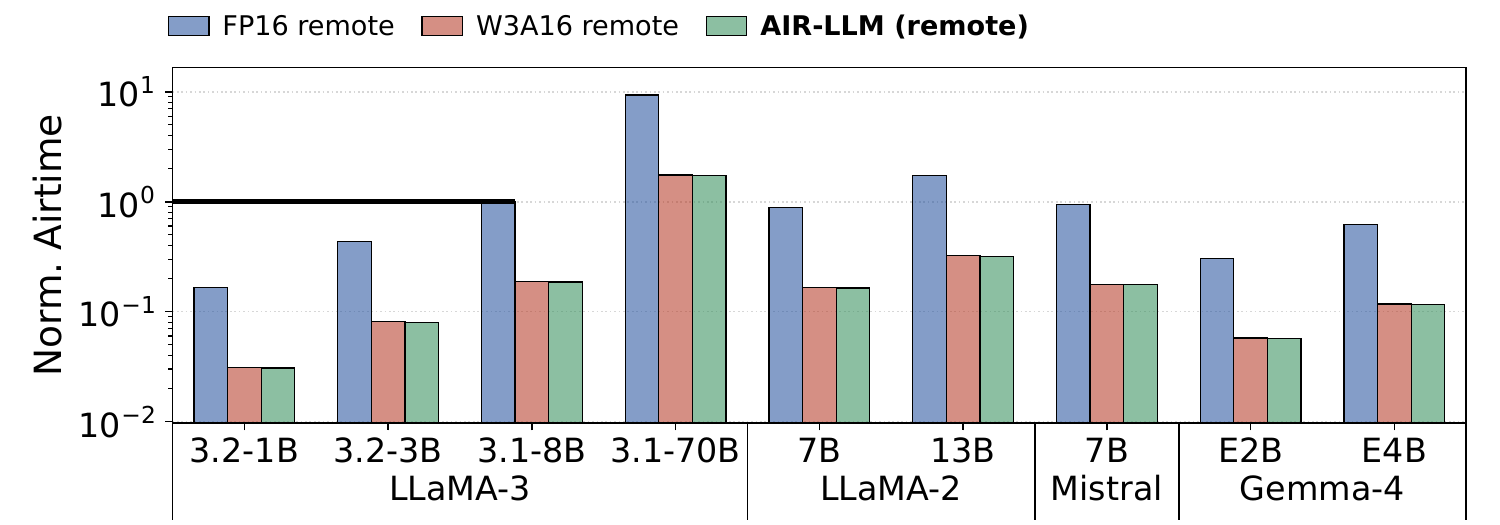}
        \par\vspace{-4pt}
        \subcaption{Airtime per token}
        \label{fig:llm-airtime-munich}
        \vspace{-7pt}
    \end{minipage}
    \caption{Figs.~\ref{fig:llm-energy} and~\ref{fig:llm-airtime} in Munich: the normalized energy (a) and airtime (b) per token over the nine LLMs.}
    \vspace{-12pt}
    \label{fig:llm-munich}
\end{figure}

\begin{figure*}[t]
  \centering
  \begin{minipage}{0.49\linewidth}
      \centering
      \includegraphics[width=\linewidth]{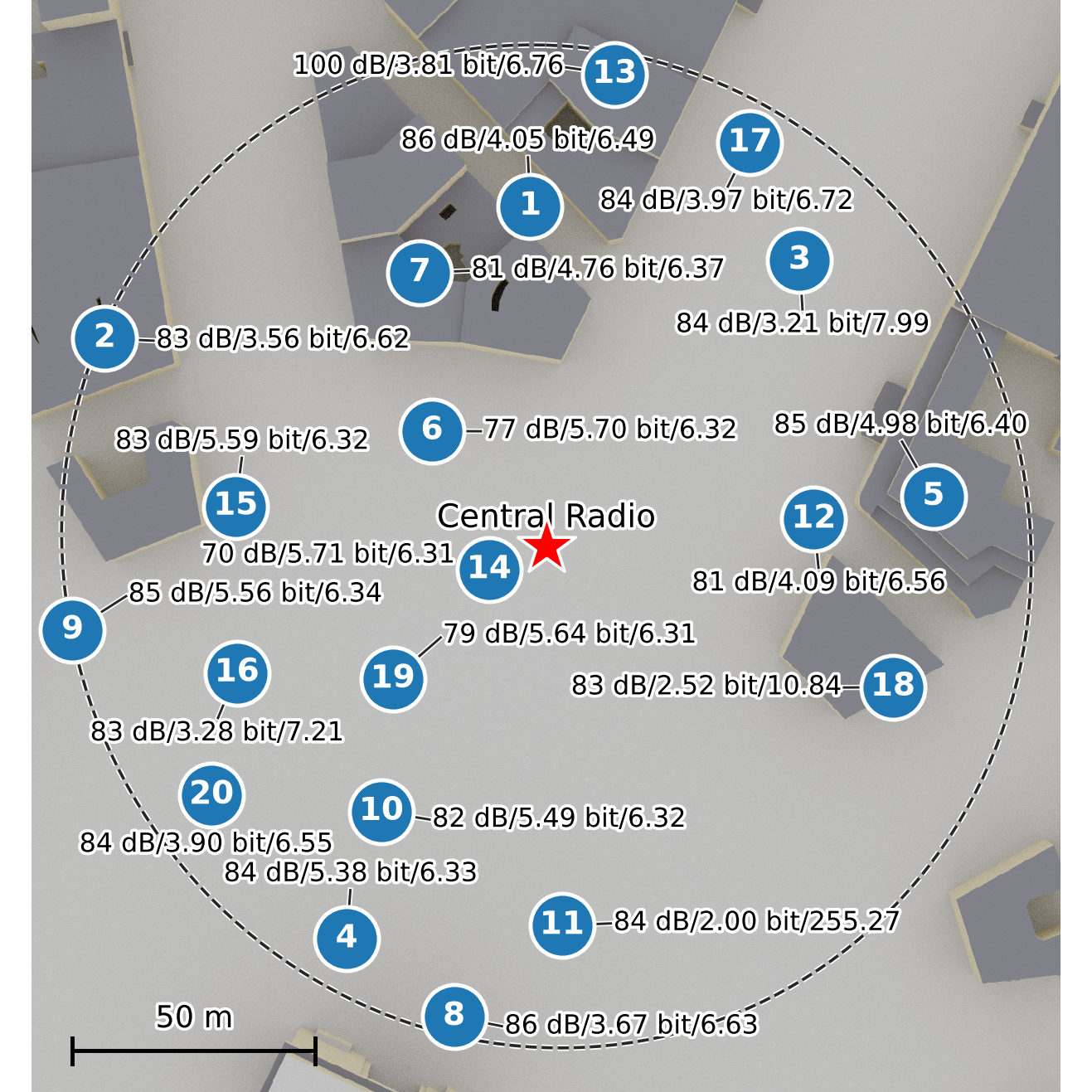}
      \subcaption{Paris}
  \end{minipage}
  \hfill
  \begin{minipage}{0.49\linewidth}
      \centering
      \includegraphics[width=\linewidth]{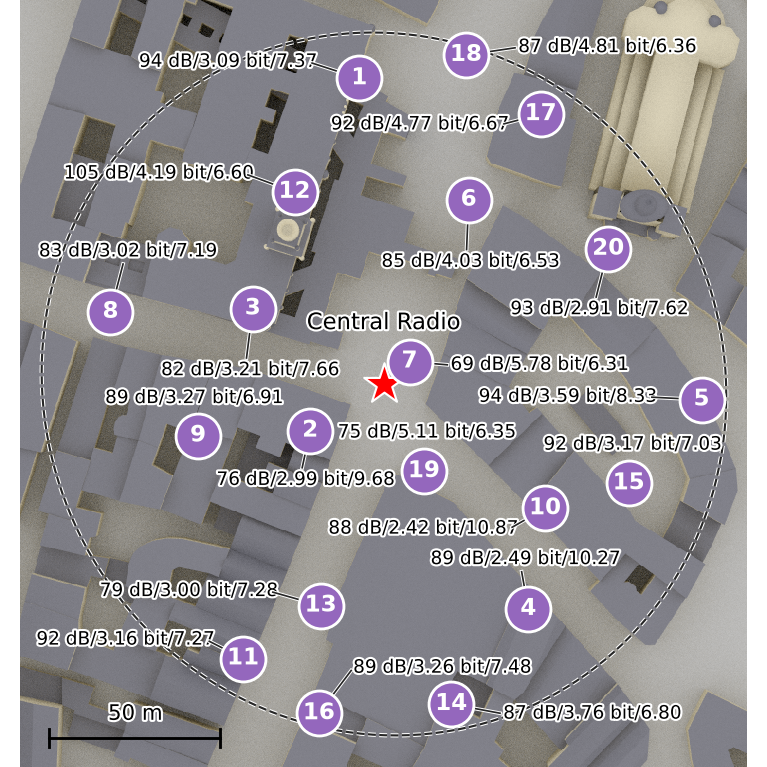}
      \subcaption{Munich}
  \end{minipage}
  \caption{The {20} users of Fig.~\ref{fig:llm-user} in Paris (a) and Munich (b), each labeled with its PL, GEMV ENOB and WikiText-2 PPL.}
  \vspace{-12pt}
  \label{fig:llm-map}
\end{figure*}

\begin{table*}[t]
    \centering
    \caption{What the twenty devices of Paris in Fig.~\ref{fig:llm-map} generate from the shared prompt ``The Place de l'Étoile is a square in Paris, France, best known for\ldots'', which the FP16 model continues as ``\ldots~the Arc de Triomphe, located at its center. The place is named for the star-shaped design of the round''. PL is in dB, ENOB in bits, and PPL is measured on WikiText-2; $\hookleftarrow$ marks a line break inside a continuation.}
    \label{tab:llm-map-etoile}
    \small
    \setlength{\tabcolsep}{3pt}
    \begin{tabular}{cccc@{\hspace{6pt}}l}
    \toprule
    \textbf{Device} & \textbf{PL} & \textbf{ENOB} & \textbf{PPL} & \textbf{Generated continuation} \\
    \midrule
    14 & 70.0 & 5.71 & 6.31 & {\scriptsize \ldots~its statue of the Marianne, a symbol of the French Republic. The square is located in the 16th arr} \\
    6 & 76.6 & 5.70 & 6.32 & {\scriptsize \ldots~its central fountain and the surrounding traffic circle. It is located in the 17th arrondissement, in the north} \\
    19 & 78.8 & 5.64 & 6.31 & {\scriptsize \ldots~its central fountain. The square is located in the 8th arrondissement, at the intersection of the Champs} \\
    15 & 82.5 & 5.59 & 6.32 & {\scriptsize \ldots~its market. It is located in the 18th arrondissement, in the district of Montmartre. It} \\
    9 & 85.4 & 5.56 & 6.34 & {\scriptsize \ldots~the Arc de Triomphe at its center. The name Place de l'Étoile (Star Square) comes} \\
    10 & 82.3 & 5.49 & 6.32 & {\scriptsize \ldots~the fountain in the center, which is the work of the sculptor Louis Derbré. The square is located at} \\
    4 & 84.5 & 5.38 & 6.33 & {\scriptsize \ldots~its statue of the Marianne, a symbol of the French Republic. The square is located in the 16th arr} \\
    5 & 85.5 & 4.98 & 6.40 & {\scriptsize \ldots~its market. It is located in the 18th arrondissement, in the district of Montmartre. It} \\
    7 & 80.8 & 4.76 & 6.37 & {\scriptsize \ldots~its central monument, the Arc de Triomphe, which is located at the center of the Place de l'É} \\
    12 & 81.3 & 4.09 & 6.56 & {\scriptsize \ldots~its central location in the city's 1st arrondissement, and for its metro station, Étoile,} \\
    1 & 86.3 & 4.05 & 6.49 & {\scriptsize \ldots~its fountain, the Fontaine de l'Étoile. The square is located in the 16th arrond} \\
    17 & 84.5 & 3.97 & 6.72 & {\scriptsize \ldots~its landmark, the Arc de Triomphe. It is located at the center of the Place de l'Éto} \\
    20 & 84.5 & 3.90 & 6.55 & {\scriptsize \ldots~the fountain in the middle, the Étoile, which is the center of the square. The square is located in} \\
    13 & 99.9 & 3.81 & 6.76 & {\scriptsize \ldots~its association with the 18th century French Revolution. It is located in the 20th arrondissement of Paris} \\
    8 & 86.0 & 3.67 & 6.63 & {\scriptsize \ldots~its central location and its proximity to the Champs-Élysées. The Place de l'Étoile is} \\
    2 & 83.2 & 3.56 & 6.62 & {\scriptsize \ldots~its equestrian statue. It is located in the 17th arrondissement, at the intersection of the bou} \\
    16 & 82.6 & 3.28 & 7.21 & {\scriptsize \ldots~its central fountain, which is the largest in Paris. The Place de l'Étoile is located in the} \\
    3 & 84.1 & 3.21 & 7.99 & {\scriptsize \ldots~its fountain, which is the oldest in the city. The fountain was built in 1772, and is the oldest} \\
    18 & 82.6 & 2.52 & 10.84 & {\scriptsize \ldots~its shopping mall. $\hookleftarrow$ The Place de l'Étoile is a square in Paris, France, best known for its} \\
    11 & 84.0 & 2.00 & 255.27 & {\scriptsize \ldots~a reason a patient. I a re a piece a piece a patient a a piece a patient a piece a piece a} \\
    \bottomrule
    \end{tabular}
\end{table*}

\begin{table*}[t]
    \centering
    \caption{What the twenty devices of Munich in Fig.~\ref{fig:llm-map} generate from the shared prompt ``Munich is a city in southern Germany, best known for\ldots'', which the FP16 model continues as ``\ldots~its beer and Oktoberfest, the world's largest beer festival. The city is also home to a number of museums,''. PL is in dB, ENOB in bits, and PPL is measured on WikiText-2; $\hookleftarrow$ marks a line break inside a continuation.}
    \label{tab:llm-map-munich}
    \small
    \setlength{\tabcolsep}{3pt}
    \begin{tabular}{cccc@{\hspace{6pt}}l}
    \toprule
    \textbf{Device} & \textbf{PL} & \textbf{ENOB} & \textbf{PPL} & \textbf{Generated continuation} \\
    \midrule
    7 & 68.6 & 5.78 & 6.31 & {\scriptsize \ldots~its annual Christmas market. The city is also known for its beer gardens, which are open year-round. The city is} \\
    19 & 75.4 & 5.11 & 6.35 & {\scriptsize \ldots~its annual Christmas market. It is the capital of the state of Bavaria and is the third largest city in Germany after} \\
    18 & 87.5 & 4.81 & 6.36 & {\scriptsize \ldots~its annual Oktoberfest, a celebration of all things beer. It is also the home of the world's largest beer hall} \\
    17 & 91.7 & 4.77 & 6.67 & {\scriptsize \ldots~its annual Christmas market. The market is a tradition that dates back to the 14th century, and is one of} \\
    12 & 105.0 & 4.19 & 6.60 & {\scriptsize \ldots~its ability to produce beer. The city is the capital of the state of Bavaria. The city is the home of} \\
    6 & 85.0 & 4.03 & 6.53 & {\scriptsize \ldots~its beer and Oktoberfest. It is also the home of the famous Hofbrau brewery, which is the oldest brewery} \\
    14 & 86.9 & 3.76 & 6.80 & {\scriptsize \ldots~its annual Oktoberfest beer festival. It is also home to the world's largest beer hall, the Hofbräu} \\
    5 & 94.3 & 3.59 & 8.33 & {\scriptsize \ldots~its work in the production of the 1984 film The Never Ending Story. The city is the capital of the Land} \\
    9 & 88.8 & 3.27 & 6.91 & {\scriptsize \ldots~its annual Christmas market and its beer and Oktoberfest. It is also the capital of the state of Bavaria and the} \\
    16 & 88.9 & 3.26 & 7.48 & {\scriptsize \ldots~its beautiful old town, which is a UNESCO World Heritage Site. The city is also home to the world's largest beer} \\
    3 & 82.1 & 3.21 & 7.66 & {\scriptsize \ldots~its beer and the Oktoberfest, the world's largest beer festival. But there is much more to the city than just} \\
    15 & 92.0 & 3.17 & 7.03 & {\scriptsize \ldots~being the birthplace of the famous composer Richard Wagner. It is also a major centre for the automotive industry, and is} \\
    11 & 91.8 & 3.16 & 7.27 & {\scriptsize \ldots~its beer gardens and its beer gardens. The city is also known for its beer gardens and its beer gardens. The city} \\
    1 & 94.1 & 3.09 & 7.37 & {\scriptsize \ldots~its beer and its football club. It is also the home of the world's largest beer festival, Oktoberfest. The} \\
    8 & 82.7 & 3.02 & 7.19 & {\scriptsize \ldots~its Oktoberfest. The city is also the home of the FC Bayern Munich soccer team. The team is the most successful} \\
    13 & 78.7 & 3.00 & 7.28 & {\scriptsize \ldots~its annual Oktoberfest, a celebration of beer and all things Bavarian. $\hookleftarrow$ Munich is also the home of FC} \\
    2 & 76.4 & 2.99 & 9.68 & {\scriptsize \ldots~the Volkswagen factory, the University of Munich, the famous opera house, and the Oktoberfest beer festival. The city} \\
    20 & 92.7 & 2.91 & 7.62 & {\scriptsize \ldots~its beer and its annual Oktoberfest. The city is also home to the famous Hofbräuhaus, which is} \\
    4 & 88.5 & 2.49 & 10.27 & {\scriptsize \ldots~its $\hookleftarrow$ munich $\hookleftarrow$ Munich is a city in southern Germany, best known for its $\hookleftarrow$ munich $\hookleftarrow$ M} \\
    10 & 88.2 & 2.42 & 10.87 & {\scriptsize \ldots~its annual beer festival, but also for its many fine museums. The city is also a great place to visit for its} \\
    \bottomrule
    \end{tabular}
\end{table*}

\myparatight{Energy and airtime per token in Munich}
Fig.~\ref{fig:llm-munich} repeats Figs.~\ref{fig:llm-energy} and~\ref{fig:llm-airtime} on the selected device of Munich, whose roughly {3}\thinspace{bit} of end-to-end resolution reads it against the W3A16 baselines.
The energy conclusion of \S\ref{sec:eval-llm-baseline} carries over unchanged: on LLaMA-3.1-8B, {\name} costs {165.5}$\times$ and {32.1}$\times$ less energy than the remote FP16 and W3A16 baselines, and {51.5}$\times$ and {10.7}$\times$ less than the two local ones, holding at {28.3--35.3}$\times$ and {10.0--11.1}$\times$ against the two W3A16 baselines over all nine LLMs.
That the gain against the local FP16 baseline is exactly the {51.5}$\times$ of Paris is no coincidence: {\name}'s per-token energy is set by its own DAC, digital pipeline, and unit number, none of which depends on the channel, and a local baseline pays data movement rather than airtime, so neither side of that ratio moves with the scene.
The airtime does move, and it moves in {\name}'s favor: the remote FP16 baseline takes {5.37--5.42}$\times$ {\name}'s airtime here against {5.11}$\times$ in Paris, since the worse link costs the communication baseline capacity while {\name}'s signal duration is fixed by $\band$ and $\antNum$ alone {\eqref{eq: gemv airtime per flop}}.
The quantized baseline, in contrast, closes to {1.01--1.02}$\times$ from the {1.28}$\times$ that W4A16 leaves in Paris: a quarter fewer bits per weight is enough to offset the capacity its own link costs it.

\myparatight{What each edge device generates}
Fig.~\ref{fig:llm-map} and Tabs.~\ref{tab:llm-map-etoile} and~\ref{tab:llm-map-munich} put the PPLs of \S\ref{sec:eval-llm-user} next to what the model actually writes: each of the {20} devices per scene continues one shared prompt, with every GEMV of LLaMA-3.1-8B carried over that device's own ray-traced channel.
The rows fall by ENOB, and the degradation along them is graded rather than abrupt.
Above about {3.5}\thinspace{bit}, which {16} of the {20} devices reach in Paris and {8} in Munich, the continuations are fluent and name the city's real landmarks, differing from the FP16 reference in which true fact they pick rather than in whether they pick one at all.
Below it they stay grammatical but drift into invented detail and begin to repeat themselves, e.g., device {3} of Paris at {3.21}\thinspace{bit} dates a fountain to {1772} and device {11} of Munich at {3.16}\thinspace{bit} loops on its beer gardens, while their PPLs are still only {7.99} and {7.27}.
Only under about {2.5}\thinspace{bit} do they collapse: device {18} of Paris restarts the prompt verbatim at {2.52}\thinspace{bit}, device {4} of Munich does the same at {2.49}\thinspace{bit}, and device {11} of Paris, the near-singular channel of \S\ref{sec:eval-llm-user}, degenerates into repeated tokens at {2.00}\thinspace{bit} and the outlying PPL of {255.27}.
This is the practical reading of the CDFs in Fig.~\ref{fig:llm-user-cdf-ppl}: {\name} degrades an LLM gracefully over most of a coverage zone, and the devices it fails are the few whose channel is too poorly conditioned to invert rather than the ones that are merely far away: device {12} of Munich, the most distant of all {40} at {105.0}\thinspace{dB}, still resolves {4.19}\thinspace{bit} and reads a PPL of {6.60}.

\begin{table*}[t]
    \centering
    \caption{Over six commonsense-reasoning benchmarks, the zero-shot accuracy (\%) of LLaMA-3.1-8B by {\name} on the selected users of Paris and Munich vs. the FP16, WoQ, and analog computing baselines.}
    \label{tab:zeroshot_acc}
    \small
    \setlength{\tabcolsep}{3pt}
    \vspace{-6pt}
    \begin{tabular}{c|c|cccccc|c}
    \toprule
    \textbf{Precision} & \textbf{Method} & \textbf{PIQA} & \textbf{ARC-e} & \textbf{ARC-c} & \textbf{BoolQ} & \textbf{HellaSwag} & \textbf{WinoGrande} & \textbf{Avg.} \\
    \midrule
    \textbf{FP16} & - & 81.01 & 81.10 & 53.50 & 81.96 & 78.89 & 73.56 & 75.00 \\
    \midrule
    \multirow{3}{*}{\textbf{W4A16}} & RTN & 80.47 & 78.41 & 52.22 & 81.22 & 77.93 & 73.72 & 74.00 \\
    & GPTQ & 80.58 & 78.75 & 51.45 & 80.95 & 78.07 & 72.85 & 73.77 \\
    & AWQ & 80.74 & 80.18 & 52.99 & 81.19 & 78.29 & 73.64 & 74.51 \\
    \midrule
    \multirow{3}{*}{\textbf{ENOB 4}} & AWGN & 52.07 & 27.48 & 23.72 & 47.25 & 27.90 & 49.49 & 37.98 \\
    & HyFlexPIM & 80.79 & 80.51 & 53.16 & 80.92 & 78.37 & 73.72 & 74.58 \\
    & \methodEtoile & \cellEtoile{75.57} & \cellEtoile{72.81} & \cellEtoile{47.18} & \cellEtoile{81.13} & \cellEtoile{77.85} & \cellEtoile{57.77} & \cellEtoile{68.72} \\
    \midrule
    \multirow{3}{*}{\textbf{W3A16}} & RTN & 76.06 & 66.96 & 42.41 & 72.78 & 70.25 & 67.56 & 66.00 \\
    & GPTQ & 76.77 & 69.49 & 42.41 & 75.87 & 74.52 & 72.22 & 68.55 \\
    & AWQ & 77.31 & 74.87 & 46.08 & 77.98 & 72.65 & 70.88 & 69.96 \\
    \midrule
    \multirow{3}{*}{\textbf{ENOB 3}} & AWGN & 50.44 & 26.18 & 24.83 & 40.70 & 26.00 & 49.64 & 36.30 \\
    & HyFlexPIM & 79.65 & 76.47 & 50.51 & 77.25 & 76.49 & 71.67 & 72.01 \\
    & \methodMunich & \cellMunich{73.78} & \cellMunich{68.94} & \cellMunich{43.60} & \cellMunich{75.20} & \cellMunich{75.39} & \cellMunich{57.14} & \cellMunich{65.67} \\
    \bottomrule
    \end{tabular}
    \vspace{-8pt}
\end{table*}

\begin{table}[t]
    \centering
    \caption{The WikiText-2 PPL of LLaMA-3.1-8B when the weight signal is repeated {1} to {16} times and averaged, with $\infty$ as the distortion floor.}
    \label{tab:ppl_repeat}
    \small
    \setlength{\tabcolsep}{3pt}
    \vspace{-6pt}
    \begin{tabular}{c|cccccc}
    \toprule
    \multirow{2}{*}{\textbf{Edge device}} & \multicolumn{6}{c}{\textbf{Repetitions of the weight signal}} \\
    \cline{2-7}
    & 1 & 2 & 4 & 8 & 16 & $\infty$ \\
    \midrule
    \cellEtoile{Paris} & \cellEtoile{6.49} & \cellEtoile{6.49} & \cellEtoile{6.49} & \cellEtoile{6.49} & \cellEtoile{6.49} & \cellEtoile{6.49} \\
    \cellMunich{Munich} & \cellMunich{7.37} & \cellMunich{7.27} & \cellMunich{7.22} & \cellMunich{7.20} & \cellMunich{7.19} & \cellMunich{7.17} \\
    \bottomrule
    \end{tabular}
    \vspace{-8pt}
\end{table}

\begin{table*}[t]
    \centering
    \caption{Tab.~\ref{tab:ppl_txpower} repeated at the ISM ({0.915}\thinspace{GHz}) and mmWave ({28}\thinspace{GHz}) bands, each read against its own regulatory ceiling.}
    \label{tab:ppl_band}
    \small
    \setlength{\tabcolsep}{5pt}
    \vspace{-6pt}
    \begin{tabular}{c|c|ccccccccc}
    \toprule
    \multirow{2}{*}{\textbf{Band}} & \multirow{2}{*}{\textbf{Edge device}} & \multicolumn{9}{c}{\textbf{Central radio transmit power (dBm)}} \\
    \cline{3-11}
    & & 10 & 15 & 20 & 25 & 30 & 35 & 40 & 45 & 50 \\
    \midrule
    \multirow{2}{*}{\textbf{915\thinspace{MHz}}} & \cellEtoile{Paris} & \cellEtoile{9.21} & \cellEtoile{7.56} & \cellEtoile{7.02} & \cellEtoile{6.81} & \cellEtoile{6.72} & \cellEtoile{6.70} & \cellEtoile{6.69} & \cellEtoile{6.69} & \cellEtoile{6.69} \\
    & \cellMunich{Munich} & \cellMunich{$5.9{\times}10^{5}$} & \cellMunich{$3.7{\times}10^{3}$} & \cellMunich{28.14} & \cellMunich{9.15} & \cellMunich{7.28} & \cellMunich{6.74} & \cellMunich{6.54} & \cellMunich{6.47} & \cellMunich{6.44} \\
    \midrule
    & & 60 & 65 & 70 & 75 & 80 & 85 & 90 & 95 & 100 \\
    \midrule
    \multirow{2}{*}{\textbf{28\thinspace{GHz}}} & \cellEtoile{Paris} & \cellEtoile{9.31} & \cellEtoile{7.65} & \cellEtoile{7.22} & \cellEtoile{7.10} & \cellEtoile{7.07} & \cellEtoile{7.06} & \cellEtoile{7.05} & \cellEtoile{7.06} & \cellEtoile{7.06} \\
    & \cellMunich{Munich} & \cellMunich{$3.9{\times}10^{5}$} & \cellMunich{$4.3{\times}10^{5}$} & \cellMunich{$2.9{\times}10^{5}$} & \cellMunich{$3.1{\times}10^{4}$} & \cellMunich{170.87} & \cellMunich{21.48} & \cellMunich{12.59} & \cellMunich{10.76} & \cellMunich{9.98} \\
    \bottomrule
    \end{tabular}
    \vspace{-8pt}
\end{table*}

\begin{table}[t]
    \centering
    \caption{Tab.~\ref{tab:ppl_unit} repeated at the ISM and mmWave bands, each at the transmit power where it saturates in Tab.~\ref{tab:ppl_band}.}
    \label{tab:ppl_band_unit}
    \small
    \setlength{\tabcolsep}{4pt}
    \renewcommand{\arraystretch}{1.1}
    \vspace{-6pt}
    \begin{tabular}{c@{\hspace{6pt}}cc|c@{\hspace{6pt}}cc}
    \toprule
    \multicolumn{3}{c}{\textbf{Ant \# = {4}}} & \multicolumn{3}{c}{\textbf{Ant \# = {8}}} \\
    \cmidrule(r){1-3} \cmidrule(l){4-6}
    \textbf{Unit \#} & \cellEtoile{Paris} & \cellMunich{Munich} & \textbf{Unit \#} & \cellEtoile{Paris} & \cellMunich{Munich} \\
    \midrule
    \multicolumn{6}{c}{\textbf{915\thinspace{MHz}}: {30}\thinspace{dBm} at $\antNum={4}$ and {40}\thinspace{dBm} at $\antNum={8}$} \\
    \midrule
      2 & \cellEtoile{$3.3{\times}10^{5}$} & \cellMunich{$4.2{\times}10^{5}$} &   8 & \cellEtoile{79.04} & \cellMunich{94.78} \\
      4 & \cellEtoile{77.08} & \cellMunich{27.44} &  16 & \cellEtoile{7.25} & \cellMunich{6.51} \\
      6 & \cellEtoile{8.81} & \cellMunich{7.71} &  24 & \cellEtoile{6.43} & \cellMunich{6.54} \\
      8 & \cellEtoile{6.73} & \cellMunich{7.32} &  32 & \cellEtoile{6.32} & \cellMunich{6.62} \\
     12 & \cellEtoile{6.36} & \cellMunich{7.29} &  48 & \cellEtoile{6.32} & \cellMunich{6.68} \\
     16 & \cellEtoile{6.35} & \cellMunich{7.35} &  64 & \cellEtoile{6.32} & \cellMunich{6.72} \\
    \midrule
    \multicolumn{6}{c}{\textbf{28\thinspace{GHz}}: {80}\thinspace{dBm} at $\antNum={4}$ and {90}\thinspace{dBm} at $\antNum={8}$} \\
    \midrule
      2 & \cellEtoile{$2.3{\times}10^{5}$} & \cellMunich{$2.3{\times}10^{5}$} &   8 & \cellEtoile{$7.0{\times}10^{4}$} & \cellMunich{$1.5{\times}10^{5}$} \\
      4 & \cellEtoile{$1.2{\times}10^{5}$} & \cellMunich{$8.8{\times}10^{4}$} &  16 & \cellEtoile{$1.5{\times}10^{3}$} & \cellMunich{$6.0{\times}10^{3}$} \\
      6 & \cellEtoile{115.89} & \cellMunich{$1.7{\times}10^{3}$} &  24 & \cellEtoile{14.35} & \cellMunich{374.36} \\
      8 & \cellEtoile{7.07} & \cellMunich{198.95} &  32 & \cellEtoile{8.59} & \cellMunich{402.40} \\
     12 & \cellEtoile{6.48} & \cellMunich{758.98} &  48 & \cellEtoile{6.86} & \cellMunich{$1.7{\times}10^{3}$} \\
     16 & \cellEtoile{6.35} & \cellMunich{$1.4{\times}10^{3}$} &  64 & \cellEtoile{6.78} & \cellMunich{$4.1{\times}10^{3}$} \\
    \bottomrule
    \end{tabular}
    \vspace{-8pt}
\end{table}

\myparatight{Zero-shot accuracy using LLaMA-3.1-8B}
We evaluate {\name} and the same set of baselines on LLaMA-3.1-8B's inference on six commonsense-reasoning benchmarks, as shown in Tab.~\ref{tab:zeroshot_acc}.
This metric is harsher on analog computing than the PPL, because of how it is scored: every candidate continuation is concatenated to the same stem and run through its own inference pass, and the answer is the candidate with the highest log-likelihood.
An error that is deterministic given the weights, the quantization grid of WoQ or the frozen conductance spread of HyFlexPIM, shifts all the candidates of a question in the same direction, so the comparison among them, and thus the argmax, largely survives.
A non-deterministic one does not: the AWGN baseline and {\name}'s thermal noise are redrawn independently in every pass, so each candidate is scored under its own realization of the error, and the choice flips whenever that perturbation exceeds the log-likelihood margin between the candidates.
AWGN pays for this in full, collapsing to the chance level on all six tasks ({37.98\%} at ENOB {4} and {36.30\%} at ENOB {3}) despite an ENOB matched to {\name}'s, whereas the weight-frozen HyFlexPIM stays within {0.42} points of the FP16 reference at ENOB {4}.
{\name} sits between the two, averaging {68.72\%} in Paris and {65.67\%} in Munich against the {75.00\%} FP16 reference, as only its thermal noise is redrawn per transmission while its channel-calibration residual and the mixer's distortion hold across the candidates of a question: Paris sits inside the W3A16 band ({66.00--69.96\%}) and Munich {0.33} points below its weakest member.
Task by task, the degradation is mild where the margins between the candidate continuations are wide ({77.85\%/75.39\%} on HellaSwag against {78.89\%}, and {81.13\%/75.20\%} on BoolQ against {81.96\%}), and the average is dominated by WinoGrande ({57.77\%/57.14\%} against {73.56\%}), whose two candidates differ in a single pronoun referent and whose log-likelihood margin is therefore the narrowest of the six, so a few bits of analog error are enough to flip the choice.

\subsection{Impact of the System Factors}
\label{app:supp-factors}

\myparatight{Impact of repeating the weight signal}
The central radio may send the same weight signal several times and let the edge device average the copies, which trades airtime for a lower thermal noise: repeating it $n$ times costs $n$ times the airtime and lowers the noise power by $10\log_{10}n$\thinspace{dB}, while leaving the receive chain's own distortion untouched.
Tab.~\ref{tab:ppl_repeat} sweeps this knob under the same LLaMA-3.1-8B setup as Tab.~\ref{tab:ppl_txpower}, where the last column measures the distortion floor that no repetition count can pass.
The Paris user does not move at all, staying at {6.49} from a single copy to the floor, which confirms the same conclusion as its flat row in Tab.~\ref{tab:ppl_txpower}: this user is already limited by the channel calibration and the RF mixer rather than by the noise.
The noise-limited Munich user does improve, but slowly: {16} copies, i.e., {16} times the airtime, buy {7.37} to {7.19}, only {2.4\%} and already within {0.3\%} of its {7.17} floor.
Repetition is therefore a much worse deal than the transmit power for the same user, which buys {7.9\%} over the same {10}\thinspace{dB} in Tab.~\ref{tab:ppl_txpower} at no airtime cost, and we keep a single copy of the weight signal throughout the other evaluations.

\myparatight{Impact of the carrier frequency}
We repeat the transmit power sweep of Tab.~\ref{tab:ppl_txpower} in two other bands, the {915}\thinspace{MHz} ISM band over {25}\thinspace{MHz} and the {28}\thinspace{GHz} mmWave band over {400}\thinspace{MHz}, on the same two edge devices at the same positions and under the same LLaMA-3.1-8B setup, as shown in Tab.~\ref{tab:ppl_band}.
The channels are re-traced at each carrier with the same scenes, whose ITU-R P.2040 material models are extrapolated below their {1}\thinspace{GHz} lower edge for {915}\thinspace{MHz}, and the noise floor follows the bandwidth of each band.
Since the path loss of the two devices drops by {12}\thinspace{dB} at {915}\thinspace{MHz} and grows by {32}\thinspace{dB} (Paris) and {41}\thinspace{dB} (Munich) at {28}\thinspace{GHz}, the power axis is shifted accordingly, with the regulatory ceilings of {36}\thinspace{dBm} EIRP in the ISM band and {75}\thinspace{dBm} per {100}\thinspace{MHz} at {28}\thinspace{GHz} inside the swept range.
In the ISM band, both devices reach the accuracy of the paper's operating point at {20}\thinspace{dB} less power: Munich reads {7.28} at {30}\thinspace{dBm} against the {7.37} that Tab.~\ref{tab:ppl_txpower} measures at {50}\thinspace{dBm}, while Paris has already saturated at {6.72}.
Both therefore stay usable inside the regulatory ceiling, reading {6.70} and {6.74} at {35}\thinspace{dBm}, and the band's own floors ({6.69} in Paris and {6.44} in Munich) even reverse the two devices' order, the lower carrier reaching the more distant device better than the nearer one.
What the band costs is airtime: its {25}\thinspace{MHz} is a quarter of the bandwidth $\band$ that {\eqref{eq: design airtime mvm}} divides, so every GEMV occupies four times the air.

At {28}\thinspace{GHz} the trade reverses.
Paris needs {80}\thinspace{dBm} to saturate, {30}\thinspace{dB} more than at {3.5}\thinspace{GHz}, and saturates {9\%} worse, at a floor of {7.05} against {6.48}; this gap is a distortion floor that no transmit power removes, the sparser multipath of the band leaving each per-subcarrier channel matrix worse conditioned for the calibration units to invert.
Munich, another {17}\thinspace{dB} of path loss away (a gap the band itself widens, from {7.9}\thinspace{dB} at {3.5}\thinspace{GHz} to {16.7}\thinspace{dB} here), never becomes usable at all: {170.87} at the regulatory ceiling and still {9.98} at {100}\thinspace{dBm}, i.e., {19}\thinspace{dB} beyond what the band is allowed to radiate.
Its {400}\thinspace{MHz} does cut the airtime fourfold, but only for a device close enough to the central radio to spend it.
Across the three bands, therefore, the carrier frequency trades airtime against coverage almost one for one, and {\name} is the more comfortable the lower the band, the opposite of where the spectrum for it is easiest to find.

\myparatight{Calibration units in the other bands}
Tab.~\ref{tab:ppl_band_unit} repeats the antenna and calibration unit sweep of Tab.~\ref{tab:ppl_unit} in both other bands, on the same two devices and with the same two antenna blocks.
Each block is read at the transmit power where its own band saturates these devices, taken off Tab.~\ref{tab:ppl_band} rather than carried over from the {3.5}\thinspace{GHz} table, so that the unit count is again the only quantity the table varies: {30}\thinspace{dBm} at $\antNum={4}$ and {40}\thinspace{dBm} at $\antNum={8}$ in the ISM band, and {80}\thinspace{dBm} and {90}\thinspace{dBm} at {28}\thinspace{GHz}, each following the same $+{10}$\thinspace{dB} step between the two antenna blocks that the {3.5}\thinspace{GHz} table uses.

\subsection{Broadcasting to Massive Edge Users}
\label{app:supp-users}

\begin{figure}[t]
    \centering
    \includegraphics[width=\linewidth]{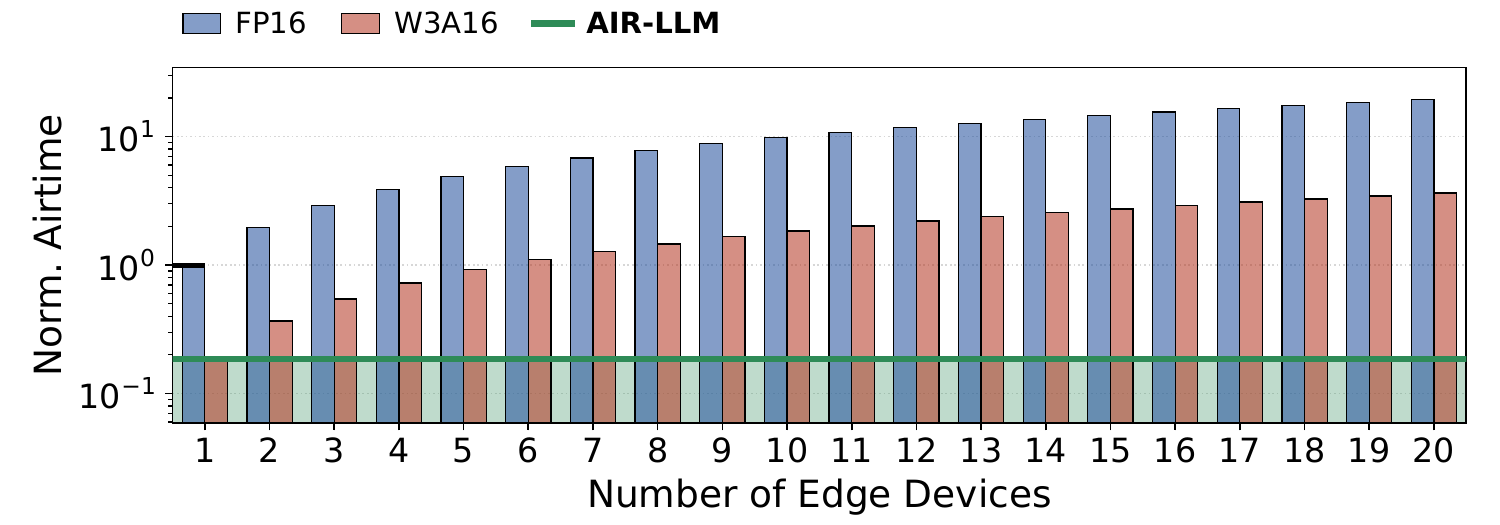}
    \caption{Fig.~\ref{fig:llm-user-airtime} in Munich, against the same {3}-bit baseline.}
    \vspace{-12pt}
    \label{fig:llm-user-airtime-munich}
\end{figure}

\myparatight{Airtime over the edge users in Munich}
Fig.~\ref{fig:llm-user-airtime-munich} repeats Fig.~\ref{fig:llm-user-airtime} on the {20} devices of Munich, against that scene's W3A16 baseline.
The picture is the one of \S\ref{sec:eval-llm-user}, if anything sharper: the unicast baselines still grow linearly in the user number $\userNum$ while {\name}'s single broadcast stays flat, so at $\userNum={20}$ it takes {104.2}$\times$ and {19.5}$\times$ less airtime than the FP16 and the W3A16 baselines.
Note that the gain against FP16 is the same {104}$\times$ as in Paris even though every individual link here is worse: both sides of that ratio are read on the same channels, so the scene cancels out of it, and what remains is the $\userNum$-fold duplication that a unicast pays and a broadcast does not.
The gain against the quantized baseline is smaller than the {26.0}$\times$ that W4A16 leaves in Paris only because W3A16 carries a quarter fewer bits per weight, and not because the broadcast advantage has weakened.

\clearpage

\end{document}